\documentclass[aps,prd,twocolumn,superscriptaddress,nofootinbib,floatfix,longbibliography]{revtex4-2}

\usepackage{amsmath,amssymb}
\usepackage{graphicx}
\graphicspath{{Figures/}}
\usepackage{bm}
\usepackage{comment}
\usepackage[colorlinks=true,allcolors=blue]{hyperref}
\usepackage{placeins}

\newcommand{\zGB}{\zeta_{\rm GB}}
\newcommand{\zCS}{\zeta_{\rm CS}}
\newcommand{\aGB}{\alpha_{\rm GB}}
\newcommand{\aCS}{\alpha_{\rm CS}}
\newcommand{\GB}{\mathcal{G}}
\newcommand{\pRR}{{}^{\ast}\!RR}
\newcommand{\Oone}{\mathcal{O}(\zGB\,\zCS)}

\begin{document}

\title{Mixed-Order Parity-Violating Corrections to Rotating Black Holes in
Combined Scalar Gauss--Bonnet and Dynamical Chern--Simons Gravity}

\author{Abhishek Rout}
\email{arout@mailbox.sc.edu}

\author{Brett Altschul}
\email{altschul@mailbox.sc.edu}

\affiliation{Department of Physics and Astronomy,\\
University of South Carolina, Columbia, South Carolina 29208, USA}

\date{\today}

\begin{abstract}
We construct the leading mixed-order $\Oone$ deformation of rotating black holes
in a combined scalar Gauss--Bonnet and dynamical Chern--Simons effective
theory, the sector in which the parity-even and parity-odd corrections act
simultaneously. Using a pseudospectral solution of the coupled mixed scalar and
metric equations, we compute horizon and equatorial-geodesic observables---the
horizon angular velocity, surface gravity, area, and Wald entropy; the innermost
stable circular orbit and photon-ring radii; and photon-ring eikonal estimates of
the ringdown---at fixed mass and angular momentum over the spin range
$0.1\le a\le0.9$, where the solution is numerically convergent and the
perturbative expansion is controlled for sufficiently small dimensionless
couplings. We verify the exact Kerr and Schwarzschild limits, reproduce available
pure-sector benchmarks, and quantify spectral, residual, and extraction
uncertainties. The mixed solution satisfies every component of the field
equations, including the one that is not imposed in its construction, and its
Wald entropy satisfies the first law to about one part in $10^{5}$. The mixed
coefficients exhibit spin dependence qualitatively distinct from the individual
sectors within this EFT truncation. All four horizon quantities, and several
orbital and eikonal ones, change sign as functions of spin. The ringdown results
are leading photon-ring eikonal estimates; a full coupled finite-$\ell$
perturbation calculation is left to future work.
\end{abstract}

\maketitle

\section{Introduction}
\label{sec:intro}

The direct detection of gravitational waves from compact binary
coalescences~\cite{GW150914,GWTC3} has opened the strong-field, highly dynamical
regime of gravity to precision observation. Together with very-long-baseline
imaging of supermassive black holes~\cite{EHT2019}, these observations provide
increasingly stringent tests of general relativity (GR) in the vicinity of
black hole horizons~\cite{LIGOtestGR,Berti2015}. A central question in this
program is whether the Kerr geometry---the unique stationary, axisymmetric,
vacuum solution of GR---receives corrections from new physics at the curvature
scales probed by these systems.

In the gravitational-wave sector, the post-merger ringdown offers a
particularly clean probe; black hole spectroscopy---measurements of multiple
quasinormal frequencies of the remnant---tests the no-hair structure of the Kerr
geometry directly~\cite{Dreyer2004,BertiCardosoWill2006}, and analyses of
GW150914 and subsequent events have begun to realize this
program~\cite{Isi2019}. In the electromagnetic sector, horizon-scale imaging of
the photon ring by the Event Horizon Telescope provides a complementary,
strong-field test of the Kerr hypothesis~\cite{Psaltis2019}.

A broad and theoretically-well-motivated class of such corrections arises in
effective field theories (EFTs) that augment the Einstein--Hilbert action with
higher-curvature scalars coupled to light dynamical fields~\cite{YunesYagiPretorius2016}.
Two curvature-squared invariants play distinguished roles in four dimensions.
The Gauss--Bonnet invariant
$\GB = R_{\mu\nu\rho\sigma}R^{\mu\nu\rho\sigma}
- 4R_{\mu\nu}R^{\mu\nu} + R^2$
is parity even, and when coupled to a dynamical scalar (dilaton) field defines
scalar Gauss--Bonnet (sGB) gravity~\cite{Kanti1996,SotiriouZhou2014}. The
gravitational Pontryagin density
$\pRR = \tfrac12\,\epsilon^{\mu\nu\rho\sigma}
R_{\mu\nu\alpha\beta}R_{\rho\sigma}{}^{\alpha\beta}$
is parity odd, and when coupled to a dynamical pseudoscalar (axion) defines
dynamical Chern--Simons (dCS) gravity~\cite{JackiwPi2003,AlexanderYunes2009}.
Both invariants are topological on their own and therefore modify the field
equations only through their couplings to the spacetime-dependent spinless fields. Crucially,
each theory admits black hole solutions that carry scalar hair and deviate from the
Kerr form, and both can arise naturally in the low-energy limit of heterotic string
compactifications~\cite{Gross1987,Campbell1992}, where the dilaton and axion
appear together.

Because the curvature invariants in the sGB and dCS sectors have opposite parities, they leave complementary
imprints on black hole spacetimes. The sGB correction modifies the
even-parity metric structure and sources monopolar scalar hair even for
non-rotating black holes; the dCS correction alters the
(gravitomagnetic) sector and requires rotation to source its pseudoscalar field.
These effects have been studied extensively in the cases where only a single
new term is present in the action---both using the slow-rotation expansion~\cite{Yunes2009dCS,Pani2011}
and, more recently, for rapidly rotating black holes via spectral
methods~\cite{LamChungYunes2025}. The latter work constructs rotating sGB and
dCS black holes to leading order in each coupling across nearly the full range
of possible spins and provides high-accuracy metric functions and observables that we will
take as our benchmarks.

However, comparatively little attention has been paid to the regime in which
both couplings are present simultaneously. This is the physically generic
situation---as string-inspired actions can contain the dilaton and axion fields simultaneously---and
it is qualitatively richer than the sum of its parts. At leading order the two
sectors are independent; the total correction to any observable is simply the
sum of its sGB and dCS pieces, each linear in the respective coupling. The first
genuinely new physics appears at the \emph{mixed} order $\Oone$, quadratic in
the couplings and bilinear in the two parity channels. At this order the
two types of deformations interfere; the metric correction
$h^{(1,1)}_{\mu\nu}$ is sourced by cross terms between the sGB and dCS fields and
cannot be obtained from either pure sector. Observables built from
$h^{(1,1)}$ therefore vanish identically unless both couplings are
nonzero, making them a clean, mixed-coupling diagnostic within the combined sGB--dCS effective theory for the simultaneous
presence of both types of corrections.

This paper will be concerned with computations in this mixed sector. Building on the spectral framework
of Ref.~\cite{LamChungYunes2025}, we solve the coupled scalar and linearized
metric equations to second order in the couplings, obtaining the scalar
fields and the metric correction $h^{(1,1)}_{\mu\nu}$ for rotating black
holes across $0.1 \le a \le 0.9$. From these we extract a comprehensive set of horizon and orbital observables:
the horizon angular velocity $\Omega_H^{(1,1)}$ and surface gravity
$\kappa^{(1,1)}$; the horizon area $A^{(1,1)}$ and Wald entropy $S^{(1,1)}$; the
innermost stable circular orbit (ISCO) $r_{\rm ISCO}^{(1,1)}$ and photon-ring radius shifts
$r_{\rm ph}^{(1,1)}$ for prograde and retrograde orbits;
and the leading (eikonal) quasinormal-mode (QNM) frequency and damping-rate corrections
$\Omega_c^{(1,1)}$ and $\lambda^{(1,1)}$. The surface gravity is obtained from a
covariant first-principles construction, the entropy from the Wald
formula~\cite{Wald1993,IyerWald1994}, and the ringdown observables follow
from the Cardoso--Berti correspondence between the photon ring and the eikonal
quasinormal frequencies~\cite{Cardoso2009,Berti2009}. The QNM spectrum of
these modified-gravity backgrounds is itself under
active development; the scalar and gravitational perturbation spectra of rotating
sGB and dCS black holes have recently been computed on spectrally constructed backgrounds~\cite{ChungLamYunes2025},
and the quasinormal spectra of black holes more generally have a long history as probes of strong-field
gravity~\cite{KonoplyaZhidenko2011}. Our eikonal estimate provides
the leading large-$\ell$ contribution to this spectrum in the mixed sector, prior to a full coupled perturbation calculation.

Our construction sits within a broader landscape of second-order, beyond-Kerr
studies, from which it is worth distinguishing it. Extensions that are quadratic in the spin
have been built for the dCS sector in the slow-rotation
expansion~\cite{YagiYunesTanaka2012}, capturing the second order in the spin
but within a single parity sector; our expansion is instead second order in the
\emph{couplings} and valid at arbitrary spin. In the ringdown context, the
parametrized QNM framework of McManus,~\textit{et al.} treats precisely
the coupled, second-order-in-coupling corrections that arise when distinct
perturbation sectors mix~\cite{McManus2019}, and QNM corrections in
higher-derivative gravity have been computed for rotating
backgrounds~\cite{Cano2022}; these address the perturbation spectrum on a
\emph{fixed} background, complementary to our construction of the corrected
\emph{background} itself, whose mixed sector itself supplies the leading input to such a
spectrum. Finally, fully nonperturbative rotating solutions in the pure sGB
sector~\cite{KleihausKunzRadu2011} mark the endpoint of the same coupling
expansion in the regime where it ceases to converge---the near-extremal,
strong-coupling limit we discuss in section~\ref{sec:eftcontrol}.

Our central physical result is that the mixed-order corrections are
often qualitatively different from those in either pure sector. The
mixed-order corrections to the horizon angular velocity and the surface gravity
change sign as functions of spin---near $a\approx0.61$ and $a\approx0.71$,
respectively---at spins where neither pure-sector correction does. The reversal
of $\Omega_H^{(1,1)}$ is driven by the linear response to $h^{(1,1)}$, which
changes sign near $a\approx0.57$. The horizon area and the
Wald entropy reverse as well, near $a\approx0.89$ and $a\approx0.76$. These
spin-dependent sign changes are a a distinctive fingerprint of the simultaneous presence of the two couplings.
In the ringdown calculations, we find that the mixing raises the prograde
QNM frequency for $a\gtrsim0.23$ and lengthens its damping time for
$a\gtrsim0.65$, with a pronounced prograde-retrograde asymmetry---an effect of potentially
direct relevance to ringdown tests of GR.

We place particular emphasis on validation, since the mixed sector has no
preexisting benchmarks. Every observable is checked against its exact Kerr limit
(Bardeen ISCO and photon-ring radii, the Gauss--Bonnet topological invariant,
the Schwarzschild eikonal frequency), against pure-sector values from
Ref.~\cite{LamChungYunes2025} where available, and for truncation and
resolution convergence. The mixed solution itself is tested in three ways that
need no benchmark: the $r\chi$-component of the field equations, which is not
imposed, is satisfied to the same accuracy as the components that are;
$\Omega_H$ and $\kappa$ are constant over the horizon; and the Wald entropy
agrees with the value required by the first law of black hole thermodynamics. We further identify and resolve
a severe ill-conditioning of the mixed-order metric system at high spin,
which we trace to the disparity in the magnitudes of the collocated equations and cure by equilibration of
the spectral operator; this is essential for obtaining reliable near-extremal
results.

The remainder of this paper is organized as follows. Section~\ref{sec:theory}
presents the action, the coupling bookkeeping, and the perturbative expansion.
Section~\ref{sec:method} describes the spectral method, the two-stage solution methodology,
the conditioning fix, and our validation methods.
Sections~\ref{sec:horizon}, \ref{sec:orbital}, and~\ref{sec:qnm} present the
horizon, orbital, and ringdown observables, respectively; the Wald entropy and
the first law are treated in Sec.~\ref{sec:entropy}.
Section~\ref{sec:discussion} discusses the physical interpretation of the sign
changes, the approach to the extremal limit, and the error budget. We conclude
in Sec.~\ref{sec:conclusion} with a summary and a discussion of future
directions, including the full coupled quasinormal spectrum. Throughout we adopt
the metric signature $(-,+,+,+)$ and geometrized units $G=c=1$.

\section{Theoretical Framework}
\label{sec:theory}

\subsection{Action and field equations}
\label{sec:action}

We consider the four-dimensional effective theory in which the
Einstein--Hilbert term is supplemented by the two independent
curvature-squared invariants, each coupled to a dynamical field. The
parity-even dilaton $\varphi$ couples to the Gauss--Bonnet invariant and the
parity-odd axion $\vartheta$ couples to the gravitational Pontryagin density.
The action is~\cite{LamChungYunes2025}
\begin{equation}
\begin{aligned}
    S = \frac{1}{16\pi}\!\int\! d^4x\,&\sqrt{-g}
\big( R - \tfrac12\nabla_\mu\varphi\nabla^\mu\varphi
- \tfrac12\nabla_\mu\vartheta\nabla^\mu\vartheta\\
&+ \aGB\,\varphi\,\GB + \aCS\,\vartheta\,\pRR \big) + S_{\rm mat},
\end{aligned}
\label{eq:action}
\end{equation}
where
\begin{align}
\GB &= R_{\mu\nu\rho\sigma}R^{\mu\nu\rho\sigma}
- 4R_{\mu\nu}R^{\mu\nu} + R^2, \\
\pRR &= {}^{\ast}\!R^{\mu\nu\rho\sigma}R_{\mu\nu\rho\sigma}
\equiv \tfrac12\,\epsilon^{\rho\sigma\alpha\beta}
R^{\mu\nu}{}_{\alpha\beta}R_{\mu\nu\rho\sigma},
\label{eq:dualriemann}
\end{align}
$\epsilon^{\mu\nu\rho\sigma}$ is the Levi--Civita tensor, and $\aGB,\aCS$ are
coupling constants of dimension $(\mathrm{length})^2$. With every term placed
under the common prefactor $(16\pi)^{-1}$, the overall factor cancels from all
field equations, leaving unit coefficients on the curvature-coupling tensors, a
factor $\tfrac12$ on the scalar stress-energy, and source-normalized scalar
wave equations. In four dimensions both $\GB$ and $\pRR$ are topological when
they appear without multiplicative couplings to scalars, and they therefore contribute to the dynamics
only through their couplings to $\varphi$ and $\vartheta$.

Varying eq.~\eqref{eq:action} with respect to $g^{\mu\nu}$, $\varphi$, and
$\vartheta$ yields the coupled system
\begin{align}
G_{\mu\nu} + \aGB\,\mathcal{H}_{\mu\nu} + \aCS\,\mathcal{C}_{\mu\nu}
&= \tfrac12\, T^{(\varphi)}_{\mu\nu} + \tfrac12\, T^{(\vartheta)}_{\mu\nu},
\label{eq:einstein}\\
\Box\varphi &= -\,\aGB\,\GB,
\label{eq:phiEOM}\\
\Box\vartheta &= -\,\aCS\,\pRR,
\label{eq:thetaEOM}
\end{align}
where $\mathcal{H}_{\mu\nu}$ is the tensor obtained from the variation of
$\varphi\,\GB$, $\mathcal{C}_{\mu\nu}$ is the parity-odd Cotton-type tensor from
the variation of $\vartheta\,\pRR$~\cite{JackiwPi2003,AlexanderYunes2009}, and
$T^{(\varphi)}_{\mu\nu}$ and $T^{(\vartheta)}_{\mu\nu}$ are the canonical scalar
stress-energy tensors. The scalar equations~\eqref{eq:phiEOM} and
\eqref{eq:thetaEOM} show that $\varphi$ is sourced by the (parity-even)
Gauss--Bonnet density and $\vartheta$ by the (parity-odd) Pontryagin density; in
particular $\pRR$ vanishes for a non-rotating black hole, so $\vartheta$---and
hence the entire dCS sector---requires rotation to be excited.

\subsection{Coupling hierarchy and order counting}
\label{sec:counting}

Because the invariants enter eqs.~\eqref{eq:phiEOM} and \eqref{eq:thetaEOM} at
linear order in the couplings, the scalar fields are first order in $\aGB$ and
$\aCS$: $\varphi = \mathcal{O}(\aGB)$ and $\vartheta = \mathcal{O}(\aCS)$.
The metric back-reaction in eq.~\eqref{eq:einstein} is then quadratic in
the couplings; the scalar stress-energy is $\mathcal{O}(\aGB^2)$ or
$\mathcal{O}(\aCS^2)$, and the curvature-coupling tensors
$\aGB\mathcal{H}_{\mu\nu}$, $\aCS\mathcal{C}_{\mu\nu}$ are of the same
order once the leading scalar profiles are inserted. It is therefore natural to
organize the metric expansion according to the dimensionless parameters
\begin{equation}
\zGB \equiv \frac{\aGB^2}{M^4}, \qquad \zCS \equiv \frac{\aCS^2}{M^4},
\label{eq:zetadef}
\end{equation}
so that the leading metric corrections are $\mathcal{O}(\zGB)$ and
$\mathcal{O}(\zCS)$. In this bookkeeping the scalars carry half-integer powers,
$\varphi = \mathcal{O}(\zGB^{1/2})$ and $\vartheta = \mathcal{O}(\zCS^{1/2})$.
Throughout we label a quantity's order by the pair of powers $(m,n)$ of $(\zGB,\zCS)$ it carries;
the leading scalars are thus $\varphi^{(1/2,0)}$ and $\vartheta^{(0,1/2)}$, while the leading
metric corrections are $h^{(1,0)}$ (sGB) and $h^{(0,1)}$ (dCS). In the numerical calculations we set
$M = 1$, so the dimensionless parameters reduce numerically to
$\zeta_{\rm GB} = \alpha^2_{\rm GB}$ and $\zeta_{\rm CS} = \alpha^2_{\rm CS}$.

\subsection{The mixed sector}
\label{sec:mixed}

Expanding all fields around the Kerr solution $g^{(0)}_{\mu\nu}$,
\begin{align}
g_{\mu\nu} &= g^{(0)}_{\mu\nu} + \zGB\, h^{(1,0)}_{\mu\nu}
+ \zCS\, h^{(0,1)}_{\mu\nu} + \zGB\zCS\, h^{(1,1)}_{\mu\nu} + \dots,
\label{eq:gexp}\\
\varphi &= \zGB^{1/2}\big[\varphi^{(1/2,0)} + \zCS\,\varphi^{(1/2,1)} + \dots\big],\\
\vartheta &= \zCS^{1/2}\big[\vartheta^{(0,1/2)} + \zGB\,\vartheta^{(1,1/2)} + \dots\big],
\end{align}
and inserting these into eqs.~(\ref{eq:einstein}--\ref{eq:thetaEOM}), one
obtains a hierarchy of linear problems ordered by $(m,n)$. At leading order the
two sectors decouple: $h^{(1,0)}$ sourced entirely by the sGB fields and
$h^{(0,1)}$ entirely by the dCS fields, reproducing the pure-sector solutions of
Ref.~\cite{LamChungYunes2025}. Any observable evaluated to
$\mathcal{O}(\zGB) + \mathcal{O}(\zCS)$ is simply the sum of its independent sGB
and dCS pieces.

The lowest order at which the two parity channels communicate is the mixed order
$\Oone$. Collecting all terms proportional to $\zGB\zCS$ in
eq.~\eqref{eq:einstein} yields a linear equation for the mixed metric
correction $h^{(1,1)}_{\mu\nu}$,
\begin{equation}
\mathcal{E}\!\left[h^{(1,1)}\right]_{\mu\nu} = S^{(1,1)}_{\mu\nu},
\label{eq:mixedEQ}
\end{equation}
where $\mathcal{E}$ is the linearized Einstein operator on the Kerr background,
and the source $S^{(1,1)}_{\mu\nu}$ is built from bilinear cross terms between
the sGB and dCS pure-sector fields together with the mixed scalars
$\varphi^{(1/2,1)}$ and $\vartheta^{(1,1/2)}$. Schematically,
\begin{equation}
\begin{aligned}
    S^{(1,1)} \sim & \,\,\tfrac12 T^{(1,1)} - \aGB\mathcal{H}^{(1,1)}
- \aCS\mathcal{C}^{(1,1)}\\
&- 2\,G^{(2)}\!\big[h^{(1,0)},h^{(0,1)}\big],
\end{aligned}
\label{eq:source}
\end{equation}
where $G^{(2)}$ denotes the second-order Einstein tensor evaluated on the
product of the two pure-sector metric corrections. The mixed scalars themselves
obey linear equations sourced by the pure-sector fields, e.g.\
$\Box^{(0)}\varphi^{(1/2,1)}$ is driven by the dCS deformation of the
Gauss--Bonnet density.

Two features of eq.~\eqref{eq:mixedEQ} are central to what follows. First,
$h^{(1,1)}_{\mu\nu}$ cannot be obtained from either pure sector; it represents genuinely
new information requiring the simultaneous solution of both. Second, any
observable constructed from $h^{(1,1)}$ vanishes identically unless
both couplings are nonzero, since it is proportional to $\zGB\zCS$. The
mixed-order observables computed in this work are therefore a diagnostic of
the simultaneous presence of both couplings
within the combined sGB--dCS effective theory.

\subsection{Metric ansatz and boundary conditions}
\label{sec:ansatz}

At each order the metric correction is parametrized by a set of scalar functions
of the Boyer--Lindquist-type coordinates $(r,\chi)$, with $\chi = \cos\theta$.
Following Ref.~\cite{LamChungYunes2025} we write the correction in the form
\begin{align}
h_{\mu\nu}\,dx^\mu dx^\nu
&= H_1\,dt^2
- 2H_2\,\frac{2 a r}{\Sigma}(1-\chi^2)\,dt\,d\phi \nonumber\\
&\quad + H_3\,\frac{\Sigma}{\Delta}\,dr^2
+ H_3\,\frac{\Sigma}{1-\chi^2}\,d\chi^2 \nonumber\\
&\quad + H_4\,(1-\chi^2)\!\Big[r^2+a^2 \nonumber\\
&\qquad +\frac{2a^2 r}{\Sigma}(1-\chi^2)\Big]d\phi^2.
\label{eq:ansatz}
\end{align}
Here, $\Sigma = r^2 + a^2\chi^2$, $\Delta = r^2 - 2r + a^2$ (with $M=1$), and the
four functions $H_i(r,\chi)$ carry the appropriate order labels $(m,n)$.
Under reflection through the equatorial plane ($\chi\to-\chi$) both coupling terms
in the action~\eqref{eq:action} are even: the dilaton coupling $\varphi\,\GB$
because $\varphi$ and $\GB$ are separately even, and the axion coupling
$\vartheta\,\pRR$ because the pseudoscalar $\vartheta$ and the parity-odd
Pontryagin density $\pRR$ are \emph{each} odd, so that their product is even. The
metric corrections are therefore even in $\chi$, and the circularity of the
spacetime (discussed in appendix~\ref{app:gauge-circ}) is preserved at each order,
so that the $tr$-, $t\chi$-, $r\phi$-, and
$\chi\phi$-components vanish identically. The remaining off-diagonal component,
$h_{r\chi}$, is set to zero by a choice of gauge.
(See appendix~\ref{app:gauge} for further details.)

The functions $H_i$ are fixed by regularity at the horizon $r=r_+$ together with
asymptotic (``peeling'') boundary conditions at spatial infinity. The latter are
imposed so that the Arnowitt--Deser--Misner (ADM) mass $M$ and angular momentum
$J$ are unchanged by the corrections at every order; that is, the
deformations redistribute the interior structure of the black hole at fixed
asymptotic charges. This choice fixes the physical meaning of the corrections
and is essential for the first law discussed in section~\ref{sec:entropy}. Writing
$H_i = H_{i,\infty} + H_{i,1}/r + \mathcal{O}(r^{-2})$ at large $r$, asymptotic
flatness requires $H_{1,\infty}=0$ and $H_{4,\infty}=H_{3,\infty}$, and the
charges of the corrected metric are then
\begin{equation}
\begin{aligned}
M_{\rm ADM} &= \sqrt{1+H_{3,\infty}}
\left[M + \frac{H_{3,1}}{2\,(1+H_{3,\infty})}\right],\\
J_{\rm ADM} &= \sqrt{1+H_{3,\infty}}\,\big(1+H_{2,\infty}\big)\, J,
\end{aligned}
\label{eq:adm}
\end{equation}
where the square roots arise from the rescaling of $r$ to the areal radius. At
first order in either coupling, $M_{\rm ADM}=M$ and $J_{\rm ADM}=J$ reduce to the
conditions of Ref.~\cite{LamChungYunes2025}, $H_{1,\infty}=0$,
$H_{2,\infty}=H_{3,1}/(2M)$, and $H_{3,\infty}=H_{4,\infty}=-H_{3,1}/M$. The
pure-sector solutions have $H_{3,\infty}\neq0$, so at mixed order
eq.~\eqref{eq:adm} gives the inhomogeneous conditions,
\begin{equation}
\begin{aligned}
H^{(1,1)}_{1,\infty} &= 0, \\
H^{(1,1)}_{2,\infty} - \frac{H^{(1,1)}_{3,1}}{2M} &= c_\infty, \\
H^{(1,1)}_{3,\infty} + \frac{H^{(1,1)}_{3,1}}{M} &= -\frac{c_\infty}{2}, \\
H^{(1,1)}_{4,\infty} + \frac{H^{(1,1)}_{3,1}}{M} &= -\frac{c_\infty}{2},
\end{aligned}
\label{eq:peeling11}
\end{equation}
with $c_\infty \equiv H^{(1,0)}_{3,\infty}H^{(0,1)}_{3,\infty}$. Imposing the
homogeneous first-order conditions on $h^{(1,1)}_{\mu\nu}$ instead would give a
black hole whose mass and angular momentum are shifted at $\Oone$ by
$\delta M/M = c_\infty/4$ and $\delta J/J = -3c_\infty/4$. The spectral
representation of the $H_i$ is described in Sec.~\ref{sec:method}.

\section{Numerical Method}
\label{sec:method}

\subsection{Spectral representation}
\label{sec:spectral}

We solve the field equations with a pseudospectral collocation method. The
radial domain $r\in[r_+,\infty)$ is compactified through
\begin{equation}
z = \frac{2r_+}{r} - 1 \in [-1,+1],
\label{eq:zmap}
\end{equation}
which maps the horizon $r=r_+$ to $z=+1$ and spatial infinity to $z=-1$. Each
metric function $H_i(r,\chi)$ and each scalar field is expanded in Chebyshev
polynomials $T_j(z)$ along the compactified radius and Legendre polynomials
$P_\ell(\chi)$ in the polar variable $\chi=\cos\theta$,
\begin{equation}
H_i(z,\chi) = \sum_{n=0}^{N_z}\sum_{\ell=0}^{N_\chi}
c^{(i)}_{j\ell}\, T_j(z)\, P_\ell(\chi).
\label{eq:spectral}
\end{equation}
As established following eq.~\eqref{eq:ansatz}, the corrections are even in
$\chi$ under equatorial reflection, restricting the expansion to even $\ell$ and
halving the number of angular modes. Radial and angular derivatives are computed analytically from the
spectral coefficients via the Chebyshev and Legendre recurrence relations, so
that no finite-difference errors enter the differential operators. The equations are imposed at the
Chebyshev--Gauss--Lobatto nodes in both $z$ and
$\chi$. The exponential convergence of the expansion allows a
modest resolution to resolve the smooth corrections; we use $N_z=N_\chi=30$
throughout, and $N_z=N_\chi=50$ for the convergence tests of
section~\ref{sec:validation}. All computations are performed in IEEE double-precision arithmetic, for which
the unit round-off is $\epsilon_{\rm mach}\approx2.2\times10^{-16}$. Where we
state below that a quantity vanishes at the level of round-off, we mean at this
level multiplied by the amplification of the differentiation operators involved.

\subsection{Pure-sector inputs and conventions}
\label{sec:inputs}

The mixed-order construction takes the leading pure-sector solutions as fixed
inputs: the metric deformation functions $H_i^{(1,0)}$ (sGB) and $H_i^{(0,1)}$
(dCS), and the leading scalars $\varphi^{(1/2,0)}$ and $\vartheta^{(0,1/2)}$.
These are obtained from the spectral solutions of Ref.~\cite{LamChungYunes2025},
generated by that work's independent Mathematica pipeline and exported as the
values of the eight metric functions and two scalars at the
collocation nodes. We import these node values directly and
fit them to the spectral representation~\eqref{eq:spectral}; no regeneration,
re-solution, or interpolation onto a different grid is performed, so the
pure-sector inputs are used exactly as provided, to the accuracy
($\lesssim10^{-8}$) at which Ref.~\cite{LamChungYunes2025} reports them.

Because these inputs are used directly, our calculation necessarily adopts the
conventions of Ref.~\cite{LamChungYunes2025}, which we summarize here for
completeness and reproducibility:
\begin{itemize}
\item \emph{Signature and units:} metric signature $(-,+,+,+)$, geometrized
units $G=c=1$, and mass normalization $M=1$, so that the dimensionless spin
$a$ coincides with the Kerr parameter and the horizon lies at
$r_+ = 1+\sqrt{1-a^2}$.
\item \emph{Coordinates:} Boyer--Lindquist-type coordinates
$(t,r,\chi,\phi)$ with $\chi=\cos\theta$, compactified radially through
$z = 2r_+/r - 1 \in[-1,+1]$.
\item \emph{Curvature:} Riemann tensor
$R^{\rho}{}_{\sigma\mu\nu} = \partial_\mu\Gamma^{\rho}{}_{\nu\sigma}
- \partial_\nu\Gamma^{\rho}{}_{\mu\sigma} + \dots$, Ricci tensor
$R_{\mu\nu}=R^{\rho}{}_{\mu\rho\nu}$, and the dual Riemann tensor defined with
the Levi--Civita \emph{tensor} $\epsilon^{\mu\nu\rho\sigma}=
[\mu\nu\rho\sigma]/\sqrt{-g}$ as in eq.~\eqref{eq:dualriemann}.
\item \emph{Action normalization:} the common prefactor $(16\pi)^{-1}$ of
eq.~\eqref{eq:action}, canonical scalar kinetic terms
$-\tfrac12(\nabla\varphi)^2-\tfrac12(\nabla\vartheta)^2$, and linear scalar
couplings $\alpha_{\rm GB}\varphi\,\GB$ and $\alpha_{\rm CS}\vartheta\,\pRR$.
\item \emph{Coupling normalization:} the dimensionless expansion parameters
$\zGB=\alpha_{\rm GB}^2/M^4$ and $\zCS=\alpha_{\rm CS}^2/M^4$ of
eq.~\eqref{eq:zetadef}.
\end{itemize}
The mutual consistency of these conventions between the imported inputs and our
own curvature and field-equation machinery is not merely assumed; it is verified
by the pure-sector field-equation residuals of appendix~\ref{app:exp-audit},
which are $\sim\!10^{-11}$ of the correction tensors themselves for
$a\le0.76$. A mismatch in any of
the sign or normalization conventions above---for instance in the dual-Riemann
or Levi--Civita definition---would produce a residual of order unity, so this
check simultaneously validates the input provenance and the convention alignment.

\subsection{Two-stage solution}
\label{sec:stages}

The mixed-order problem separates into two linear stages. In \emph{Stage~I} the
mixed scalar fields $\varphi^{(1/2,1)}$ and $\vartheta^{(1,1/2)}$ are obtained
by inverting the scalar wave operator $\Box^{(0)}$ on the Kerr background,
\begin{equation}
\Box^{(0)}\varphi^{(1/2,1)} = \mathcal{S}_\varphi, \qquad
\Box^{(0)}\vartheta^{(1,1/2)} = \mathcal{S}_\vartheta,
\end{equation}
where the sources $\mathcal{S}_\varphi$ and $\mathcal{S}_\vartheta$ are built from
the pure-sector fields---$\mathcal{S}_\varphi$ from the dCS deformation of the
Gauss--Bonnet density and of the wave operator acting on $\varphi^{(1/2,0)}$,
and $\mathcal{S}_\vartheta$ from the corresponding sGB deformations of the
Pontryagin density and of the wave operator acting on $\vartheta^{(0,1/2)}$.

\begin{figure}[htbp]
\centering
\includegraphics[width=\columnwidth]{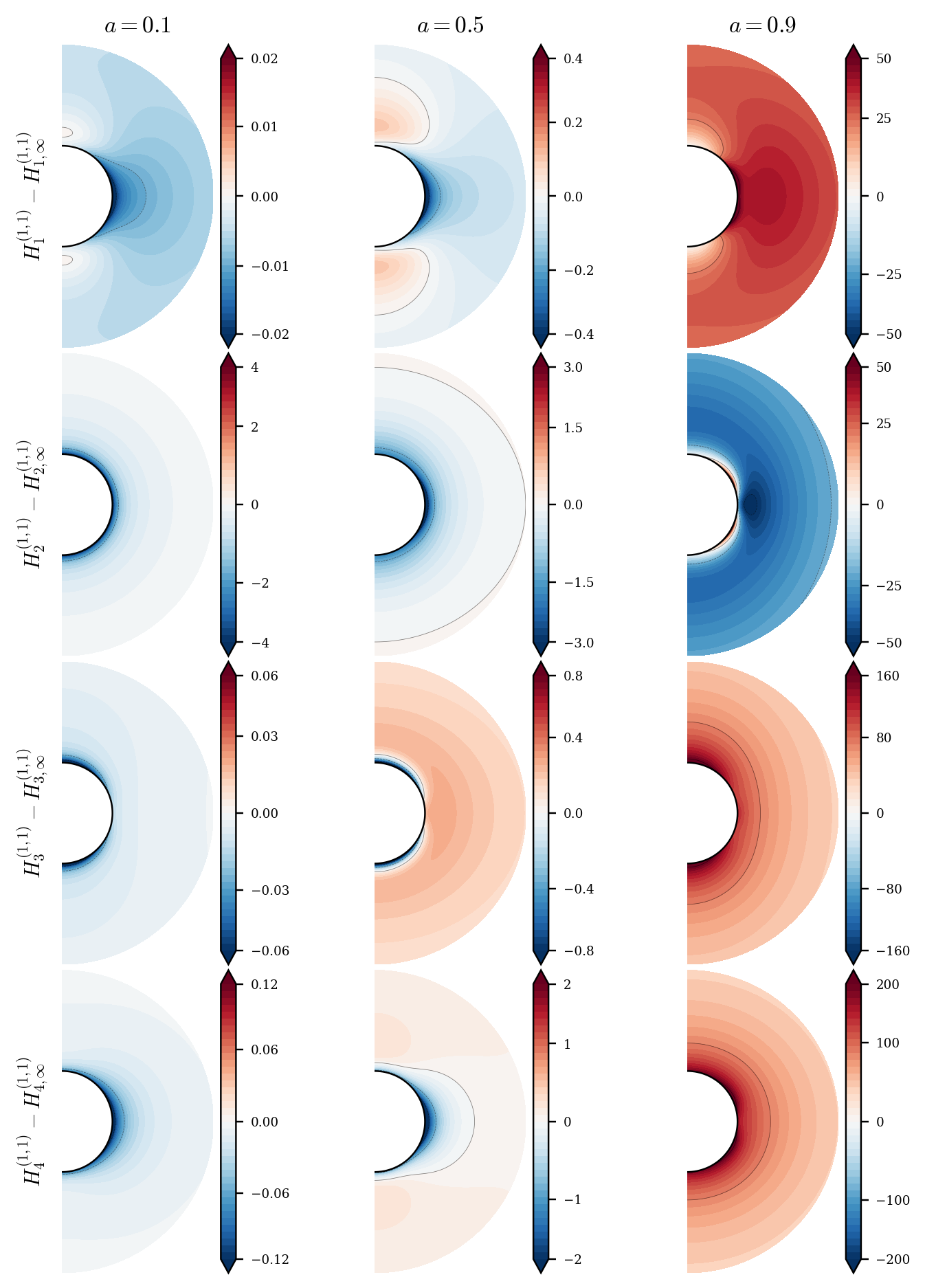}
\caption{Meridional cross-sections of the mixed-order metric functions with
their asymptotic values subtracted, $H_i^{(1,1)}-H_{i,\infty}^{(1,1)}$, for
spins $a=0.1$, $0.5$, $0.9$ (left to right), from the horizon (solid line) to
$3r_+$; the rotation axis is vertical. $H_{1,\infty}^{(1,1)}$ vanishes, and the
other three constants are fixed by Eq.~\eqref{eq:peeling11}. Each panel carries
its own colour scale, set by the 99th percentile of the plotted field; the
amplitude grows by one to three orders of magnitude across the range. Thin
contours mark zero and half the colour scale. The $H_i$ are coordinate
dependent; the physical content of $h^{(1,1)}_{\mu\nu}$ is carried by the
observables discussed in sections~\ref{sec:horizon}--\ref{sec:qnm}.}
\label{fig:meridional}
\end{figure}

In \emph{Stage~II}
the mixed metric correction $h^{(1,1)}_{\mu\nu}$ is obtained by solving the
linearized Einstein equation~\eqref{eq:mixedEQ},
\begin{equation}
\mathcal{E}\!\left[h^{(1,1)}\right]_{\mu\nu} = S^{(1,1)}_{\mu\nu},
\end{equation}
with the source~\eqref{eq:source} assembled from the pure-sector metric
functions $h^{(1,0)}$, $h^{(0,1)}$ and the scalars from Stage I. The pure-sector
inputs and their associated scalars are taken from the leading-order spectral solutions of
Ref.~\cite{LamChungYunes2025}. Figure~\ref{fig:meridional} shows the resulting
mixed-order metric functions at $a=0.1$, $0.5$, and $0.9$.

Substituting the expansion~\eqref{eq:spectral} into each stage converts the
differential equations into a linear system for the spectral coefficients.
Writing the vector of unknown coefficients as $\mathbf{c}$, the collocated
system takes the form $\mathbf{A}\mathbf{c} = \mathbf{b}$, where $\mathbf{A}$
encodes the operator evaluated at the collocation nodes together with the
boundary rows, and $\mathbf{b}$ contains the source and boundary data, including
the right-hand sides of eq.~\eqref{eq:peeling11}. The
system is overdetermined once the horizon regularity and asymptotic
peeling conditions from section~\ref{sec:ansatz} are appended, and is solved in the
least-squares sense. The five even-parity components
$(tt,\,t\phi,\,rr,\,\chi\chi$, and $\phi\phi)$ of eq.~\eqref{eq:mixedEQ} are
collocated and solved simultaneously. The $r\chi$-component is not collocated;
it follows from the other five through the contracted Bianchi identity, provided
the source is covariantly conserved, and we use it as an independent check of
the solution. (See section~\ref{sec:validation}.)

\subsection{Conditioning at high spin}
\label{sec:conditioning}

The Stage II system is well behaved at low and moderate spin but becomes
severely ill-conditioned as $a\to1$. The source $S^{(1,1)}_{\mu\nu}$ grows
sharply near the horizon---becauase the metric perturbation contains factors of
$\Sigma/\Delta$ that diverge as $\Delta(r_+)\to0$---so the collocation rows
nearest $z=+1$ have magnitudes many orders larger than those in the bulk. The
condition number of the naive operator $\mathbf{A}$ reaches $\kappa(\mathbf{A})
\sim 10^{14}$ by $a=0.9$, at the edge of double-precision resolvability. In this
regime the least-squares solution is dominated by the near-horizon rows and is
unreliable. It leaves the field equations at large radius unsatisfied, and the
extracted mixed observables are wrong by factors of order unity
($\Omega_H^{(1,1)}\approx6$ at $a=0.9$, versus $18.07$ after equilibration),
even though they vary little with the spectral truncation.

We resolve this by equilibrating the operator before inversion. Each row of
$\mathbf{A}$ is rescaled by its inverse maximum magnitude and each column by its
inverse norm,
\begin{equation}
\tilde{\mathbf{A}} = \mathbf{D}_r\mathbf{A}\mathbf{D}_c, \qquad
\tilde{\mathbf{c}} = \mathbf{D}_c^{-1}\mathbf{c},
\end{equation}
(with diagonal matrices $\mathbf{D}_r$ and $\mathbf{D}_c$), so that no single equation or
coefficient dominates the fit. This standard preconditioning lowers the
condition number to $\kappa(\tilde{\mathbf{A}})\sim10^{5}$ and yields a solution that
is stable under truncation and converges with
resolution; after equilibration the
mixed observables are stable under changes of truncation, and the $30\times30$ and $50\times50$
solutions agree to better than one part in
$10^{4}$ at $a=0.8$ and $a=0.9$. Equilibration is
therefore essential for reliable results at high spin, and all values quoted in
this work employ it.

\subsection{Validation}
\label{sec:validation}

The mixed sector has no preexisting external benchmark to compare with,
so we validate the calculation using exact limits, available pure-sector benchmarks,
observable-dependent residual, truncation, resolution, and extraction tests, and
three consistency conditions that the mixed solution must satisfy.
\begin{itemize}
\item \emph{Exact limits:} Each observable is checked against its analytically
known Kerr or Schwarzschild value: the Bardeen ISCO and photon-ring
radii~\cite{Bardeen1972}, the Gauss--Bonnet topological invariant
$\oint\mathcal{R}^{(2)}\sqrt{h}\,d^2x = 8\pi$, and the Schwarzschild eikonal
frequency $\Omega_c=\lambda=1/(3\sqrt3)$. These are reproduced to machine or
near-machine precision.
\item \emph{Pure-sector agreement:} The leading-order surface gravity and
horizon angular velocity produced by our machinery reproduce the pure-sector
values of Ref.~\cite{LamChungYunes2025} to several significant digits.
\item \emph{Truncation stability:} Mixed observables are required to be stable
under changes of the spectral truncation at fixed resolution.
\item \emph{Resolution convergence:} Mixed observables computed at $30\times30$
and $50\times50$ are required to agree; we verify this explicitly at $a=0.8$
and $a=0.9$, where every mixed observable agrees to better than one part in
$10^{4}$.
\item \emph{Unimposed field equation:} The $r\chi$-component of
eq.~\eqref{eq:mixedEQ} is not part of the collocation solution, so its residual tests the source and
the solution together. Relative to the largest component of
$S^{(1,1)}_{\mu\nu}$, the residual of the five collocated components is below
$10^{-8}$ for $a\le0.76$ and reaches $10^{-7}$ at $a=0.9$; that of the $r\chi$-component
is below $3\times10^{-8}$ for $a\le0.76$ and reaches
$5\times10^{-6}$ at $a=0.9$. For $50\times50$ both are $\sim 10^{-9}$ at $a=0.9$.
\item \emph{Zeroth law:} $\Omega_H^{(1,1)}$ and $\kappa^{(1,1)}$ must be
constant over the horizon. Their variation with $\chi$ is at most
$2\times10^{-7}$ for $a\le0.76$ and below $3\times10^{-5}$ at $a=0.9$.
\item \emph{First law:} The Wald entropy $S^{(1,1)}$ must equal the value
implied by $\kappa$ and $\Omega_H$ at fixed $M$ and $J$. The two agree to one
part in $10^{5}$ (section~\ref{sec:entropy}).
\end{itemize}
Results that fail these tests are excluded from our quoted values. In practice
this affects only the near-extremal endpoint $a=0.99$, where the mixed metric
correction grows without bound and the perturbative expansion itself ceases to
be valid; we discuss this point separately in section~\ref{sec:discussion}.

\section{Horizon Observables}
\label{sec:horizon}

We begin with quantities defined on the event horizon, which characterize the
intrinsic geometry and mechanics of the corrected black hole. Each observable is
computed to mixed order $\Oone$ from the metric functions of
section~\ref{sec:ansatz}, evaluated at $r=r_+$ ($z=+1$). We evaluate coordinate-independent
horizon and geodesic quantities in the gauge
fixed by the stated asymptotic mass, angular momentum, and boundary conditions
(appendix~\ref{app:gauge}); the metric functions $H_i$ themselves are
coordinate dependent.

\subsection{Horizon angular velocity}
\label{sec:omega}

The horizon angular velocity is $\Omega_H = -g_{t\phi}/g_{\phi\phi}$ evaluated at
$r_+$. Since the $t$- and $\phi$-deformations enter multiplicatively in the
ansatz~\eqref{eq:ansatz}, $\Omega_H$ is a ratio of corrected metric functions,
and its mixed-order piece is \emph{not} simply the linear response to
$h^{(1,1)}$; the nonlinearity of the ratio generates cross terms between the two
pure sectors. Expanding to $\Oone$ (appendix~\ref{app:omega}),
\begin{align}
\Omega_H^{(1,1)} = \frac{a}{2 r_+}\Big[
&\big(H_2^{(1,1)} - H_4^{(1,1)}\big)
- H_2^{(1,0)} H_4^{(0,1)} \nonumber\\
&- H_2^{(0,1)} H_4^{(1,0)}
+ 2\,H_4^{(1,0)} H_4^{(0,1)} \Big]_{r_+}.
\label{eq:omega11}
\end{align}
The first bracketed term is the linear response to the mixed metric; the
remaining products are the parity-mixing cross terms. As required by the zeroth
law, $\Omega_H^{(1,1)}$ is independent of $\chi$, which we verify numerically and
use as an internal consistency check.

\begin{figure*}[htbp]
\centering
\includegraphics[width=0.95\textwidth]{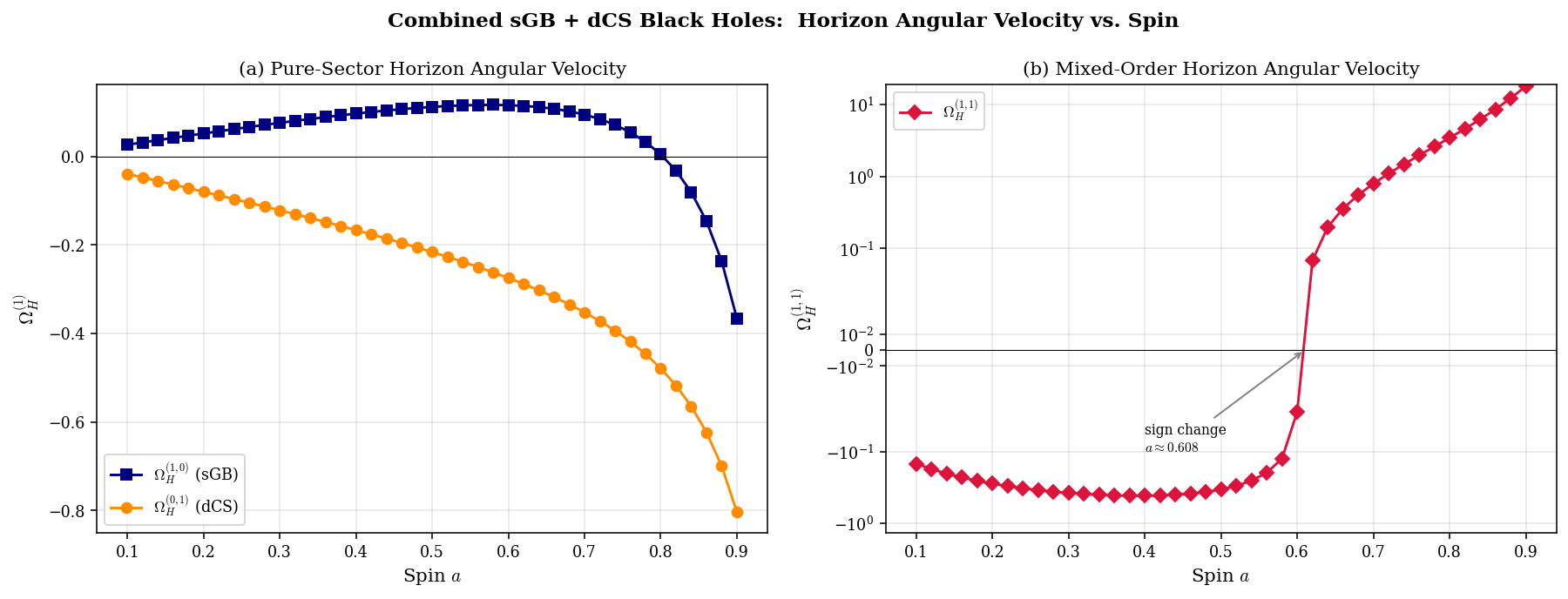}
\caption{Horizon angular velocity corrections versus spin $a$. (a) Pure-sector
corrections $\Omega_H^{(1,0)}$ (sGB, blue squares) and $\Omega_H^{(0,1)}$ (dCS,
orange circles). (b) Mixed-order correction $\Omega_H^{(1,1)}$ (red diamonds), on
a symmetric logarithmic scale. The polar ($\chi$) spread, which vanishes by the
zeroth law, is below $10^{-5}$ and smaller than the symbols.}
\label{fig:omega}
\end{figure*}

Figure~\ref{fig:omega} shows $\Omega_H^{(1,1)}$ as a function of spin. It is
negative at low spin, reaches a minimum of $-0.413$ near $a\approx0.39$,
\emph{changes sign} at $a_\times\approx0.608$, and grows rapidly thereafter, to
$+18.1$ at $a=0.9$. The pure sectors behave differently: $\Omega_H^{(0,1)}$ is
negative and monotonic, while $\Omega_H^{(1,0)}$ is positive up to
$a\approx0.80$, where it changes sign. The reversal of the mixed correction is
driven by the linear term of eq.~\eqref{eq:omega11}, which changes sign at
$a\approx0.57$ on the equator (appendix~\ref{app:omega-numerics});
the cross terms are negative at every spin and move the zero crossing to
slightly higher spin.

\subsection{Surface gravity}
\label{sec:kappa}

The surface gravity follows from the horizon-generating Killing vector
$\xi = \partial_t + \Omega_H\,\partial_\phi$ through
$\kappa^2 = -\tfrac12(\nabla_\mu\xi_\nu)(\nabla^\mu\xi^\nu)$, evaluated at the
horizon. For the metric of Sec.~\ref{sec:ansatz}, whose horizon remains at
$r=r_+$, this definition reduces on the horizon to
\begin{equation}
\kappa = \frac12\left[\frac{\Delta'(r_+)\,\partial_r D}
{(1+H_3)\,\Sigma\,g_{\phi\phi}}\right]^{1/2}_{r=r_+},
\quad D \equiv g_{t\phi}^2 - g_{tt}\,g_{\phi\phi},
\label{eq:kappahor}
\end{equation}
where $D$ vanishes at $r_+$ and $H_3$ is the full correction function. This
expression is exact in the $H_i$ and uses only their values and first radial
derivatives on the horizon, which the spectral representation supplies without
any near-horizon extrapolation (appendix~\ref{app:kappa}). The mixed-order
correction $\kappa^{(1,1)}$ is extracted by finite differencing
eq.~\eqref{eq:kappahor} in the two couplings. As a benchmark, the pure-sector
values $\kappa^{(1,0)}$ and $\kappa^{(0,1)}$ obtained this way reproduce those
of Ref.~\cite{LamChungYunes2025} to five significant digits at every spin
compared, including $a=0.9$.

\begin{figure*}[htbp]
\centering
\includegraphics[width=0.95\textwidth]{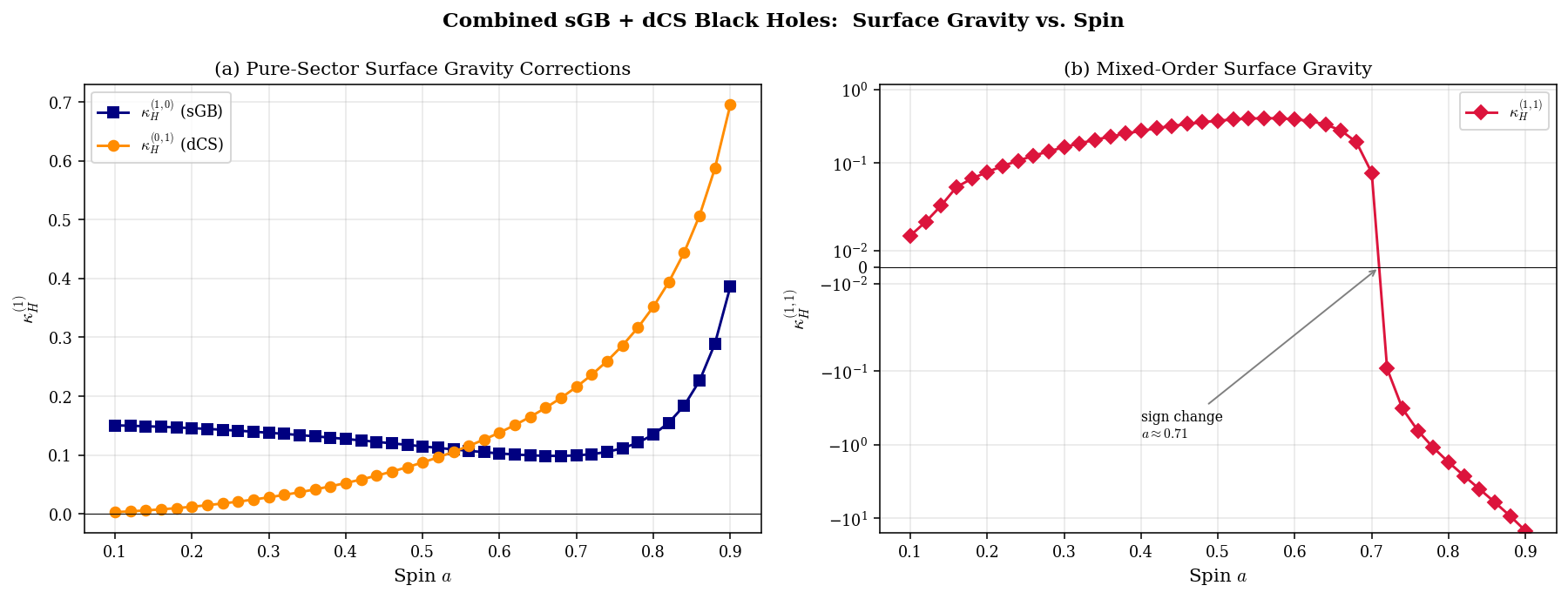}
\caption{Surface gravity corrections versus spin $a$. (a) Pure-sector
corrections $\kappa^{(1,0)}$ (sGB, blue squares) and $\kappa^{(0,1)}$ (dCS,
orange circles). (b) Mixed-order correction $\kappa^{(1,1)}$ (red diamonds), on a
symmetric logarithmic scale. The polar ($\chi$) spread, which vanishes by the
zeroth law, is below $3\times10^{-5}$ and smaller than the symbols. The sign
change is at $a_\times\approx0.710$.}
\label{fig:kappa}
\end{figure*}

Figure~\ref{fig:kappa} shows $\kappa^{(1,1)}$ versus spin. It rises to a maximum
of $+0.407$ near $a\approx0.57$ and then \emph{changes sign} at
$a_\times\approx0.710$, becoming increasingly negative toward high spin
($-14.8$ at $a=0.9$). Thus the mixed correction to the horizon temperature
switches from raising it, at low spin, to lowering it, at high spin. Both
pure-sector corrections, by contrast, are positive at every spin. The appearance
of a sign change in \emph{both} $\Omega_H^{(1,1)}$ and $\kappa^{(1,1)}$---at
distinct but comparable spins, not near where any pure sector expression changes sign---is a
qualitative signature of the combined theory.

\subsection{Horizon area}
\label{sec:area}

The horizon area is
$A = \oint \sqrt{g_{\chi\chi}\,g_{\phi\phi}}\;d\chi\,d\phi$ evaluated at $r_+$.
The integrand is finite and smooth despite the coordinate singularity of
$g_{\chi\chi}$ at the poles, since the divergence cancels against the vanishing
of $g_{\phi\phi}$. We evaluate the integral spectrally, reproducing the Kerr
value $A^{(0)} = 8\pi r_+$ to machine precision. The mixed-order correction
$A^{(1,1)}$ follows by finite differencing in the couplings.

\begin{figure}[htbp]
\centering
\includegraphics[width=0.45\textwidth]{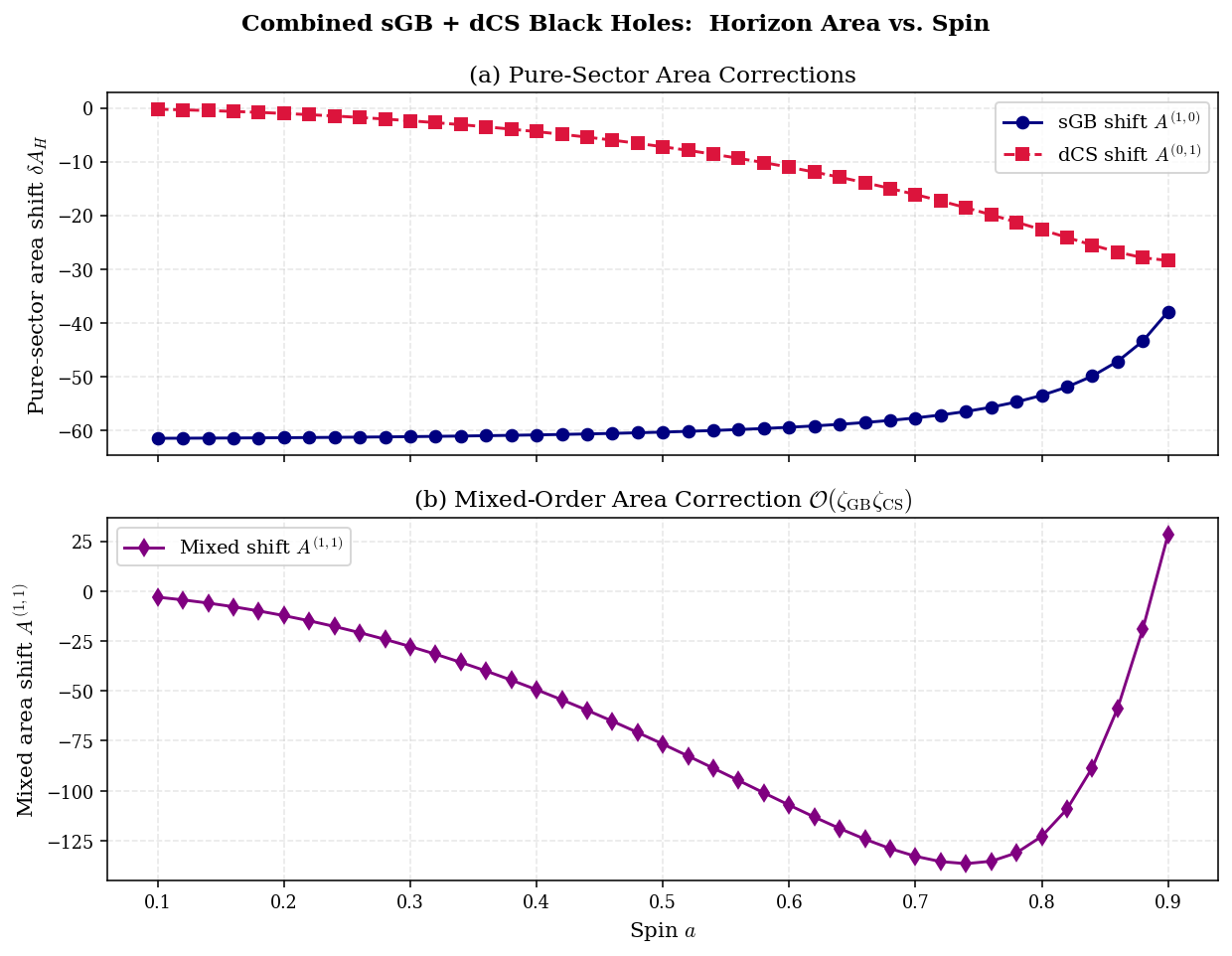}
\caption{Horizon area corrections versus spin $a$. (a) Pure-sector corrections
$A^{(1,0)}$ (sGB, blue circles) and $A^{(0,1)}$ (dCS, red squares).
(b) Mixed-order correction $A^{(1,1)}$ (purple diamonds), negative for
$a\lesssim0.89$ with a minimum near $a\approx0.74$, and positive at $a=0.9$.}
\label{fig:area}
\end{figure}

Figure~\ref{fig:area} shows the pure and mixed area corrections. The two pure
sectors behave differently as $a\to0$: the dCS correction $A^{(0,1)}$ vanishes,
since a non-rotating black hole is unmodified in dCS gravity, whereas the sGB
correction $A^{(1,0)}$ tends to a nonzero constant, as sGB gravity deforms even
the Schwarzschild horizon. The mixed correction $A^{(1,1)}$ is negative at low
and moderate spin, reaches a minimum of $-136.5$ near $a\approx0.74$, then rises
steeply, changing sign at $a_\times\approx0.888$ and reaching $+28.3$ at
$a=0.9$. The turnover is reproduced
at $50\times50$ resolution; being an integral over the horizon rather than a local
quantity, the area calculation is well conditioned, and the $30\times30$ and $50\times50$
values agree to one part in $10^{5}$ at $a=0.9$.

\subsection{Wald entropy and the first law}
\label{sec:entropy}

In a theory with nonstandard curvature couplings the black-hole entropy is not $A/4$ but the
Wald entropy~\cite{Wald1993,IyerWald1994}. For the action~\eqref{eq:action},
\begin{equation}
\begin{aligned}
S ={}& \frac{A}{4}
+ \frac{\aGB}{2}\oint_{\mathcal{H}}\varphi\,\mathcal{R}^{(2)}\sqrt{h}\,d^2x\\
&- \frac{\aCS}{4}\oint_{\mathcal{H}}\vartheta\;
{}^{\ast}\!R^{\mu\nu\rho\sigma}\epsilon_{\mu\nu}\epsilon_{\rho\sigma}
\sqrt{h}\,d^2x ,
\end{aligned}
\label{eq:wald}
\end{equation}
where $\mathcal{R}^{(2)}$ is the intrinsic Ricci scalar of the horizon
cross-section $\mathcal{H}$, $h$ is the determinant of its induced metric, and
$\epsilon_{\mu\nu}$ is the binormal, normalized to
$\epsilon_{\mu\nu}\epsilon^{\mu\nu}=-2$. Since
$\aGB\varphi=\zGB\,[\varphi^{(1/2,0)}+\zCS\,\varphi^{(1/2,1)}+\dots]$ for $M=1$,
and similarly for the axion, the mixed-order coefficient is
\begin{equation}
S^{(1,1)} = \tfrac14\,A^{(1,1)} + \tfrac12\,I_{\rm GB}^{(0,1)}
- \tfrac14\,I_{\rm CS}^{(1,0)} ,
\label{eq:S11}
\end{equation}
where $I_{\rm GB}^{(0,1)}$ is the $\mathcal{O}(\zCS)$ part of the Gauss--Bonnet
integral in Eq.~\eqref{eq:wald}, generated by the dCS deformation $h^{(0,1)}$ of
the horizon geometry and by the mixed dilaton $\varphi^{(1/2,1)}$, and
$I_{\rm CS}^{(1,0)}$ is the $\mathcal{O}(\zGB)$ part of the Chern--Simons
integral, generated by $h^{(1,0)}$ and the mixed axion $\vartheta^{(1,1/2)}$.
The entropy therefore draws on both stages of the calculation. The integrals are
evaluated spectrally on the horizon, as for the area; and the Chern--Simons
integrand is extrapolated to $r_+$ from the nearest radial nodes.

The entropy can also be obtained without any horizon integral, from the first
law $dM = T\,dS + \Omega_H\,dJ$ at fixed $\aGB$ and $\aCS$, with
$T=\kappa/2\pi$. Because the couplings have dimensions of length squared,
$S = M^2 s(a,\zGB,\zCS)$ with $a=J/M^2$, and a change of $M$ at fixed $J$,
$\aGB$, and $\aCS$ changes $a$ and both $\zeta$. The first law then implies
the Smarr-type relation
\begin{equation}
M = 2TS + 2\Omega_H J
- 4T\left(\zGB\frac{\partial S}{\partial\zGB}
+ \zCS\frac{\partial S}{\partial\zCS}\right),
\label{eq:smarr}
\end{equation}
with the derivatives taken at fixed $M$ and $J$. Expanding in the couplings and
setting $M=1$,
\begin{equation}
\begin{aligned}
S^{(1,0)} &= a\left[\frac{\Omega_H}{T}\right]^{(1,0)}
- \frac12\left[\frac1T\right]^{(1,0)},\\
S^{(1,1)} &= \frac{a}{3}\left[\frac{\Omega_H}{T}\right]^{(1,1)}
- \frac16\left[\frac1T\right]^{(1,1)},
\end{aligned}
\label{eq:Sfirstlaw}
\end{equation}
where $[X]^{(m,n)}$ is the coefficient of $\zGB^m\zCS^n$ in $X$, and
$S^{(0,1)}$ follows from the first line with $(1,0)\to(0,1)$.
Equation~\eqref{eq:Sfirstlaw} derives the entropy coefficients from $\kappa$ and
$\Omega_H$ alone, with no spin derivatives.

\begin{figure}[htbp]
\centering
\includegraphics[width=0.45\textwidth]{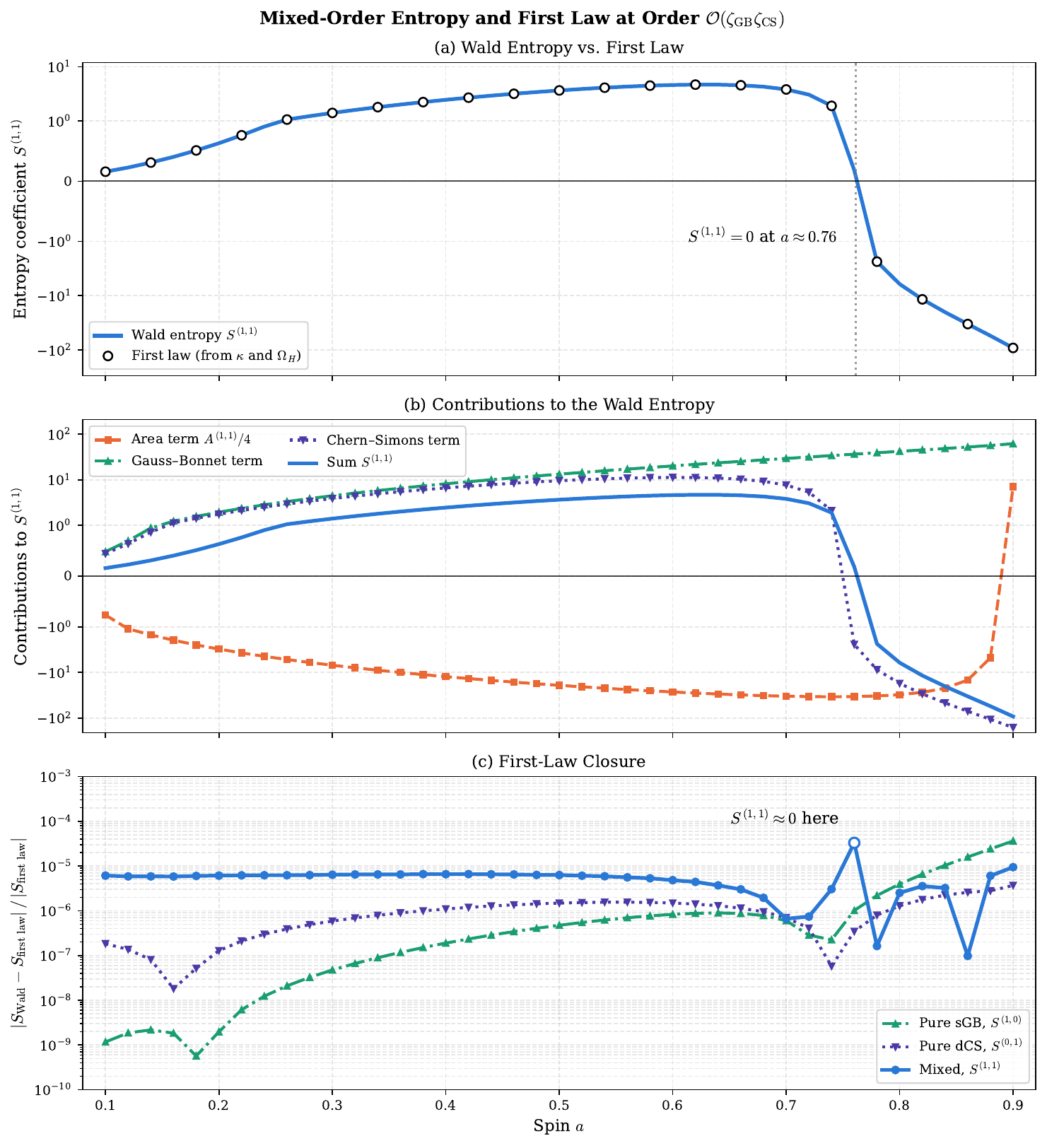}
\caption{Mixed-order entropy versus spin $a$. (a): Wald entropy $S^{(1,1)}$ from
eq.~\eqref{eq:S11} (line) and the value implied by the first law,
eq.~\eqref{eq:Sfirstlaw} (circles). (b): the three contributions to
eq.~\eqref{eq:S11}---area term (squares), Gauss--Bonnet term (upward
triangles), and Chern--Simons term (downward triangles)---and their sum (solid
line). (c): relative difference between the Wald and first-law values for the
pure sGB, pure dCS, and mixed coefficients; the open circle marks the spin
nearest the zero of $S^{(1,1)}$, where the relative difference is inflated. The
top two panels use a symmetric logarithmic scale.}
\label{fig:entropy}
\end{figure}

Figure~\ref{fig:entropy} shows the result. $S^{(1,1)}$ is positive at low spin,
where it grows as $a^2$, reaches a maximum of $+4.70$ near $a\approx0.63$,
\emph{changes sign} at $a_\times\approx0.762$, and falls to $-92.2$ at $a=0.9$.
The Wald and first-law values agree to about one part in $10^{5}$ at every spin away
from the zero of $S^{(1,1)}$, where their absolute difference is
$6\times10^{-6}$; the same comparison in the pure sectors agrees to
$4\times10^{-5}$ or better. Since eq.~\eqref{eq:Sfirstlaw} involves only
$\kappa$ and $\Omega_H$, whereas eq.~\eqref{eq:S11} involves horizon integrals of
the metric and of both mixed scalars, this agreement tests the mixed metric, the
mixed scalars, and the asymptotic conditions~\eqref{eq:peeling11} together.

The area term alone is a poor guide to the entropy correction. $A^{(1,1)}/4$ is negative for $a\lesssim0.89$,
and for $a\le0.8$ it is at least
five times larger in magnitude than $S^{(1,1)}$ and is largely canceled by
the two scalar terms: at $a=0.5$ the three
contributions to Eq.~\eqref{eq:S11} are $-19.2$, $+13.3$, and $+9.5$, summing to
$+3.66$. The sign change of $S^{(1,1)}$ is driven by the Chern--Simons term,
which turns negative near $a\approx0.75$.

\section{Orbital Observables}
\label{sec:orbital}

We now turn to observables defined by equatorial circular orbits in the
modified geometry, and then to the non-equatorial photon orbits
(section~\ref{sec:sphph}). Reflection symmetry keeps equatorial orbits equatorial, so
each equatorial observable depends only on the components $g_{tt}$,
$g_{t\phi}$, $g_{\phi\phi}$, and their radial derivatives. Throughout we
distinguish prograde and retrograde orbits. The radii quoted in this section are
values of the coordinate $r$ of the ansatz~\eqref{eq:ansatz}, at fixed $M$ and
$J$. Because $H_{3,\infty}\neq0$, this coordinate differs from the areal radius
even asymptotically, so the radius shifts are not invariant on their own; the
invariant content of the photon orbit is carried by its frequency and Lyapunov
exponent, described in Sec.~\ref{sec:qnm}.

\subsection{Innermost stable circular orbit}
\label{sec:isco}

For a circular geodesic at radius $r$, the orbital angular velocity is
\begin{equation}
\Omega_\pm(r) = \frac{-\,g_{t\phi}' \pm
\sqrt{\,g_{t\phi}'^2 - g_{tt}'\,g_{\phi\phi}'}}{g_{\phi\phi}'},
\label{eq:orbfreq}
\end{equation}
with $\pm$ denoting prograde and retrograde motion and primes radial
derivatives. The specific energy $E(r)$ of the circular orbit follows from
$\Omega_\pm$ and the metric, and ISCO is
the marginally stable orbit, located where $dE/dr = 0$. We solve this condition
for the corrected metric and extract the mixed-order shift
$r_{\rm ISCO}^{(1,1)}$ by finite differencing in the two couplings, separately
for each orbital sense (appendix~\ref{app:orbits}). As a validation, setting the
couplings to zero recovers the Bardeen ISCO radii~\cite{Bardeen1972} to four
digits at all spins.

\begin{figure*}[htbp]
\centering
\includegraphics[width=0.95\textwidth]{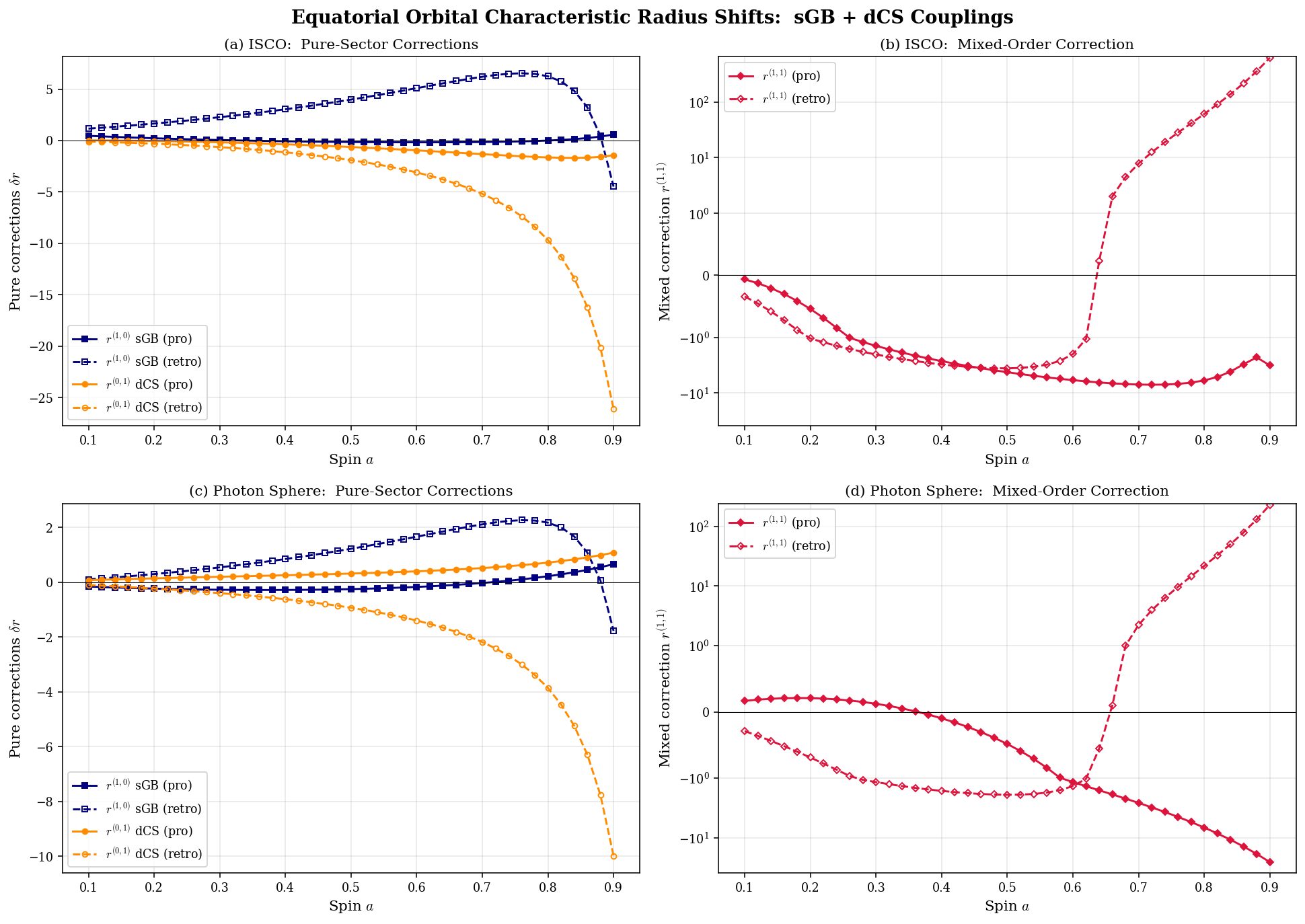}
\caption{Equatorial orbital radius corrections as a function of black-hole spin
$a$, for the ISCO (top) and the photon ring (bottom). Left panels (a), (c): the
pure-sector corrections $r^{(1,0)}$ (sGB, blue) and $r^{(0,1)}$ (dCS, orange).
Right panels (b), (d): the mixed-order corrections $r^{(1,1)}$, shown on a
symmetric logarithmic scale. Filled symbols and solid lines denote prograde
orbits, open symbols and dashed lines retrograde. The mixed corrections show a
pronounced prograde/retrograde asymmetry: the retrograde shifts change sign
near $a\approx0.64$ (ISCO) and $a\approx0.66$ (photon ring) and grow steeply
toward high spin, while the prograde ISCO shift stays negative. At zero
coupling the Bardeen radii~\cite{Bardeen1972} are recovered to four digits at
all spins.}
\label{fig:orbits}
\end{figure*}

Figure~\ref{fig:orbits} shows the pure and mixed ISCO shifts. The prograde and
retrograde corrections differ markedly in both magnitude and sign. The
retrograde shift $r_{\rm ISCO}^{(1,1)}$ is negative at low spin, with a minimum
of $-3.60$ near $a\approx0.49$, changes sign at $a_\times\approx0.637$,
and grows steeply thereafter, to $+60.8$ at $a=0.8$ and $+621$ at $a=0.9$. The
prograde shift is negative at every spin; it falls to $-7.09$ near $a\approx0.72$, decreases in magnitude to $-2.3$ at
$a=0.88$, and turns down again to $-3.2$ at $a=0.9$. The
$30\times30$ and $50\times50$ values agree to $2\times10^{-5}$ or better at
$a=0.8$ and $a=0.9$ for both orbital senses. Because the ISCO is located by a
derivative condition, $dE/dr=0$, its mixed coefficient carries a root-finding
noise of order $10^{-3}$ in absolute terms (section~\ref{sec:errors}), which is
significant only close to the retrograde zero.

\subsection{Photon ring}
\label{sec:photon}

The photon ring is set by the equatorial circular null geodesic (photon sphere),
located where
\begin{equation}
g_{tt} + 2\,\Omega_\pm\, g_{t\phi} + \Omega_\pm^2\, g_{\phi\phi} = 0,
\label{eq:photoncond}
\end{equation}
with $\Omega_\pm$ from eq.~\eqref{eq:orbfreq}. This algebraic condition
determines the prograde and retrograde photon radii, whose mixed-order shifts
$r_{\rm ph}^{(1,1)}$ we again extract by finite differencing. At zero coupling
the exact Kerr photon-sphere radii are recovered to four digits.

Figure~\ref{fig:orbits} also shows these results. As for the ISCO, the mixed photon
shift displays a strong prograde-retrograde asymmetry. The prograde shift
$r_{\rm ph}^{(1,1)}$ is positive at low spin, with a maximum of $+0.21$ near
$a\approx0.18$; changes sign at $a_\times\approx0.364$; and falls to
$-25.6$ at $a=0.9$. The retrograde shift is negative at low spin, with a minimum
of $-1.90$ near $a\approx0.50$, changes sign at $a_\times\approx0.657$, and
reaches $+231$ at $a=0.9$. Because the photon-ring
condition~\eqref{eq:photoncond} is algebraic rather than derivative-based, its
extraction is free of the noise that affects the ISCO, and the $30\times30$ and
$50\times50$ values agree to $10^{-6}$ at $a=0.8$ and $a=0.9$. The photon orbit
controls the size and asymmetry of the black-hole shadow; its impact parameter
$b=1/\Omega_c$ (section~\ref{sec:qnm}) fixes the displacement of the shadow's
extremal edges, while a full two-dimensional silhouette additionally
requires the non-equatorial photon orbits, to which we now turn.

\subsection{Non-equatorial photon orbits}
\label{sec:sphph}
 
In Kerr spacetime, the photon ring is the equatorial representative of a larger family of spherical
photon orbits, null geodesics at constant $r$ that oscillate in latitude while
precessing in azimuth~\cite{Teo2003}. Their existence rests on the Carter
constant~\cite{Carter1968}, which separates the radial and polar motion. The
corrected spacetime has no such constant; the pure-sector and mixed corrections
spoil the separability, and the usual construction of spherical orbits from the
roots of the radial potential is no longer available. We therefore integrate the
null geodesic equations numerically in the full metric $g_{\mu\nu}=g^{\rm Kerr}_{\mu\nu}
+h_{\mu\nu}(\zGB,\zCS)$, built from the solutions of section~\ref{sec:method}
and the pure-sector inputs, at the intermediate spin $a=0.5$.
 
Two symmetries replace the Carter constant. The metric is invariant under
reflection through the equatorial plane and under the simultaneous reversal
$(t,\phi)\to(-t,-\phi)$, so a geodesic that leaves the equator with $p_r=0$ and
reaches its polar turning point with $p_r=0$ is periodic in $(r,\theta)$. The
energy and axial angular momentum remain conserved, and the orbits are
labelled by the inclination $i$, defined through the polar turning angle
$\theta_{\rm t}$ as $i=90^\circ-\theta_{\rm t}$ for prograde and $i=90^\circ+\theta_{\rm t}$
for retrograde orbits, so that $i=0^\circ$ and $i=180^\circ$ are the prograde and
retrograde photon rings. For each $i$ the turning radius is found by shooting
(appendix~\ref{app:orbits-sph}). The resulting orbits are not exactly spherical, as
the radius at the turning latitude, $r_{\rm turn}$, differs from the radius at the
equator, $r_{\rm eq}$, by an amount that vanishes identically in the Kerr metric. We extract
the pure-sector and mixed coefficients of $r_{\rm eq}$ and of
$\Delta r=r_{\rm turn}-r_{\rm eq}$ by the same finite-difference procedure as for the
equatorial orbits.
 
\begin{figure*}[htbp]
\centering
\includegraphics[width=0.75\textwidth]{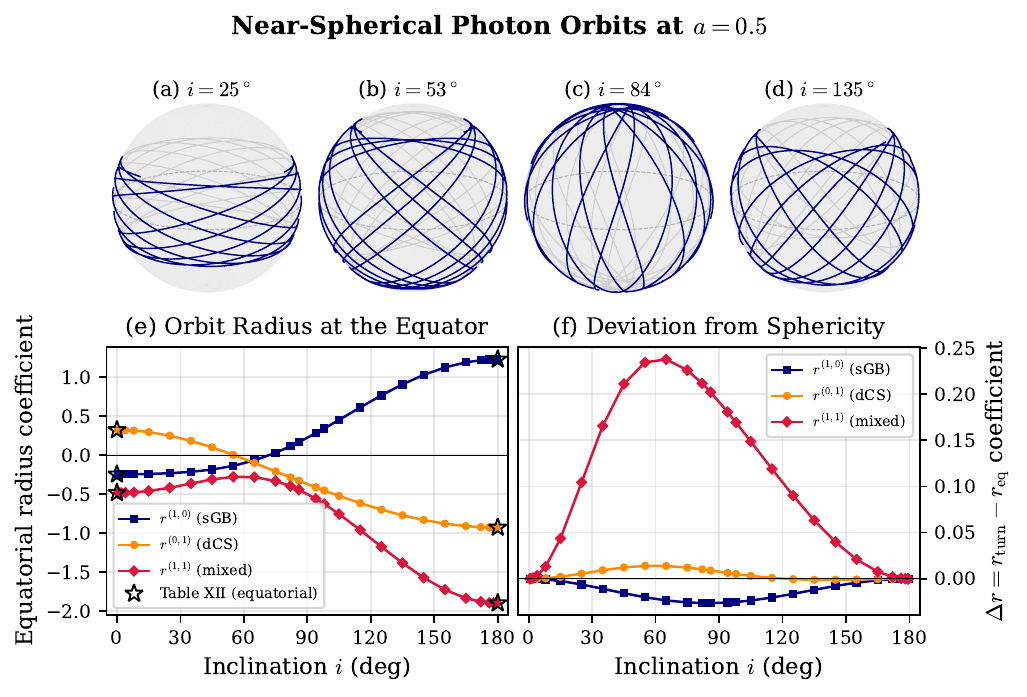}
\caption{Near-spherical photon orbits at $a=0.5$. In the corrected spacetime
there is no Carter constant, and photon orbits are not exactly spherical; they are
found by shooting from the polar turning point and imposing $p_r=0$ at the
equator (appendix~\ref{app:orbits-sph}). (a)--(d) Orbits at inclinations
$i=25^\circ$, $53^\circ$, $84^\circ$ (prograde), and $135^\circ$ (retrograde),
computed with $\zGB=\zCS=0.02$ and drawn on the sphere of radius $r_{\rm eq}$; at
this coupling the departure from sphericity is below $10^{-3}$ and is not
visible. (e) Pure-sector and mixed coefficients of the orbit radius at the
equator versus inclination ($i=0^\circ$, prograde equatorial; $i=180^\circ$,
retrograde equatorial); the stars are the photon-ring values of
Table~\ref{tab:photondata}. (f) Coefficients of the departure from sphericity,
$\Delta r=r_{\rm turn}-r_{\rm eq}$, which vanishes identically in the Kerr background.}
\label{fig:sphph}
\end{figure*}

Figure~\ref{fig:sphph} shows the results; panels (a)--(d) illustrate the
geometry. Two features stand out. First,
at both ends of the family the equatorial radius reproduces the photon-ring
coefficients of Table~\ref{tab:photondata} to the digits quoted, in every sector,
an independent check of the photon-ring results, since the shooting method uses
the full off-equatorial metric and no circular-orbit condition. Between the ends
the mixed coefficient $r_{\rm eq}^{(1,1)}$ interpolates smoothly from $-0.484$
(prograde) to $-1.897$ (retrograde), with a shallow maximum of $-0.28$ near
$i\approx55^\circ$. Second, the departure from sphericity is dominated by the
mixed sector. The mixed coefficient $\Delta r^{(1,1)}$ rises to $+0.24$ near
$i\approx65^\circ$, while the pure-sector coefficients remain below $0.03$ in
magnitude at every inclination; $\Delta r$ vanishes at both equatorial ends, as
it must. Like the other radii of this section, $\Delta r$ is defined in the
coordinates of the ansatz~\eqref{eq:ansatz}, in which the Kerr spherical orbits
have $\Delta r=0$. At the illustrative coupling of panels (a)--(d) the departure
from sphericity is below $10^{-3}$, so the orbits are visually indistinguishable
from their Kerr counterparts; its physical significance is that the breakdown of
Carter separability, which is weak in each pure theory, is an order of magnitude
stronger at mixed order.
 
\section{Photon-Ring Eikonal Estimates for Ringdown}
\label{sec:qnm}

The QNM spectrum governs the ringdown following a black-hole
merger and is a primary target of gravitational-wave tests of the Kerr
hypothesis. A full QNM calculation in the combined theory is beyond our scope;
the corrected background is not of Petrov type~D, so the perturbation equations
do not separate, and for a rotating black hole the gravitational perturbations of each parity
are coupled to the perturbations of both scalar fields.
We instead compute the leading, eikonal contribution, which is directly
accessible from the photon-ring data from section~\ref{sec:photon}. Accordingly,
throughout this section $\omega_{\ell n}$ denotes a
large-$\ell$ estimate of the ringdown frequency, not a solution of the coupled
finite-$\ell$ perturbation problem.

In the eikonal (large-$\ell$) limit, the quasinormal frequencies are governed by
the equatorial photon orbit through the correspondence~\cite{Cardoso2009}
\begin{equation}
\omega_{\ell n} \simeq \Omega_c\left(\ell + \tfrac12\right)
- i\left(n + \tfrac12\right)|\lambda|,
\label{eq:eikonal}
\end{equation}
where $\Omega_c$ is the orbital frequency of the photon orbit---the real part of
the mode eigenvalue---and $\lambda$ is its Lyapunov exponent, which sets the damping rate.
The frequency is $\Omega_c = \Omega_\pm(r_{\rm ph})$, evaluated at the photon
radius of eq.~\eqref{eq:photoncond}, and the Lyapunov exponent is
\begin{equation}
\lambda = \sqrt{\frac{R''(r_{\rm ph})}{2\,\dot t^2}}\,,
\label{eq:lyapunov}
\end{equation}
where $R(r) \equiv \dot r^2$ is the radial potential of the null geodesic and
$\dot t$ its conserved-energy-normalized time component (appendix~\ref{app:qnm}).
At zero coupling this construction reproduces the exact Schwarzschild value
$\Omega_c = \lambda = 1/(3\sqrt3)$ and the known Kerr spin dependence---in
particular the characteristic decrease of the prograde damping rate toward the
extremal limit---providing a stringent check on the machinery.

We follow common usage in calling this orbit the photon ring, but
electromagnetic propagation plays no role in the argument. In general
relativity, short-wavelength gravitational perturbations propagate along null
geodesics of the background, so the unstable circular null orbit controls how
such perturbations are trapped and how they leak away; the null geodesic is a
proxy for the propagation of the gravitational excitations themselves. In the
modified theories considered here the gravitational perturbations are coupled
to the scalar fields, and the eikonal modes of such coupled systems are governed
by the extremum of an effective potential that need not coincide with the one
defining the null geodesics of the corrected
metric~\cite{GlampedakisSilva2019,SilvaGlampedakis2020,Bryant2021}. The
quantities $\Omega_c^{(1,1)}$ and $\lambda^{(1,1)}$ computed below are
therefore the contribution of the corrected background geometry to the eikonal
frequencies.

\begin{figure}[htbp]
\centering
\includegraphics[width=0.45\textwidth]{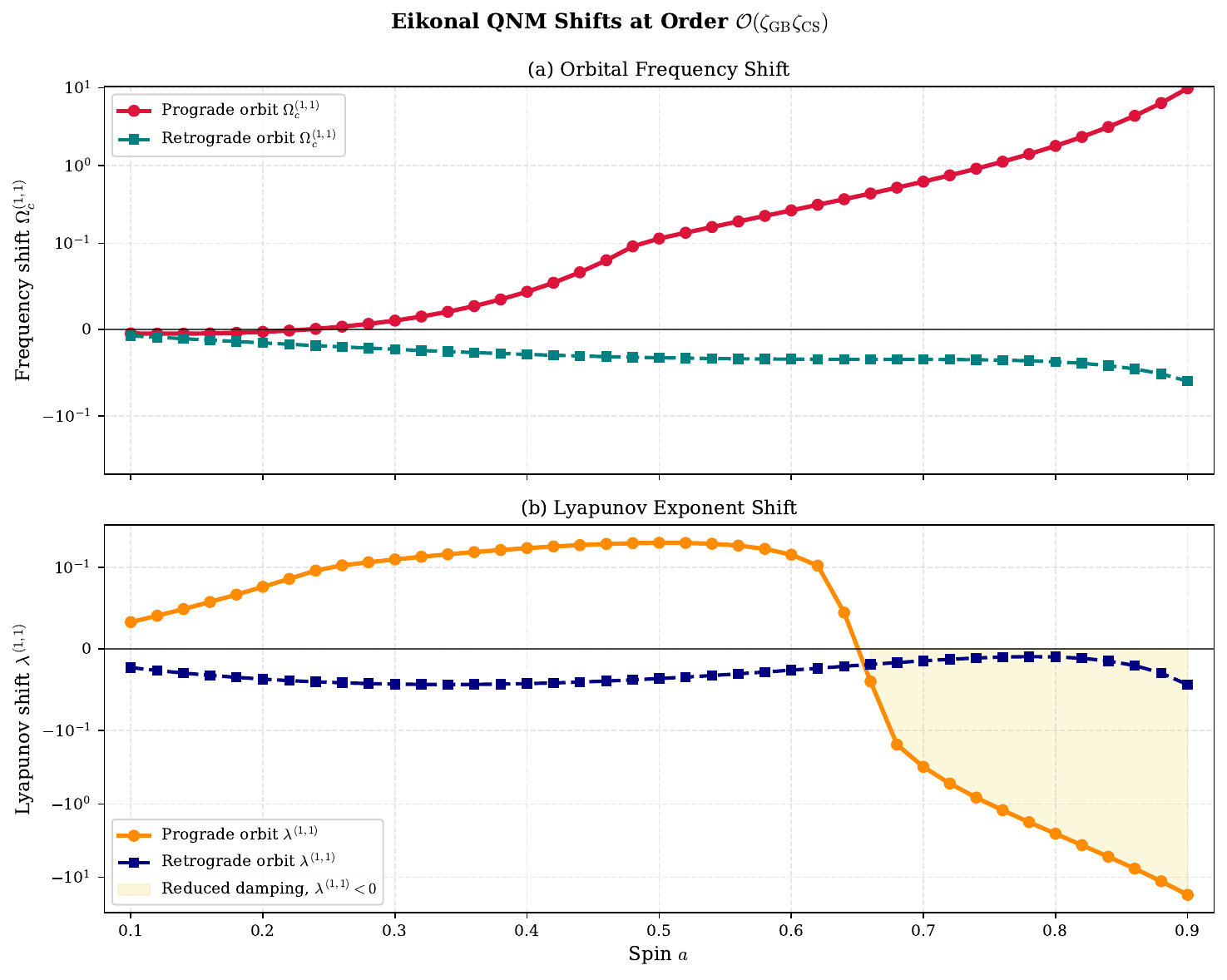}
\caption{Mixed-order corrections to the eikonal ringdown estimates versus spin
$a$, on a symmetric logarithmic scale: (a) the photon-orbit frequency
$\Omega_c^{(1,1)}$ and (b) the Lyapunov exponent $\lambda^{(1,1)}$
(damping rate). Circles and solid lines, prograde; squares and dashed
lines, retrograde. The prograde $\Omega_c^{(1,1)}$ changes sign at
$a_\times\approx0.231$ and the prograde $\lambda^{(1,1)}$ at
$a_\times\approx0.651$.}
\label{fig:qnm}
\end{figure}

We extract the mixed-order corrections $\Omega_c^{(1,1)}$ and $\lambda^{(1,1)}$
by finite differencing eqs.~(\ref{eq:eikonal}--\ref{eq:lyapunov}) in the two
couplings, for prograde and retrograde orbits. Figure~\ref{fig:qnm} shows the
results. The mixed correction to the prograde frequency is small and negative
below $a\approx0.23$ and positive above, growing steeply with spin, to $+0.62$
at $a=0.7$ and $+9.8$ at $a=0.9$. The retrograde correction is negative and
small at every spin [$|\Omega_c^{(1,1)}|\le0.06$], producing a pronounced
prograde-retrograde asymmetry. The prograde damping correction
$\lambda^{(1,1)}$ is positive at low and moderate spin, with a maximum of
$+0.22$ near $a\approx0.51$; changes sign at $a_\times\approx0.651$; and
becomes strongly negative at high spin ($-17.3$ at $a=0.9$); for
$a\gtrsim0.65$ the mixed term lowers the prograde damping rate and so
lengthens the damping time, while at lower spin it shortens it. The
retrograde $\lambda^{(1,1)}$ is negative and small throughout
[$|\lambda^{(1,1)}|\le0.05$]. Since the eikonal frequencies inherit the
algebraic photon-ring condition, the mixed shifts converge cleanly under grid
refinement; at $a=0.8$ and $a=0.9$ the $30\times30$ and $50\times50$ values
agree to $10^{-6}$ or better for $\Omega_c^{(1,1)}$ and the prograde
$\lambda^{(1,1)}$, and to $5\times10^{-5}$ for the retrograde $\lambda^{(1,1)}$.

\subsection{Fidelity of the eikonal correspondence}
\label{sec:eikfidelity}

\begin{figure}[htbp]
\centering
\includegraphics[width=\columnwidth]{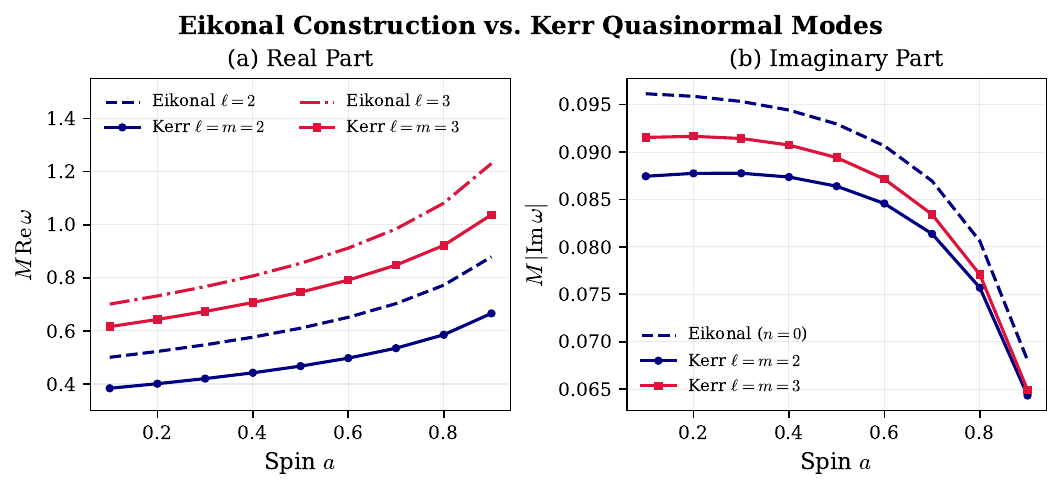}
\caption{Fidelity of the zero-coupling eikonal construction against the exact
fundamental ($n=0$) Kerr quasinormal modes~\cite{BertiCardosoWill2006}.
(a) real part: eikonal $\Omega_c(\ell+\tfrac12)$ (dashed) versus the gravitational
$\ell=m=2$ and $\ell=m=3$ frequencies (solid). (b) imaginary part:
eikonal $(n+\tfrac12)\lambda$ (dashed) versus the same modes. The eikonal
overestimates the real part by $\approx30\%$ ($\ell=2$) and $\approx15\%$
($\ell=3$) and the imaginary part by $\approx6$--$10\%$, with fidelity improving
with $\ell$; it nonetheless reproduces the spin dependence faithfully.}
\label{fig:eikfidelity}
\end{figure}

Because the observationally relevant modes are the low multipoles $\ell=2,3$,
whereas the eikonal correspondence is asymptotic in $\ell$, we quantify its
fidelity by comparing the zero-coupling construction against the exact Kerr
quasinormal frequencies. Figure~\ref{fig:eikfidelity} shows the eikonal real
part $\Omega_c(\ell+\tfrac12)$ and imaginary part $(n+\tfrac12)\lambda$ against
the fundamental ($n=0$) gravitational $\ell=m=2$ and $\ell=m=3$ Kerr
modes, using the fits of Berti, Cardoso, and Will~\cite{BertiCardosoWill2006}.
The eikonal overestimates the real part by $\approx\!30\%$ for $\ell=2$ and by
$\approx\!15\%$ for $\ell=3$, and the imaginary part by $\approx\!6$--$10\%$;
the fidelity improves systematically with $\ell$, as expected for a large-$\ell$
approximation, and the imaginary part---set by the orbital instability rate---is
reproduced more accurately than the real part throughout.

Two features are important for interpreting our mixed-order results. First, the
eikonal reproduces the spin dependence of the exact modes faithfully
(figure~\ref{fig:eikfidelity}): the offset is a slowly-varying multiplicative
factor rather than a spin-dependent distortion. Since our mixed-order results are
fractional corrections, a spin-independent multiplicative offset largely
cancels in the ratio $\delta\omega/\omega$, so the fractional shifts
$\Omega_c^{(1,1)}/\Omega_c^{(0)}$ and $\lambda^{(1,1)}/\lambda^{(0)}$ are expected
to be considerably more robust than the absolute eikonal frequencies. Second,
the correspondence is known to degrade for near-extremal co-rotating modes, where
the prograde photon orbit approaches the horizon and the prograde damping tends
to zero ($\lambda\to0$); although our highest spin $a=0.9$ is not near-extremal,
the prograde imaginary part is already dropping steeply
[figure~\ref{fig:eikfidelity} b], and the prograde $\lambda^{(1,1)}$ at $a=0.9$
should accordingly be read as the least-controlled of our ringdown estimates.

A quantitative calibration of the eikonal estimate against the observed modes
would require solving the coupled perturbation equations at finite $\ell$ on the
corrected backgrounds, the program pursued for the pure sectors in
Ref.~\cite{ChungLamYunes2025}; our results furnish the leading, large-$\ell$
contribution to that spectrum in the mixed sector.

\section{Results and Discussion}
\label{sec:discussion}

Taken together, the observables of sections~\ref{sec:horizon}--\ref{sec:qnm}
provide a consistent picture of how the combined sGB--dCS theory deforms a
rotating black hole at mixed order. We collect the physical implications here.

\subsection{Sign changes as a signature of the mixed sector}
\label{sec:signs}

The most striking feature of the mixed sector is how its coefficients change
sign as functions of spin. All four horizon quantities do so: the horizon
angular velocity $\Omega_H^{(1,1)}$ at $a_\times=0.608$, the surface gravity
$\kappa^{(1,1)}$ at $0.710$, the Wald entropy $S^{(1,1)}$ at $0.762$, and the
horizon area $A^{(1,1)}$ at $0.888$. So do the prograde eikonal damping rate
$\lambda^{(1,1)}$ ($0.651$) and frequency $\Omega_c^{(1,1)}$ ($0.231$), the
retrograde ISCO radius ($0.637$), and both photon-ring radii ($0.364$ prograde,
$0.657$ retrograde). Most of these reversals fall in the interval
$0.6\lesssim a\lesssim0.76$. By contrast, the only pure-sector horizon
coefficient that changes sign is $\Omega_H^{(1,0)}$, at $a\approx0.80$. For
$\Omega_H^{(1,1)}$ the decomposition of eq.~\eqref{eq:omega11} identifies the
origin of the reversal; the linear response to the mixed metric $h^{(1,1)}$
changes sign at $a\approx0.57$ on the equator, while the bilinear cross terms formed from the
parity-even (sGB) and parity-odd (dCS) pure-sector fields are negative at every
spin. The two contributions therefore do not compete at low spin, where both are
negative; the reversal is a property of the mixed metric itself.

The crossing spins are located by cubic-spline interpolation of each coefficient
on the $\Delta a=0.02$ grid of the sweep. Their uncertainty is set by this
interpolation: linear interpolation between the bracketing spins moves every
crossing by less than $10^{-3}$, and we take $\pm0.001$ as the interval. The
other error sources are negligible by comparison: the polar spread of the
horizon quantities, the resolution dependence of the coefficients, and the
extraction noise of the ISCO (section~\ref{sec:errors}) each displacing the
crossings by less than $10^{-4}$.

Since a mixed observable is proportional to $\zGB\zCS$, it vanishes identically
unless both couplings are present. The pattern described above is therefore specific to the mixed order $\Oone$.
Establishing that such a
pattern could not arise in a broader class of modified-gravity theories would
require a dedicated theory-space comparison, which we do not attempt here.
Nevertheless, the spin-dependent sign reversal is a
distinctive, mixed-coupling diagnostic within the combined sGB--dCS effective theory of the
\emph{simultaneous} presence of the parity-even and parity-odd
curvature invariants in the strong-field regime, and constitutes the central qualitative physical
result of this work.

\subsection{Prograde-retrograde asymmetry}
\label{sec:asymmetry}

The orbital and ringdown observables show a pronounced asymmetry between
prograde and retrograde motion. It is most visible in the eikonal ringdown.
For prograde orbits the mixed correction raises the frequency at all but the
lowest spins and changes the damping rate by an amount that grows steeply with
spin, reaching $\Omega_c^{(1,1)}=+9.8$ and $\lambda^{(1,1)}=-17.3$ at $a=0.9$.
For retrograde orbits both coefficients stay below $0.06$ in magnitude across
the whole range. Such a differential imprint on co- and counter-rotating modes
is a potential discriminator in ringdown tests, where multiple modes of a single
remnant can in principle be measured independently.

\subsection{Approach to extremality and EFT validity}
\label{sec:extremal}

All quoted results lie in the range $0.1 \le a \le 0.9$, where the mixed
observables are converged and validated. As the spin increases beyond this
range the mixed metric correction grows rapidly: at $30\times30$ the largest
$|H_i^{(1,1)}|$ increases by more than two orders of magnitude between $a=0.9$
and $a=0.99$, from $\approx150$ to $\approx3\times10^{4}$. At
$a=0.99$ the correction reaches magnitudes at which the
perturbative expansion~\eqref{eq:gexp} can no longer be regarded as small for
any but the tiniest couplings, and the EFT description breaks
down. We therefore treat $a=0.99$ as an EFT-breakdown endpoint rather than a
predictive data point; although the Stage-II system remains well posed there,
the underlying expansion is not controlled, and we exclude it from
our quoted observables. This near-extremal growth is consistent with the general
expectation that higher-curvature corrections become large where the horizon
curvature is greatest.

\subsection{Error budget}
\label{sec:errors}

Three sources of uncertainty affect the mixed observables.

\emph{The mixed solution.} The accuracy of $h^{(1,1)}$ is measured by the
field-equation residuals from section~\ref{sec:validation}. For $a\le0.76$ all six
components, including the uncollocated $r\chi$-component, are satisfied to
$3\times10^{-8}$ or better relative to the largest source component; and at
$a=0.9$ the residuals are at the $10^{-7}$ (collocated) and $5\times10^{-6}$
($r\chi$) levels. The effect on the observables is bounded by the resolution
comparison of Table~\ref{tab:resconv}, so that at $a=0.8$ and $a=0.9$ every mixed
observable agrees between the $30\times30$ and $50\times50$ grid calculations to
a fractional precision of $5\times10^{-5}$ or
better. The relative polar ($\chi$) spread of $\Omega_H^{(1,1)}$ and $\kappa^{(1,1)}$,
which must vanish by the zeroth law, is at most $3\times10^{-5}$. The
pure-sector inputs are used at the accuracy ($\lesssim10^{-8}$) at which
Ref.~\cite{LamChungYunes2025} reports them.

\emph{Extraction.} The mixed coefficients are obtained by taking finite differences of
the full nonlinear observables in the two couplings, with a step size $\epsilon$; this leaves
leaves a truncation error of order $\epsilon^2$ that grows with the size of the
coefficients. For the surface gravity and the first law we use
$\epsilon=10^{-4}$, for which this error is below $10^{-5}$ at every spin. For
the orbital and eikonal quantities, where root finding limits how small the
step can be, we use $\epsilon=10^{-3}$; comparison with $\epsilon=2\times10^{-3}$
shows a relative truncation error below $10^{-3}$ for $a\le0.8$, rising at
$a=0.9$ to about $0.5\%$ for the prograde $r_{\rm ISCO}^{(1,1)}$ and for the
retrograde $\Omega_c^{(1,1)}$ and $\lambda^{(1,1)}$, and to $0.1\%$ or less for
the remaining coefficients. Quantities
defined through radial derivatives are more delicate than those fixed by
algebraic conditions. The ISCO is located by $dE/dr=0$, and the resulting
root-finding noise in $r_{\rm ISCO}^{(1,1)}$ is about $4\times10^{-4}$
(prograde) and up to $5\times10^{-3}$ (retrograde) in absolute terms, which is
significant only near the retrograde zero. The Lyapunov exponent requires
$R''(r_{\rm ph})$, which we evaluate with a five-point stencil
(appendix~\ref{app:qnm}).

\emph{Crossing spins.} These are limited by interpolation on the
$\Delta a=0.02$ grid, to $\pm0.001$. (See section~\ref{sec:signs}.)

The largest uncertainties in our set are thus the $\approx0.5\%$ truncation
error at $a=0.9$ in the prograde $r_{\rm ISCO}^{(1,1)}$ and the retrograde
$\Omega_c^{(1,1)}$ and $\lambda^{(1,1)}$, and the root-finding noise of the
retrograde $r_{\rm ISCO}^{(1,1)}$ near its zero; every other mixed coefficient
is accurate to a part in $10^{3}$ or better over the quoted range. The first-law test in
section~\ref{sec:entropy}, which closes to one part in $10^{5}$, is independent of
all of the above and supports the validity of the error budget as a whole.

It is also worth commenting on the completeness of the ansatz we have used.
The solver collocates the five even-parity components
$(tt,\,t\phi,\,rr,\,\chi\chi,\,\phi\phi)$ against the four
functions $H_{1\ldots4}$, with $H_3$ shared between $g_{rr}$ and $g_{\chi\chi}$.
This is not an over-restriction; for a stationary, axisymmetric, circular
perturbation the $r\chi$-block carries three components ($rr$, $r\chi$, $\chi\chi$)
and two gauge functions ($\xi^{r}$, $\xi^{\chi}$), which can be used to set
$h_{r\chi}=0$ and
$h_{rr}/g_{rr}^{(0)}=h_{\chi\chi}/g_{\chi\chi}^{(0)}\equiv H_3$.
(See appendix~\ref{app:gauge} for details.) The six nontrivial field equations for the four
remaining functions are then related by the two nontrivial components of the
contracted Bianchi identity. That a solution exists within this ansatz is shown
directly by the residuals quoted above: all six equations are satisfied---the
sixth indepdently, without being imposed. The fit of the imported node data to the spectral
representation is exact to round-off ($\sim\!10^{-15}$) at both $30\times30$
and $50\times50$ resolutions.

\subsection{Perturbative control and the domain of validity}
\label{sec:eftcontrol}

The exclusion of the near-extremal regime, stated qualitatively above, can be
made quantitative. Since the metric ansatz~\eqref{eq:ansatz} is multiplicative,
the deformation functions $H_i$ measure the fractional correction to the
corresponding metric components; a natural control parameter is therefore their
maximum amplitude over the exterior,
\begin{equation}
A^{(m,n)}(a) \equiv \max_{r,\chi}\,\big|H_i^{(m,n)}(r,\chi)\big|,
\end{equation}
and the total fractional deformation is bounded by
\begin{equation}
\epsilon_{\rm pert}(a) =
\zGB\,A^{(1,0)} + \zCS\,A^{(0,1)} + \zGB\zCS\,A^{(1,1)}.
\label{eq:epspert}
\end{equation}

\begin{figure}[htbp]
\centering
\includegraphics[width=\columnwidth]{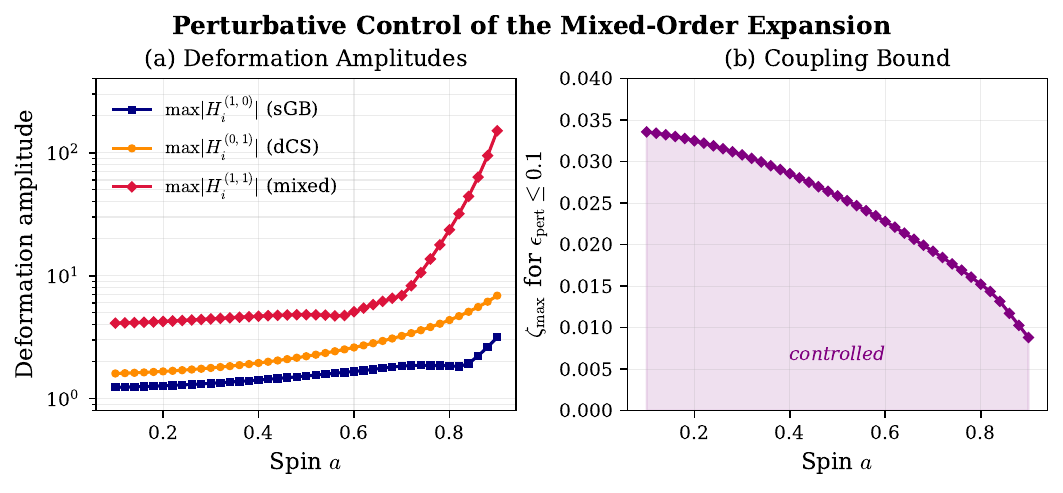}
\caption{Perturbative control of the mixed-order expansion. (a) Maximum
amplitude $A^{(m,n)}=\max_{r,\chi}|H_i^{(m,n)}|$ of the deformation functions
for the two pure sectors and the mixed sector. The mixed amplitude exceeds both
pure-sector amplitudes at every spin and grows steeply for $a\gtrsim0.7$,
identifying the mixed correction as the first to leave the
perturbative regime as the black hole moves toward extremality. (b) Maximum admissible coupling
$\zeta_{\max}$ (with $\zGB=\zCS=\zeta$) for a total fractional deformation
$\epsilon_{\rm pert}\leq0.1$; the shaded region is perturbatively controlled.}
\label{fig:eft}
\end{figure}

Figure~\ref{fig:eft} a shows the three amplitudes. The pure-sector amplitudes
grow only mildly across the spin range, from $A^{(1,0)}\approx1.2$ to $3.2$ and
$A^{(0,1)}\approx1.6$ to $6.9$. The mixed amplitude behaves very differently; it
exceeds both pure-sector amplitudes at every spin, is nearly flat at
$A^{(1,1)}\approx4$--$5$ up to $a\approx0.6$, and then grows steeply, to
$\approx24$ at $a=0.8$ and $\approx151$ at $a=0.9$. It is thus the mixed
correction that first threatens the perturbative ordering as extremality is
approached; by $a=0.99$ the mixed amplitude reaches $\sim\!3\times10^{4}$, and
the expansion is no longer controlled for any but the tiniest couplings. At low
spin $A^{(1,1)}$ is set by $H_2^{(1,1)}$ on the horizon, which tends to a
nonzero constant as $a\to0$. Since $H_2$ multiplies $g^{(0)}_{t\phi}\propto a$,
the corresponding metric deformation still vanishes in that limit, so
$A^{(1,1)}$ overstates the mixed deformation at low spin.

Requiring $\epsilon_{\rm pert}\le0.1$ as a representative threshold and setting
$\zGB=\zCS=\zeta$ for definiteness, eq.~\eqref{eq:epspert} yields the maximum
admissible coupling shown in figure~\ref{fig:eft} b: $\zeta_{\max}$ decreasing
monotonically from $\approx0.034$ at $a=0.1$ to $\approx0.009$ at $a=0.9$. This
is a quantitative description of the controlled domain within which our results
should be interpreted. We stress that the threshold $0.1$ is a chosen
convention for perturbative control, not a universal EFT bound, and that
$\epsilon_{\rm pert}$ is a coordinate measure; a fully invariant diagnostic
would require fractional corrections to curvature scalars, which we have not
computed here.

Finally, we emphasize the distinction between numerical solvability and
perturbative control. The equilibrated Stage-II system
(appendix~\ref{app:conditioning}) remains numerically well posed even at
$a=0.99$, where the residual of the collocated equations is $\sim\!10^{-9}$.
That the perturbative expansion is
nonetheless uncontrolled there is a statement about the physics of the
expansion, not about the numerics.

\subsection{Physical magnitude and observability}
\label{sec:observability}

\begin{figure}[htbp]
\centering
\includegraphics[width=0.35\textwidth]{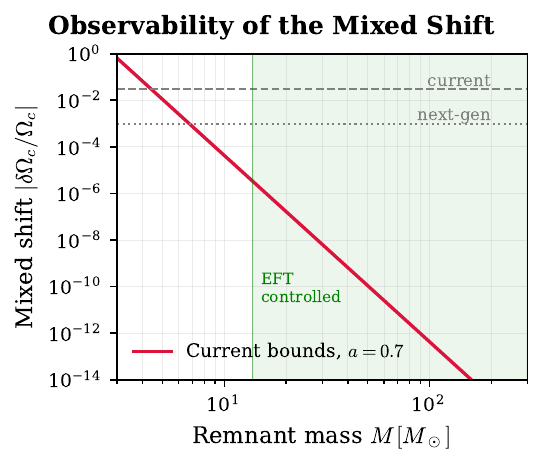}
\caption{Fractional mixed-order correction to the eikonal ringdown frequency
versus remnant mass, at $a=0.7$, for current coupling bounds. The shift scales
as $M^{-8}$. Horizontal lines mark representative ringdown-frequency precisions
(current and next-generation); the shaded region ($M\gtrsim14\,M_\odot$) is
where the dCS coupling at its current bound remains perturbatively controlled.
Within the controlled region the mixed effect lies well below any foreseeable
detection threshold.}
\label{fig:obs}
\end{figure}

To assess the physical size of the mixed corrections we restore dimensions. A
fractional shift in an observable $O$ is
\begin{equation}
\frac{\delta O}{O} = \zGB\,\zCS\,\frac{O^{(1,1)}}{O^{(0)}}
= \frac{\alpha_{\rm GB}^{2}\alpha_{\rm CS}^{2}}{M^8}\frac{O^{(1,1)}}{O^{(0)}},
\end{equation}
so that the mixed shift falls as $M^{-8}$ with black-hole mass, as the
product of two separately small corrections. Adopting representative
current bounds $\sqrt{\alpha_{\rm GB}}\lesssim1.7$~km~\cite{Nair2019} and
$\sqrt{\alpha_{\rm CS}}\lesssim8.5$~km~\cite{Silva2021}, and a remnant with
$a=0.7$, the mixed correction to the eikonal ringdown frequency is
$|\delta\Omega_c/\Omega_c|\approx 7\times10^{-9}$ at $M=30\,M_\odot$ and smaller
still for heavier remnants. This is shown in figure~\ref{fig:obs}).

The steep mass dependence has a clear consequence. Requiring the dCS deformation
to remain perturbative at its current bound restricts $M\gtrsim14\,M_\odot$;
throughout this controlled range the mixed shift is $\lesssim10^{-5}$, far below
the precision of current or next-generation ringdown measurements. The shift
reaches the $\sim\!0.1\%$ level only for $M\lesssim 7\,M_\odot$, where the dCS
coupling bound no longer guarantees a controlled small-coupling expansion. The
mixed signature is therefore not a target for direct detection with present
constraints; its value is as a clean, parametrically distinct diagnostic of
\emph{simultaneous} parity-even and parity-odd corrections, whose amplitude is
fixed by the product of the two couplings and could become relevant only if
future observations were to constrain both couplings much closer to the
small-coupling boundary, or in hypothetical low-mass systems. We stress that a
meaningful search for the mixed effect presupposes prior detection or tight
constraint of the individual pure sectors.

\subsection{Low-spin parity scaling}
\label{sec:lowspin}

Because the dynamical Chern--Simons sector is sourced only by rotation---the
Pontryagin density vanishing for a static background---every quantity that
depends on the dCS sector must vanish as $a\to0$ with a definite power of the
spin. This provides an independent check on the spin-dependence of the source
assembly, complementary to the exact-limit and residual tests, since it probes
the analytic structure of the corrections rather than their value at a single
spin.

\begin{figure}[htbp]
\centering
\includegraphics[width=0.35\textwidth]{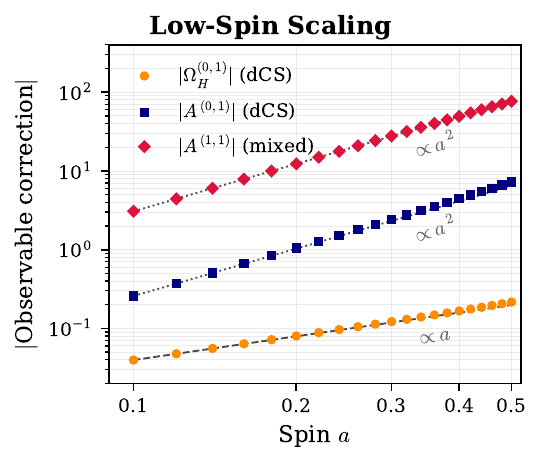}
\caption{Low-spin scaling of representative pure-dCS and mixed corrections. The
pure-sector dCS horizon angular velocity scales as $\propto a$ and the dCS
horizon area as $\propto a^2$; the mixed horizon area inherits the $\propto a^2$
suppression of the spin-sourced dCS sector. Dashed and dotted lines show the
reference power laws $a$ and $a^2$. Fitted low-spin slopes are $1.02$, $2.03$,
and $2.00$ respectively.}
\label{fig:lowspin}
\end{figure}

Figure~\ref{fig:lowspin} shows the low-spin behavior of representative pure-dCS
and mixed corrections on a logarithmic scale. The pure-sector dCS corrections
follow clean power laws: the horizon angular velocity scales as
$\Omega_H^{(0,1)}\propto a$ (fitted slope $1.02$), reflecting that the fractional
dCS deformation of the gravitomagnetic metric component is finite at low spin
while $\Omega_H^{(0)}\propto a$ itself, whereas the horizon area scales as
$A^{(0,1)}\propto a^2$ (slope $2.03$). The mixed horizon area inherits the same
suppression, $A^{(1,1)}\propto a^2$ (slope $2.00$), confirming that the mixed
correction vanishes at the rate set by the spin-sourced dCS sector.

The other mixed horizon coefficients follow the same pattern. Under a reversal
of the spin, $\Omega_H$ is odd while $\kappa$, $A$, and $S$ are even, so the
mixed coefficients, which vanish at $a=0$, must do so as $a$ and $a^2$
respectively. We find $\Omega_H^{(1,1)}\propto a$ (slope $0.92$ for
$a\le0.2$), and $\kappa^{(1,1)}$ and $S^{(1,1)}$ $\propto a^2$ (slopes $1.98$
and $2.00$). The two parts of $\Omega_H^{(1,1)}$ in eq.~\eqref{eq:omega11}, the
linear response to $h^{(1,1)}$ and the pure-sector cross terms, each scale as
$a$ and are both negative at low spin. The sign change of $\Omega_H^{(1,1)}$
discussed in section~\ref{sec:signs} is therefore not a low-spin effect, as it arises
at $a\approx0.61$, after the linear response has reversed.

\section{Conclusions}
\label{sec:conclusion}

We have computed the leading mixed-order $\Oone$ corrections to rotating black
holes in the combined scalar-Gauss--Bonnet and dynamical Chern--Simons
effective theory, the regime in which the parity-even and parity-odd sectors act
simultaneously. Solving the coupled scalar and linearized metric equations to
second order in the couplings with a pseudospectral method, we obtained the
mixed scalar fields and the mixed metric correction $h^{(1,1)}_{\mu\nu}$ for
rapidly rotating black holes, and from these extracted a suite of observables spanning horizon, orbital, and ringdown
physics at fixed mass and angular momentum: the horizon angular velocity
$\Omega_H^{(1,1)}$, surface gravity $\kappa^{(1,1)}$, area $A^{(1,1)}$, and
Wald entropy $S^{(1,1)}$; the ISCO and photon-ring shifts $r_{\rm ISCO}^{(1,1)}$,
$r_{\rm ph}^{(1,1)}$ for both orbital senses; and the eikonal quasinormal
frequency and damping-rate corrections $\Omega_c^{(1,1)}$ and $\lambda^{(1,1)}$.
We verified the exact Kerr and Schwarzschild limits, reproduced
available pure-sector benchmarks, and quantified the remaining numerical uncertainties of the mixed observables.
The mixed solution satisfies all six nontrivial field equations, one of which
was not imposed in its construction, and its Wald entropy agrees with the first
law of black hole thermodynamics to about one part in $10^{5}$.

Our central finding is that the mixed sector is qualitatively distinct from
either pure theory. All four horizon coefficients change sign as functions of spin, as do several
orbital and eikonal ones. Since these mixed observables are
proportional to $\zGB\zCS$, they vanish unless both couplings are present, and
the spin-dependent sign reversal is therefore a distinctive mixed-coupling signature
within the combined sGB--dCS EFT. The
orbital and ringdown observables carry a pronounced prograde/retrograde
asymmetry; in particular, the mixing raises the prograde ringdown frequency for
$a\gtrsim0.23$ and lengthens its damping time for $a\gtrsim0.65$, an effect of
direct relevance to ringdown tests of the Kerr hypothesis. Off the equator, the breakdown of Carter separability leaves the photon orbits
nearly but not exactly spherical, and this departure from sphericity is an order of
magnitude larger at mixed order than in either pure theory. The entropy also
shows that the horizon area alone is a poor guide to the thermodynamics at this
order; for $a\lesssim0.6$ the area term of $S^{(1,1)}$ is about five times the
entropy and of opposite sign, while at $a=0.9$ it contributes less than a tenth
of it.

Along the way we identified and resolved a severe ill-conditioning problem with the
mixed-order metric system at high spin; without equilibrating the matrix operators to
be inverted, the near-extremal results
would be unreliable. We also found that holding the mass and angular momentum fixed
at mixed order requires inhomogeneous asymptotic conditions, because the
pure-sector metric functions do not vanish at infinity, and that the
perturbative expansion itself ceases to be controlled as extremality is
approached, which we could quantify through the rapid growth of $h^{(1,1)}$.

Two important directions remain open. A full QNM calculation, in which the gravitational perturbations
are coupled to both scalar fields, is the natural next step beyond the eikonal estimates obtained
here and would sharpen the connection to ringdown observations. Moreover, the
corrected rotating solution constructed in this work could describe the remnant of a
compact-binary merger, providing a route by which the mixed-order corrections
could enter gravitational-wave observations directly.

\section*{Acknowledgments}

We are grateful to Kelvin Ka-Ho Lam and  Nicol\'{a}s Yunes for their assistance with understanding their
algorithms and data.

\bibliography{references}

\appendix

\section{Field Equations and Correction Tensors}
\label{app:fieldeqs}

This appendix collects the variation of the action~\eqref{eq:action} and the
explicit form of the two curvature-coupling tensors that enter the modified
Einstein equations. We work with the matter assumed to be vacuum, $S_{\rm mat}=0$, throughout.

\subsection{Einstein--Hilbert and kinetic terms}
\label{app:fieldeqs-eh}

The variation of the Einstein--Hilbert term gives the Einstein tensor
$G_{\mu\nu} = R_{\mu\nu} - \tfrac12 g_{\mu\nu}R$ in the standard way,
\begin{equation}
\delta\!\int\! d^4x\,\sqrt{-g}\,R
= \int\! d^4x\,\sqrt{-g}\;G_{\mu\nu}\,\delta g^{\mu\nu}.
\end{equation}
Varying the canonical kinetic terms yields the scalars' stress-energy tensors.
With the convention
$T_{\mu\nu} = -\big(2/\sqrt{-g}\big)\,\delta(\sqrt{-g}\,\mathcal{L})/\delta g^{\mu\nu}$,
\begin{align}
T^{(\varphi)}_{\mu\nu} &= \nabla_\mu\varphi\nabla_\nu\varphi
- \tfrac12 g_{\mu\nu}\big(\nabla\varphi\big)^2,
\label{eq:Tphi}\\
T^{(\vartheta)}_{\mu\nu} &= \nabla_\mu\vartheta\nabla_\nu\vartheta
- \tfrac12 g_{\mu\nu}\big(\nabla\vartheta\big)^2 .
\label{eq:Ttheta}
\end{align}

\subsection{The Gauss--Bonnet sector}
\label{app:fieldeqs-gb}

Consider the dilaton Gauss--Bonnet contribution
\begin{equation}
S_{\rm GB} = \aGB\!\int\! d^4x\,\sqrt{-g}\;\varphi\,\GB .
\end{equation}
Because $\GB$ is a topological density in four dimensions, the variation of the
uncoupled term $\int d^4x\,\sqrt{-g}\,\GB$ vanishes identically; the entire
contribution to the field equations arises from derivatives of the scalar. The
variation gives
\begin{equation}
\delta S_{\rm GB} = \aGB\!\int\! d^4x\,\sqrt{-g}\;
\mathcal{H}^{(\varphi)}_{\mu\nu}\,\delta g^{\mu\nu},
\end{equation}
with
\begin{align}
\mathcal{H}^{(\varphi)}_{\mu\nu}
={}& 4R_{\mu\rho\nu\sigma}\nabla^\rho\nabla^\sigma\varphi
- 2R\,\nabla_\mu\nabla_\nu\varphi
+ 2g_{\mu\nu}R\,\Box\varphi \nonumber\\
&- 4R_{\mu\nu}\Box\varphi
+ 4R_\mu{}^\rho\nabla_\rho\nabla_\nu\varphi
+ 4R_\nu{}^\rho\nabla_\rho\nabla_\mu\varphi \nonumber\\
&- 4g_{\mu\nu}R^{\rho\sigma}\nabla_\rho\nabla_\sigma\varphi .
\label{eq:Htensor}
\end{align}
Equation~\eqref{eq:Htensor} can be written compactly in terms of the double dual
of the Riemann tensor,
${}^{\ast\ast}\!R_{\mu\rho\nu\sigma} \equiv \tfrac14\,
\epsilon_{\mu\rho\alpha\beta}R^{\alpha\beta\gamma\delta}
\epsilon_{\gamma\delta\nu\sigma}$,
\begin{equation}
\mathcal{H}^{(\varphi)}_{\mu\nu}
= -4\;{}^{\ast\ast}\!R_{\mu\rho\nu\sigma}\,\nabla^\rho\nabla^\sigma\varphi .
\label{eq:Hcompact}
\end{equation}
On a Ricci-flat spacetime
${}^{\ast\ast}\!R_{\mu\rho\nu\sigma} = -R_{\mu\rho\nu\sigma}$, every Ricci and
scalar-curvature term in eq.~\eqref{eq:Htensor} drops out, and
$\mathcal{H}^{(\varphi)}_{\mu\nu} = 4R_{\mu\rho\nu\sigma}
\nabla^\rho\nabla^\sigma\varphi$. This is the form that applies at leading order,
where the tensor is evaluated on the Kerr background. At mixed order the
Ricci-dependent terms contribute, because the metric deformation of the opposite
sector is not Ricci flat. (See appendix~\ref{app:exp-tensors}.)

\subsection{The Chern--Simons sector}
\label{app:fieldeqs-cs}

The axion-Chern--Simons contribution is determined similarly. It is
\begin{equation}
S_{\rm CS} = \aCS\!\int\! d^4x\,\sqrt{-g}\;\vartheta\,\pRR .
\end{equation}
Using the fact that the Pontryagin density is a total divergence,
$\pRR = \nabla_\mu K^\mu$ with $K^\mu$ the gravitational Chern--Simons current,
and integrating by parts (discarding the boundary term),
\begin{equation}
S_{\rm CS} = -\aCS\!\int\! d^4x\,\sqrt{-g}\;\big(\nabla_\mu\vartheta\big)K^\mu,
\end{equation}
which makes manifest that the sector contributes only when $\vartheta$ is
spacetime dependent. Varying with respect to the metric gives
\begin{equation}
\delta S_{\rm CS} = \aCS\!\int\! d^4x\,\sqrt{-g}\;
\mathcal{C}_{\mu\nu}\,\delta g^{\mu\nu},
\end{equation}
where $\mathcal{C}_{\mu\nu}$ is the Chern--Simons generalization of the Cotton tensor. According to the
conventions of Ref.~\cite{LamChungYunes2025}, which we have adopted, it is
\begin{equation}
\mathcal{C}_{\mu\nu} = -4\Big[\big(\nabla_\sigma\vartheta\big)\,
\epsilon^{\sigma\delta\alpha}{}_{(\mu|}\nabla_{\alpha} R_{|\nu)\delta}
+ \big(\nabla_\sigma\nabla_\tau\vartheta\big)\,
{}^{\ast}\!R^{\tau}{}_{(\mu\nu)}{}^{\sigma}\Big],
\label{eq:Ctensor}
\end{equation}
with the dual Riemann tensor of Eq.~\eqref{eq:dualriemann}. Each term contains
one Levi--Civita tensor, which is the origin of the parity-odd protperties of the
sector. On the Ricci-flat Kerr background the first term, which is built from
$\nabla_\alpha R_{\nu\delta}$, vanishes identically, so that only the
dual-Riemann term contributes to the leading correction.

\subsection{The coupled system}
\label{app:fieldeqs-system}

Collecting all contributions and imposing $\delta S/\delta g^{\mu\nu}=0$ gives
the modified Einstein equations quoted in the main text,
\begin{equation}
G_{\mu\nu} + \aGB\,\mathcal{H}^{(\varphi)}_{\mu\nu} + \aCS\,\mathcal{C}_{\mu\nu}
= \tfrac12\Big[T^{(\varphi)}_{\mu\nu} + T^{(\vartheta)}_{\mu\nu}\Big].
\label{eq:einsteinapp}
\end{equation}
Meanwhile, varying with respect to the scalars gives their sourced wave equations
\begin{align}
\Box\varphi &= -\,\aGB\,\GB ,
\label{eq:phiapp}\\
\Box\vartheta &= -\,\aCS\,\pRR .
\label{eq:thetaapp}
\end{align}
The parity assignment of each sector is manifest here. The Gauss--Bonnet
invariant is parity even and sources the dilaton; the Pontryagin density is
parity odd and sources the axion. Their behavior on black-hole backgrounds
differs in a way that shapes the entire calculation: $\GB$ is nonzero already for the
Schwarzschild geometry, and therefore always sources $\varphi$; whereas $\pRR$ vanishes
identically for a static background and is proportional to the spin for a Kerr black hole, so
that $\vartheta$---and with it the whole dCS sector---is excited only for rotating
black holes. This is the origin of the contrasting $a\to0$ limits observed in the
pure-sector observables in sections~\ref{sec:horizon}--\ref{sec:orbital}.

Equations~(\ref{eq:einsteinapp}--\ref{eq:thetaapp}), together with the
definitions of the invariants and of the correction tensors
\eqref{eq:Hcompact} and \eqref{eq:Ctensor}, constitute the complete system. It
is nonlinear and mutually coupled, both through the geometry and through the
higher-curvature invariants. The perturbative scheme via which we solve it is
laid out in appendix~\ref{app:expansion}.

\subsection{Numerical implementation and checks}
\label{app:fieldeqs-checks}

The correction tensors are constructed directly from
eqs.~\eqref{eq:Htensor} and~\eqref{eq:Ctensor} on the spectral grid, using
analytically differentiated Kerr curvature tensors and covariant Hessians of the
scalars. Two properties provide useful implementation checks. First,
$\mathcal{C}_{\mu\nu}$ is identically traceless, and the trace of
$\mathcal{H}^{(\varphi)}_{\mu\nu}$ is
$-4G^{\rho\sigma}\nabla_\rho\nabla_\sigma\varphi$, which vanishes on a
Ricci-flat background. We
verify $g^{\mu\nu}\mathcal{H}^{(\varphi)}_{\mu\nu}$ and
$g^{\mu\nu}\mathcal{C}_{\mu\nu}$ vanish to $\sim\!10^{-14}$ on the Kerr
background, at the level of round-off precision. Second, the normalization of each tensor
is fixed---not assumed---by requiring that the corresponding pure-sector field
equations be satisfied by the known leading-order solutions taken from
Ref.~\cite{LamChungYunes2025}. Inserting $h^{(1,0)}$ and $\varphi^{(1/2,0)}$
into the $\mathcal{O}(\zGB)$ Einstein equation, and $h^{(0,1)}$ and
$\vartheta^{(0,1/2)}$ into the $\mathcal{O}(\zCS)$ equation, we find relative
residuals of $\sim\!10^{-11}$ over most of the spin range
(appendix~\ref{app:exp-audit}), which both validates the tensors and pegs their
normalizations unambiguously.

\section{The Double-Parameter Expansion}
\label{app:expansion}

This appendix develops the perturbative scheme by which the coupled system of
appendix~\ref{app:fieldeqs} is solved. We establish the power counting, expand
each object order by order in the two couplings, and assemble the mixed-order
source.

\subsection{Power counting}
\label{app:exp-counting}

The ordering of the expansion follows directly from the field equations. The
scalars are \emph{linear} in their couplings. The wave
equations~(\ref{eq:phiapp}--\ref{eq:thetaapp}) are sourced by curvature
invariants that are already nonzero on the background, and with vanishing
background values $\varphi^{(0)}=\vartheta^{(0)}=0$ we have that
$\varphi = \mathcal{O}(\aGB)$ and $\vartheta = \mathcal{O}(\aCS)$.
The magnitudes follow from a dimensional estimate. Outside the horizon the only
length scale is $M$, so each derivative contributes a factor $1/M$ and the wave
operator scales as $\Box\sim M^{-2}$. Near
the horizon the sources are $\GB\sim M^{-4}$ and $\pRR\sim a\,M^{-4}$, with $a$
the dimensionless spin, so that $\varphi\sim\aGB/M^2$ and
$\vartheta\sim\aCS\,a/M^2$.

The metric, by contrast, is \emph{quadratic}. Every source of a metric
deformation in eq.~\eqref{eq:einsteinapp} carries two powers of a scalar,
either explicitly or through the coupling.
The curvature-coupling tensors enter as $\aGB\mathcal{H}^{(\varphi)}_{\mu\nu}$
and $\aCS\mathcal{C}_{\mu\nu}$, making them
$\mathcal{O}(\aGB^2)$ and $\mathcal{O}(\aCS^2)$, respectively.
Meanwhile, the scalar stress-energy tensors $T^{(\varphi)}_{\mu\nu}\sim(\nabla\varphi)^2$
and $T^{(\vartheta)}_{\mu\nu}\sim(\nabla\vartheta)^2$ are manifestly
$\mathcal{O}(\aGB^2)$ and $\mathcal{O}(\aCS^2)$ as well.
There is no $\mathcal{O}(\aGB)$ or $\mathcal{O}(\aCS)$ metric deformation.
In terms of the dimensionless parameters $\zGB$ and $\zCS$ from
eq.~\eqref{eq:zetadef}, the scalars therefore enter at half-integer order and
the metric at integer order.

The metric expanded about the Kerr background is
\begin{align}
&g_{\mu\nu} = g^{(0)}_{\mu\nu}
+ \zGB\,h^{(1,0)}_{\mu\nu} + \zCS\,h^{(0,1)}_{\mu\nu} \nonumber\\
&+ \zGB^2\,h^{(2,0)}_{\mu\nu} + \zGB\zCS\,h^{(1,1)}_{\mu\nu}
+ \zCS^2\,h^{(0,2)}_{\mu\nu} + \mathcal{O}(\zeta^3),
\label{eq:gexpapp}
\end{align}
with the superscripts $(m,n)$ denoting the
orders in $(\zGB,\zCS)$. The scalars, being sourced at half-integer
order, expand as
\begin{align}
\varphi &= \zGB^{1/2}\,\varphi^{(1/2,0)}
+ \zGB^{3/2}\,\varphi^{(3/2,0)}
+ \zGB^{1/2}\zCS\,\varphi^{(1/2,1)} + \dots,
\label{eq:phiexp}\\
\vartheta &= \zCS^{1/2}\,\vartheta^{(0,1/2)}
+ \zCS^{3/2}\,\vartheta^{(0,3/2)}
+ \zGB\zCS^{1/2}\,\vartheta^{(1,1/2)} + \dots.
\label{eq:thetaexp}
\end{align}

\subsection{Order of the mixed sector}
\label{app:exp-mixedorder}

That the leading parity-mixing effect appears at $\Oone$---and not earlier---is
worth establishing carefully, since it is one of the underlying premises for the entire calculation.
Because every metric source is quadratic in a scalar, and each scalar carries a
definite coupling, the metric deformation is \emph{even} in each coupling
separately. The two sectors therefore cannot communicate at any order odd in
either coupling. Nor do they communicate through the kinetic terms, since the scalar
stress-energy tensors~(\ref{eq:Tphi}--\ref{eq:Ttheta}) contain no
$\varphi$-$\vartheta$ cross term.

The sectors first communicate through the sub-leading, cross-sourced scalars.
The dilaton acquires a correction once the Gauss--Bonnet invariant feels the
Chern--Simons metric deformation $h^{(0,1)} = \mathcal{O}(\aCS^2)$, and
conversely; so $\delta\varphi = \mathcal{O}\big(\aGB\,\aCS^2\big)$ and
$\delta\vartheta = \mathcal{O}\big(\aGB^2\,\aCS\big)$.
Feeding these perturbations back into the metric sources---either through the cross term
$2\nabla_\mu\varphi\,\nabla_\nu(\delta\varphi)$ in the stress-energy, or through
$\aGB\mathcal{H}^{(\varphi)}_{\mu\nu}$ evaluated using the opposite sector's
curvature deformation---the first genuinely mixed metric deformation appears at
$\mathcal{O}\big(\aGB^2\,\aCS^2\big)
= \mathcal{O}\big(\zGB\,\zCS\big).$

\subsection{Expansion of the correction tensors}
\label{app:exp-tensors}

\subsubsection{Gauss--Bonnet sector}

Substituting eq.~\eqref{eq:phiexp} into eq.~\eqref{eq:Htensor} and retaining
terms of order $\zGB$ gives $\mathcal{H}^{(1,0)}_{\mu\nu}$ as a full
seven-term expression evaluated on the background with
$\varphi\to\varphi^{(1/2,0)}$. On the specifically Kerr background, however, $R^{(0)}=0$ and
$R^{(0)}_{\mu\nu}=0$, so every Ricci-dependent term vanishes identically,
leaving the compact result
\begin{equation}
\mathcal{H}^{(1,0)}_{\mu\nu}
= 4R^{(0)}_{\mu\rho\nu\sigma}\,
\nabla^{(0)\rho}\nabla^{(0)\sigma}\varphi^{(1/2,0)} .
\label{eq:H10}
\end{equation}
Here we have used $\aGB/\zGB^{1/2} = M^2 = 1$ to express the result in terms of
the $\zeta$-scaled fields.

The mixed-order contribution follows by expanding eq.~\eqref{eq:Htensor} to
first order in $\zCS$ about the Kerr background, since $\varphi^{(1/2,0)}$ is
already $\mathcal{O}(\zGB^{1/2})$. It separates into two parts according to
which field it carries,
\begin{equation}
\mathcal{H}^{(1,1)}_{\mu\nu}
= \mathcal{H}^{(1,1)}_{\mu\nu}\Big|_{\rm curvature}
+ \mathcal{H}^{(1,1)}_{\mu\nu}\Big|_{\rm source} .
\label{eq:Hsplit}
\end{equation}
The first contribution to eq.~\eqref{eq:Hsplit} is the mixed dilaton
$\varphi^{(1/2,1)}$ contracted with the \emph{background} Riemann tensor,
\begin{equation}
\mathcal{H}^{(1,1)}_{\mu\nu}\Big|_{\rm curvature}
= 4R^{(0)}_{\mu\rho\nu\sigma}\,
\nabla^{(0)\rho}\nabla^{(0)\sigma}\varphi^{(1/2,1)} .
\label{eq:Hcurv}
\end{equation}
This term is not known at the outset; it requires the mixed scalar, which is
itself the solution of eq.~\eqref{eq:phi11}. It becomes available once Stage~I has
been completed, and is then assembled into the Stage~II source alongside the
remaining terms.

The second term on the right-hand side of eq.~\eqref{eq:Hsplit} depends only on lower-order
solutions already in hand. For concision, in the remainder of this subsection we abbreviate
$\varphi\equiv\varphi^{(1/2,0)}$ and $h_{\mu\nu}\equiv h^{(0,1)}_{\mu\nu}$, and
all unlabeled quantities are those of the Kerr background, whose metric also raises and
lowers indices. Then
\begin{equation}
\begin{aligned}
\mathcal{H}^{(1,1)}_{\mu\nu}\Big|_{\rm source}
={}& 4R^{(0,1)}_{\mu\rho\nu\sigma}\,\nabla^{\rho}\nabla^{\sigma}\varphi\\
&+ 4R_{\mu\rho\nu\sigma}
\big(\delta\nabla^\rho\nabla^\sigma\varphi\big)^{(0,1)}\\
&- 2\,\delta R\,\nabla_\mu\nabla_\nu\varphi
+ 2g_{\mu\nu}\,\delta R\,\Box\varphi\\
&- 4\,\delta R_{\mu\nu}\,\Box\varphi
- 4g_{\mu\nu}\,\delta R^{\rho\sigma}\,\nabla_\rho\nabla_\sigma\varphi\\
&+ 4\,\delta R_\mu{}^\rho\,\nabla_\rho\nabla_\nu\varphi
+ 4\,\delta R_\nu{}^\rho\,\nabla_\rho\nabla_\mu\varphi .
\end{aligned}
\label{eq:Hsource}
\end{equation}
Here $R^{(0,1)}_{\mu\rho\nu\sigma}$ is the first-order perturbation of the
Riemann tensor (with all indices lowered), as induced by the Chern--Simons metric
deformation; and
\begin{equation}
\begin{aligned}
\big(\delta\nabla^\rho\nabla^\sigma\varphi\big)^{(0,1)}
={}& -g^{\rho\alpha}g^{\sigma\beta}\,\delta\Gamma^\lambda_{\alpha\beta}\,
\partial_\lambda\varphi\\
&- h^{\rho\alpha}\nabla_\alpha\nabla^\sigma\varphi
- h^{\sigma\alpha}\nabla^\rho\nabla_\alpha\varphi,
\label{eq:dHess}
\end{aligned}
\end{equation}
with
\begin{equation}
\delta\Gamma^\lambda_{\alpha\beta}
=\tfrac12 g^{\lambda\kappa}\big(\nabla_\alpha h_{\beta\kappa}
+ \nabla_\beta h_{\alpha\kappa} - \nabla_\kappa h_{\alpha\beta}\big),
\end{equation}
is the perturbation of the Hessian with raised indices. The first term comes
from the connection and the other two from the inverse metrics that raise the
indices. The last three lines of eq.~\eqref{eq:Hsource} are the
Ricci-dependent terms of eq.~\eqref{eq:Htensor}. They vanish on the Kerr
background but not at this order, because the Chern--Simons deformation is not
Ricci flat; instead, the trace-reversed form of eq.~\eqref{eq:einstein01} below gives
\begin{equation}
\delta R_{\mu\nu} \equiv R^{(0,1)}_{\mu\nu}
= \tfrac12\,\partial_\mu\vartheta^{(0,1/2)}\partial_\nu\vartheta^{(0,1/2)}
- \mathcal{C}^{(0,1)}_{\mu\nu},
\label{eq:dRic01}
\end{equation}
with $\delta R = g^{\mu\nu}\delta R_{\mu\nu}$,
$\delta R_\mu{}^\rho = g^{\rho\sigma}\delta R_{\mu\sigma}$, and
$\delta R^{\rho\sigma} = g^{\rho\alpha}g^{\sigma\beta}\delta R_{\alpha\beta}$.
All of these terms encode the same kind of physical effect: how the parity-odd deformation
of the geometry modifies the way the parity-even dilaton gravitates.

\subsubsection{Chern--Simons sector}

The Cotton tensor~\eqref{eq:Ctensor} may be expands analogously. Since the axion
expansion~\eqref{eq:thetaexp} begins at $\mathcal{O}(\zCS^{1/2})$, the
combination $\aCS\mathcal{C}_{\mu\nu}$ expands as
\begin{align}
\aCS\,\mathcal{C}_{\mu\nu} &= \zCS\,\mathcal{C}^{(0,1)}_{\mu\nu}
+ \zCS^{3/2}\,\mathcal{C}^{(0,3/2)}_{\mu\nu} \nonumber\\
    &\quad + \zGB\zCS\,\mathcal{C}^{(1,1)}_{\mu\nu} + \mathcal{O}(\zeta^{5/2}).
\label{eq:Cexp}
\end{align}
The background contribution vanishes since $\vartheta^{(0)}=0$, and there is no
pure Gauss--Bonnet contribution, consistent with the power counting in
sections~\ref{app:exp-counting}--\ref{app:exp-mixedorder}.

The mixed term $\mathcal{C}^{(1,1)}_{\mu\nu}$ has a structure directly analogous
to that of eqs.~(\ref{eq:Hsplit}--\ref{eq:Hsource}), with the dual Riemann tensor replacing
the Riemann tensor and the axion replacing the dilaton, and it admits the same type of
splitting,
\begin{equation}
\mathcal{C}^{(1,1)}_{\mu\nu}
= \mathcal{C}^{(1,1)}_{\mu\nu}\Big|_{\rm curvature}
+ \mathcal{C}^{(1,1)}_{\mu\nu}\Big|_{\rm source} .
\label{eq:Csplit}
\end{equation}
The first contribution involves the mixed-order axion
$\vartheta^{(1,1/2)}$ contracted with the background dual Riemann tensor,
\begin{equation}
\mathcal{C}^{(1,1)}_{\mu\nu}\Big|_{\rm curvature}
= -4\,\left[\nabla^{(0)\alpha}\nabla^{(0)\beta}\vartheta^{(1,1/2)}\right]
{}^{\ast}\!R^{(0)}_{\alpha(\mu\nu)\beta}.
\label{eq:Ccurv}
\end{equation}
As with its Gauss--Bonnet counterpart~\eqref{eq:Hcurv}, this term requires the
mixed scalar, which is obtained in Stage~I from the mixed axion equation; it is
therefore a known quantity by the time the Stage~II metric source is assembled.

The remaining terms depend only on lower-order solutions,
\begin{equation}
\begin{aligned}
\mathcal{C}^{(1,1)}_{\mu\nu}\Big|_{\rm source}
={}& -4\,\big(\delta\nabla^\alpha\nabla^\beta\vartheta\big)^{(1,0)}\,
{}^{\ast}\!R^{(0)}_{\alpha(\mu\nu)\beta}\\
&-4\,\big(\nabla^{\alpha}\nabla^{\beta}\vartheta\big)\,
{}^{\ast}\!R^{(1,0)}_{\alpha(\mu\nu)\beta}\\
&-4\,\big(\nabla_\sigma\vartheta\big)\,
\epsilon^{\sigma\delta\alpha}{}_{(\mu|}\nabla_{\alpha}R^{(1,0)}_{|\nu)\delta} ,
\end{aligned}
\label{eq:Csource}
\end{equation}
where now $\vartheta\equiv\vartheta^{(0,1/2)}$. The Hessian perturbation
$\big(\delta\nabla^\alpha\nabla^\beta\vartheta\big)^{(1,0)}$ is given by
eq.~\eqref{eq:dHess} with the replacement $\varphi\to\vartheta$ and
$h_{\mu\nu}\to h^{(1,0)}_{\mu\nu}$. The last term is the first term of
eq.~\eqref{eq:Ctensor}, which vanishes on the Kerr background but not here,
since the Gauss--Bonnet deformation has the Ricci tensor
\begin{equation}
R^{(1,0)}_{\mu\nu}
= \tfrac12\,\partial_\mu\varphi^{(1/2,0)}\partial_\nu\varphi^{(1/2,0)}
- \mathcal{H}^{(1,0)}_{\mu\nu},
\label{eq:dRic10}
\end{equation}
by eq.~\eqref{eq:einstein10} below and the tracelessness of
$\mathcal{H}^{(1,0)}_{\mu\nu}$ on a Ricci-flat background. The mirror-image structure of
eqs.~(\ref{eq:Ccurv}--\ref{eq:Csource}) and
eqs.~(\ref{eq:Hcurv}--\ref{eq:Hsource}) is the algebraic expression of the
parity mixing. In each sector, the deformation produced by the opposite parity
channel modifies how that sector's scalar gravitates.

We emphasize one point of implementation. The perturbation of the dual Riemann
tensor is \emph{not} obtained by perturbing the Riemann tensor and contracting
with the background Levi--Civita tensor, because the volume element $\sqrt{-g}$ entering
the dual responds to the metric deformation as well. In our implementation both
the Riemann tensor and its dual are perturbed by finite differencing the full
geometric construction in the deformation amplitude, which captures every
contribution automatically. We have verified that this agrees with independent
analytic expansions to $\sim\!10^{-10}$.

\subsection{Scalar equations order by order}
\label{app:exp-scalars}

Expanding the dilaton equation~\eqref{eq:phiapp} at leading order gives
\begin{equation}
\Box^{(0)}\varphi^{(1/2,0)} + \GB^{(0)} = 0,
\label{eq:phi10}
\end{equation}
with $\GB^{(0)}$ as the Gauss--Bonnet invariant on the Kerr background. At mixed
order,
\begin{equation}
\Box^{(0)}\varphi^{(1/2,1)}
+ \Box^{(0,1)}\varphi^{(1/2,0)} + \GB^{(0,1)} = 0,
\label{eq:phi11}
\end{equation}
where $\GB^{(0,1)}$ is the first-order perturbation of the Gauss--Bonnet
invariant induced by the dCS metric deformation $h^{(0,1)}_{\mu\nu}$, and
$\Box^{(0,1)}$ the corresponding perturbation of the d'Alembertian. Both source
terms are fixed by lower-order solutions, so eq.~\eqref{eq:phi11} is a linear
equation for $\varphi^{(1/2,1)}$ with the \emph{same} background operator
$\Box^{(0)}$ that appears at leading order. On the Ricci-flat Kerr background
the perturbation of the invariant is
\begin{equation}
\GB^{(0,1)} = -\tfrac12\,h^{(0,1)}\,\GB^{(0)}
+ 4R^{(0)\mu\nu\rho\sigma}\,\nabla^{(0)}_\rho\nabla^{(0)}_\nu h^{(0,1)}_{\mu\sigma},
\label{eq:GB01}
\end{equation}
with $h^{(0,1)} \equiv g^{(0)\mu\nu}h^{(0,1)}_{\mu\nu}$ being the trace of the
deformation. The axion equation~\eqref{eq:thetaapp} expands in exactly the same way, with
$\pRR$ in place of $\GB$ and the roles of the two couplings exchanged, giving a
linear equation for $\vartheta^{(1,1/2)}$ sourced by the sGB deformation.

\subsection{Einstein equations order by order}
\label{app:exp-einstein}

Substituting the expansions into eq.~\eqref{eq:einsteinapp} and collecting
powers of the couplings gives a hierarchy of linear problems. At leading order in
each sector,
\begin{align}
G^{(1,0)}_{\mu\nu} + \mathcal{H}^{(1,0)}_{\mu\nu}
&= \tfrac12\,T^{(\varphi)(1,0)}_{\mu\nu},
\label{eq:einstein10}\\
G^{(0,1)}_{\mu\nu} + \mathcal{C}^{(0,1)}_{\mu\nu}
&= \tfrac12\,T^{(\vartheta)(0,1)}_{\mu\nu},
\label{eq:einstein01}
\end{align}
where $G^{(1,0)}_{\mu\nu}$ denotes the linearized Einstein tensor acting on
$h^{(1,0)}_{\mu\nu}$, and the stress-energy tensors are quadratic in the
leading scalars, e.g.
\begin{equation}
T^{(\vartheta)(0,1)}_{\mu\nu}
= \nabla^{(0)}_\mu\vartheta^{(0,1/2)}\nabla^{(0)}_\nu\vartheta^{(0,1/2)}
- \tfrac12 g^{(0)}_{\mu\nu}\left[\nabla\vartheta^{(0,1/2)}\right]^2 .
\end{equation}
Equations~(\ref{eq:einstein10}--\ref{eq:einstein01}) determine the pure-sector
deformations and reproduce the solutions of Ref.~\cite{LamChungYunes2025}, which
we take as input.

Since $\mathcal{H}^{(\varphi)}_{\mu\nu}$ is linear in the dilaton and
$\mathcal{C}_{\mu\nu}$ depends only on the axion, their expansions satisfy
$\mathcal{H}^{(0,1)}_{\mu\nu}=0$ and $\mathcal{C}^{(1,0)}_{\mu\nu}=0$. Neither
correction tensor contributes to the opposite sector at leading order.

At mixed order, collecting all terms proportional to $\zGB\zCS$ and moving
everything determined by lower orders to the right-hand side gives the equation
quoted in the main text,
\begin{equation}
\mathcal{E}\!\left[h^{(1,1)}\right]_{\mu\nu} = S^{(1,1)}_{\mu\nu},
\end{equation}
where $\mathcal{E}$ is the same linearized Einstein operator on the Kerr
background as appears in eqs.~(\ref{eq:einstein10}--\ref{eq:einstein01}),
and
\begin{equation}
\begin{aligned}
    S^{(1,1)}_{\mu\nu} &= \tfrac12 T^{(1,1)}_{\mu\nu}
- \mathcal{H}^{(1,1)}_{\mu\nu}
- \mathcal{C}^{(1,1)}_{\mu\nu}\\
&- 2\,G^{(2)}_{\mu\nu}\!\left[h^{(1,0)},h^{(0,1)}\right].
\end{aligned}
\label{eq:sourceapp}
\end{equation}
The four contributions have distinct origins. The first is the mixed-order
stress-energy tensor. For the dilaton,
\begin{equation}
\begin{aligned}
T^{(\varphi)(1,1)}_{\mu\nu}
={}& 2\,\partial_{(\mu}\varphi^{(1/2,0)}\partial_{\nu)}\varphi^{(1/2,1)}\\
&- g^{(0)}_{\mu\nu}\,g^{(0)\rho\sigma}
\partial_\rho\varphi^{(1/2,0)}\partial_\sigma\varphi^{(1/2,1)}\\
&- \tfrac12\,h^{(0,1)}_{\mu\nu}\left[\nabla\varphi^{(1/2,0)}\right]^2\\
&+ \tfrac12\,g^{(0)}_{\mu\nu}\,h^{(0,1)\rho\sigma}\,
\partial_\rho\varphi^{(1/2,0)}\partial_\sigma\varphi^{(1/2,0)} ,
\end{aligned}
\label{eq:T11}
\end{equation}
and the axion contribution follows from taking
$\varphi\to\vartheta$, $(1/2,0)\to(0,1/2)$, $(1/2,1)\to(1,1/2)$, and
$h^{(0,1)}\to h^{(1,0)}$; $T^{(1,1)}_{\mu\nu}$ is their sum. The first two
lines in eq.~\eqref{eq:T11} are the cross terms between the leading and mixed scalars, and the last
two are the response of the leading stress-energy to the metric deformation of the
opposite sector. The second and third are the expanded
correction tensors, eqs.~\eqref{eq:Hsplit} and~\eqref{eq:Csplit}. The fourth is
the second-order Einstein tensor evaluated on the \emph{product} of the two
pure-sector deformations---the purely gravitational nonlinearity that couples the
sectors even in the absence of scalar cross terms.

\subsection{Solution strategy}
\label{app:exp-strategy}

The hierarchy is solved in two stages, each involving only background operators.
In Stage~I, the pure-sector solutions $h^{(1,0)}$, $h^{(0,1)}$,
$\varphi^{(1/2,0)}$, and $\vartheta^{(0,1/2)}$ are taken as input, and the mixed scalars
$\varphi^{(1/2,1)}$ and $\vartheta^{(1,1/2)}$ are obtained by inverting
$\Box^{(0)}$ against the sources in eq.~\eqref{eq:phi11} and its axion analogue.
With the mixed scalars in hand, in Stage II the source
$S^{(1,1)}_{\mu\nu}$ appearing in eq.~\eqref{eq:sourceapp} is assembled and the linearized
Einstein equation inverted to give $h^{(1,1)}_{\mu\nu}$ ,subject to the asymptotic conditions~\eqref{eq:peeling11}.

The essential structural feature is that each stage uses the same
background operators---$\Box^{(0)}$ and $\mathcal{E}$---differing only in the
source. The nonlinearity of the original system has been traded for a sequence of
linear problems with progressively more elaborate right-hand sides. In practice
Stage~I is computationally inexpensive while Stage~II dominates the cost, as it requires
constructing and inverting the linearized Einstein operator on the full
two-dimensional spectral grid.

\subsection{Validation of the mixed source}
\label{app:exp-audit}

Since we have no preexisting external benchmark for the mixed sector, the correctness of the
assembled source $S^{(1,1)}_{\mu\nu}$ of eq.~\eqref{eq:sourceapp} must be
verified internally. We do so through independent checks on the ingredients
from which the source is built, and through the one component of the field
equations that is not used in the solution.

\emph{Field-equation residuals.} The source is assembled from the linearized
Einstein operator $\mathcal{E}$, the two correction tensors
$\mathcal{H}_{\mu\nu}$ and $\mathcal{C}_{\mu\nu}$, and the scalar stress-energy tensor.
Each of these enters the \emph{pure}-sector
equations~(\ref{eq:einstein10}--\ref{eq:einstein01}), whose solutions are known. We
insert the pure-sector metric and scalar fields into these equations, evaluate
the difference of the two sides at the interior collocation nodes, and define
the relative residual
\begin{equation}
\varrho_{\rm GB} \equiv
\frac{\max\left|\mathcal{E}[h^{(1,0)}]_{\mu\nu} + \mathcal{H}^{(1,0)}_{\mu\nu}
- \tfrac12 T^{(\varphi)(1,0)}_{\mu\nu}\right|}
{\max\left|\mathcal{H}^{(1,0)}_{\mu\nu}\right|},
\label{eq:pureresidual}
\end{equation}
where each maximum is taken over all tensor components and all interior nodes;
$\varrho_{\rm CS}$ is defined in analogously---via eq.~\eqref{eq:einstein01},
normalized by $\mathcal{C}^{(0,1)}_{\mu\nu}$. This is a maximum-element norm of
the residual field, not an operator norm. At $30\times30$ resolution we find
$\varrho_{\rm GB}\approx(0.5$--$3)\times10^{-11}$ and
$\varrho_{\rm CS}\approx(1$--$6)\times10^{-11}$ for $a\le0.76$. Both then grow
by a factor of about three for each increment of $0.02$ in spin, reaching
$\varrho_{\rm GB}\approx9\times10^{-8}$ and
$\varrho_{\rm CS}\approx2\times10^{-7}$ at $a=0.9$. The low-spin plateau lies
some five orders of magnitude above $\epsilon_{\rm mach}$, as expected once
round-off is amplified by the second-derivative operators and the near-horizon
factors of the ansatz, whereas we attribute the growth beyond $a\approx0.76$ to the
finite resolution of the pure-sector inputs near the horizon. These residuals
validate the linearized operator, both correction tensors, and the
stress-energy components simultaneously. In particular, they fix the overall normalization
of each correction tensor; an incorrect overall coefficient in the leading dCS
correction tensor would produce a residual of order unity and is thereby excluded.

\emph{Trace-free property.} Both leading-order correction tensors are trace
free on the Kerr background. We verify that
$\big|g^{\mu\nu}\mathcal{H}^{(1,0)}_{\mu\nu}\big|$ and
$\big|g^{\mu\nu}\mathcal{C}^{(0,1)}_{\mu\nu}\big|$ are at the $\lesssim 10^{-13}$ level
relative to the largest component of each tensor, putting them at the level of round-off errors.

\emph{Perturbed curvature.} The mixed source contains the perturbations of the
Riemann and dual Riemann tensors by the pure-sector deformations,
eqs.~\eqref{eq:Hsource} and~\eqref{eq:Csource}. We compute these by finite
differencing the full geometric construction in the deformation amplitude and
cross-check against an independent analytic expansion, finding agreement to
$\sim\!10^{-10}$; this in particular confirms the correct treatment of the
volume-element response in the dual Riemann tensor.

\emph{Linearization identities.} The first-order expansion of a tensor in a
metric perturbation is an identity that holds for any perturbation, whether or
not it solves a field equation. We therefore test each linearized expression
that enters the source---$\mathcal{E}[h]$, $\Box^{(0,1)}$, $\GB^{(0,1)}$, the
corresponding perturbation of $\pRR$, eq.~\eqref{eq:Hsource}, and the first two
lines of eq.~\eqref{eq:Csource}---against a central finite difference of the
exact nonlinear expression evaluated on $g^{(0)}_{\mu\nu}\pm\epsilon\,h_{\mu\nu}$,
for a smooth synthetic $h_{\mu\nu}$ and synthetic scalars. Each agrees to
$\sim\!10^{-8}$, the accuracy of the finite difference, and the bilinear
$G_{\mu\nu}$ converges quadratically in $\epsilon$ to the exact mixed
second difference.

\emph{Reduction.} Setting either coupling to zero in the source assembly returns
the corresponding pure-sector source identically, confirming that no spurious
mixed contribution is generated at first order.

\emph{Covariant conservation.} The linearized contracted Bianchi identity gives
$\nabla^{(0)\mu}\mathcal{E}[h]_{\mu\nu}=0$ identically on the Ricci-flat
background, so a consistent source must satisfy
$\nabla^{(0)\mu}S^{(1,1)}_{\mu\nu}=0$. We do not evaluate this divergence
pointwise; the source grows as $\Delta^{-2}$ toward the horizon, and the
numerical derivative of such a field is dominated by differentiation error. The
same condition is tested, without differentiating the source, by the
$r\chi$-component of the field equations. Only the
$tt$-, $t\phi$-, $rr$-, $\chi\chi$-, and $\phi\phi$-components are collocated in the solution algorithm
(appendix~\ref{app:gauge}). Once these five hold, the $r$- and $\chi$-components
of the Bianchi identity reduce to first-order equations for the residual of the
$r\chi$-component alone, sourced by $\nabla^{(0)\mu}S^{(1,1)}_{\mu\nu}$, so that residual
vanishes only if the source is conserved. Relative to the largest component of
$S^{(1,1)}_{\mu\nu}$ it is below $3\times10^{-8}$ for $a\le0.76$ and
$5\times10^{-6}$ at $a=0.9$ at $30\times30$ resolution, and $10^{-9}$ at $a=0.9$
with $50\times50$, comparable to the residuals of the collocated components
(section~\ref{sec:validation}). The test is sensitive to every part of the source.
Omitting the inverse-metric terms of eq.~\eqref{eq:dHess}, the Ricci-dependent
terms of eqs.~\eqref{eq:Hsource} and~\eqref{eq:Csource}, the trace term of
eq.~\eqref{eq:GB01}, or the last two lines of eq.~\eqref{eq:T11}, or reversing
the sign of the Ricci-dependent terms, raises the $r\chi$-component residual to $10^{-2}$
or more. Moreover, the first law (section~\ref{sec:entropy}) provides a further, global
check of the solution.

\section{Conditioning of the Stage-II Operator}
\label{app:conditioning}

This appendix documents the ill-conditioning of the mixed-order metric system at
high spin, the equilibration procedure used to cure it, and the numerical
evidence that equilibration is both necessary and sufficient for reliable
results.

\subsection{Origin of the ill-conditioning}
\label{app:cond-origin}

Discretizing the linearized Einstein equation~\eqref{eq:mixedEQ} on the
spectral grid produces a linear system
\begin{equation}
\mathbf{A}\mathbf{c} = \mathbf{b},
\label{eq:linsys}
\end{equation}
where $\mathbf{c}$ collects the spectral coefficients $c^{(i)}_{n\ell}$ of the
four mixed metric functions $H_i^{(1,1)}$, the rows of $\mathbf{A}$ contain the
linearized Einstein operator evaluated at the interior collocation nodes
together with the boundary rows, and $\mathbf{b}$ holds the source
$S^{(1,1)}_{\mu\nu}$ and the boundary data. Since the number of collocation
constraints exceeds the number of retained coefficients, eq.~\eqref{eq:linsys}
is overdetermined and is solved in the least-squares sense.

The difficulty is that the rows of $\mathbf{A}$ are wildly disparate in
magnitude. The metric ansatz~\eqref{eq:ansatz} contains the factor
$\Sigma/\Delta$ in the $rr$-component, and the source~\eqref{eq:source} inherits
powers of the same factor through the second-order Einstein tensor. Since
$\Delta(r_+) = 0$, these factors diverge as the horizon is approached, and the
collocation rows nearest $z=+1$ acquire magnitudes many orders larger than those
sampling the bulk. At the other end of the grid the operator decays as a power
of $1/r$, and the rows at large radius are smaller still; at $a=0.9$ and
$30\times30$ resolution, the largest entries of the individual rows range from
$2\times10^{-8}$ to $2\times10^{7}$. The disparity worsens with spin; as $a\to1$ the horizon
radius $r_+ = 1+\sqrt{1-a^2}$ shrinks toward $a$, the metric functions steepen
near the horizon, and the near-horizon rows increasingly dominate the system.

Since the least-squares solution minimizes the \emph{total} residual
$\lVert\mathbf{A}\mathbf{c}-\mathbf{b}\rVert_2$, rows with the largest
magnitudes are weighted most heavily. The fit therefore expends its degrees of
freedom satisfying the near-horizon equations at the expense of the weakly
weighted ones at large radius, and the resulting coefficients are contaminated
by the amplified round-off of the dominant rows. The condition number quantifies the damage: we measure
$\kappa(\mathbf{A}) \sim 10^{14}\text{--}10^{15}$ at $a = 0.9$,
which, against the double-precision $\epsilon_{\mathrm{mach}}\approx
2\times10^{-16}$, leaves essentially no significant digits in the solution.

\subsection{Symptoms}
\label{app:cond-symptoms}

The failure of the naive solution is not signalled by erratic behavior.
Table~\ref{tab:trunc} lists the mixed horizon angular velocity
$\Omega_H^{(1,1)}$ at $a=0.9$, computed at $30\times30$ resolution as the
spectral truncation is varied. The naive values are stable to a few percent, yet
they differ from the equilibrated ones by a factor of three; the naive
$\kappa^{(1,1)}$ and $A^{(1,1)}$ are likewise $-6.1$ and $+650$, against
$-14.77$ and $+28.3$.

\begin{table}[htbp]
\caption{Mixed horizon angular velocity $\Omega_H^{(1,1)}$ at $a=0.9$,
$30\times30$ resolution, computed with the naive and equilibrated operators for
different values of the truncation parameter $t$; the $t$ highest radial modes
and the $t-1$ highest angular modes are omitted ($t=3$ in all quoted results).
The naive values are stable but wrong; the equilibrated values agree to better
than one part in $10^{4}$.}
\label{tab:trunc}
\begin{ruledtabular}
\begin{tabular}{ccccc}
 & \multicolumn{2}{c}{naive} & \multicolumn{2}{c}{equilibrated} \\
$t$ & $\kappa(\mathbf{A})$ & $\Omega_H^{(1,1)}$ &
$\kappa(\mathbf{D}_r\mathbf{A}\mathbf{D}_c)$ & $\Omega_H^{(1,1)}$ \\
\hline
2 & $1.3\times10^{15}$ & $+5.99$ & $4.5\times10^{5}$ & $+18.06736$ \\
3 & $5.6\times10^{14}$ & $+6.01$ & $3.2\times10^{5}$ & $+18.06736$ \\
4 & $3.0\times10^{14}$ & $+6.03$ & $2.6\times10^{5}$ & $+18.06739$ \\
5 & $2.0\times10^{14}$ & $+5.94$ & $2.1\times10^{5}$ & $+18.06754$ \\
6 & $1.4\times10^{14}$ & $+6.37$ & $1.8\times10^{5}$ & $+18.06856$ \\
\end{tabular}
\end{ruledtabular}
\end{table}

The naive solution fails where the rows of $\mathbf{A}$ are small. To compare
equations of very different magnitude we divide each residual by the largest
entry of its row. For the naive solution this scaled residual is of order
$10^{-5}$ for $r<2r_+$, but it exceeds the equally scaled source for
$r\gtrsim10\,r_+$ and is two orders of magnitude larger than it at
$r\sim100\,r_+$---meaning the field equations at large radius are not satisfied at all.
For the equilibrated solution the scaled residual is below $10^{-7}$ at every
node. Two global checks detect the same failure. The masses inferred from the
$1/r$ terms of $g_{tt}$ and of $g_{rr}$ differ by $5\times10^{-3}$ for the naive
solution, versus $2\times10^{-7}$ after equilibration, and the first law is
violated at order unity, with $S^{(1,1)}=+63.6$ from the Wald formula, versus
$-71.0$ from $\kappa$ and $\Omega_H$. Note that residual quoted only relative to the largest
component of the source is dominated by the near-horizon nodes and does not by
itself reveal the problem; by that measure the naive solution at $t=3$ has a
residual of $3\times10^{-8}$.

Grid refinement does not help. At $50\times50$ resolution the condition number of the naive
operator reaches $10^{16}$, about a hundred singular values fall below the
round-off threshold and are discarded, and $\Omega_H^{(1,1)}$ lies between
$+5.2$ and $+6.4$ depending on truncation. The problem is not a lack of
resolution; adding collocation points to an ill-conditioned system does not
improve the conditioning, and can worsen it by introducing additional
near-degenerate modes.

\subsection{Equilibration}
\label{app:cond-fix}

The cure is to rescale the system so that no equation or unknown dominates by
construction. We apply a two-sided diagonal scaling---row and column
equilibration~\cite{vanderSluis1969,GolubVanLoan2013}---before inversion. The
equilibrated overdetermined system is solved by the singular-value
decomposition~\cite{GolubVanLoan2013,Bjorck1996} (via the standard LAPACK
least-squares driver), which is rank-revealing and avoids the condition-number
squaring incurred by the normal equations. Each row is normalized by its largest entry,
\begin{equation}
(\mathbf{D}_r)_{jj} = \Big[\max_k \lvert A_{jk}\rvert\Big]^{-1},
\end{equation}
which places all equations on a common scale, and each column of the
row-scaled matrix is normalized by its Euclidean norm,
\begin{equation}
(\mathbf{D}_c)_{kk} =
\Big[\big\lVert (\mathbf{D}_r\mathbf{A})_{:k}\big\rVert_2\Big]^{-1},
\end{equation}
which balances the sensitivity of the fit to the different spectral
coefficients. The equilibrated system
\begin{equation}
\big(\mathbf{D}_r\mathbf{A}\mathbf{D}_c\big)\tilde{\mathbf{c}}
= \mathbf{D}_r\mathbf{b},
\qquad
\mathbf{c} = \mathbf{D}_c\tilde{\mathbf{c}},
\label{eq:equil}
\end{equation}
is solved in the least-squares sense and the solution transformed back. This is
a two-sided diagonal equilibration, not a similarity transformation. Together
with the corresponding rescaling $\mathbf{c}=\mathbf{D}_c\tilde{\mathbf{c}}$ of
the unknowns, it leaves the exact solution of the finite-dimensional linear
system unchanged, while improving the conditioning seen by the finite-precision
least-squares algorithm.

The improvement is dramatic. The condition number drops by nine orders of
magnitude,
\begin{equation}
\kappa(\mathbf{A}) \sim 10^{14}\text{--}10^{15}
\;\longrightarrow\;
\kappa(\mathbf{D}_r\mathbf{A}\mathbf{D}_c) \sim 10^{5},
\end{equation}
restoring roughly ten significant digits to the solution. The rightmost column
of Table~\ref{tab:trunc} shows the consequence: that the extracted
$\Omega_H^{(1,1)}$ varies by less than one part in $10^{4}$ with the
truncation, and by one part in $10^{6}$ between $t=2$ and $t=4$.

\subsection{Convergence after equilibration}
\label{app:cond-conv}

After equilibration the solution converges with resolution.
Table~\ref{tab:resconv} compares the mixed horizon observables computed at
$30\times30$ and $50\times50$ for $a=0.8$ and $a=0.9$. The two resolutions
agree to a few parts in $10^{5}$ or better, and the same holds for every
orbital and eikonal observable; the largest difference, $5\times10^{-5}$, is
in the retrograde $\lambda^{(1,1)}$ at $a=0.8$. At $50\times50$ the equilibrated
condition number is $\sim\!10^{6}$, and at $a=0.9$ the residuals of the
collocated and $r\chi$-equations are both $\sim 10^{-9}$.

\begin{table}[htbp]
\caption{Resolution convergence of the mixed horizon observables after
equilibration.}
\label{tab:resconv}
\begin{ruledtabular}
\begin{tabular}{lccc}
quantity & $30\times30$ & $50\times50$ & rel.\ diff. \\
\hline
$\Omega_H^{(1,1)}$, $a=0.8$ & $+3.451169$ & $+3.451169$ & $1\times10^{-7}$ \\
$\kappa^{(1,1)}$, $a=0.8$   & $-1.729117$ & $-1.729118$ & $2\times10^{-7}$ \\
$A^{(1,1)}$, $a=0.8$        & $-122.8273$ & $-122.8273$ & $1\times10^{-8}$ \\
$S^{(1,1)}$, $a=0.8$        & $-6.10852$  & $-6.10849$  & $5\times10^{-6}$ \\
$\Omega_H^{(1,1)}$, $a=0.9$ & $+18.06736$ & $+18.06736$ & $2\times10^{-7}$ \\
$\kappa^{(1,1)}$, $a=0.9$   & $-14.77431$ & $-14.77430$ & $9\times10^{-7}$ \\
$A^{(1,1)}$, $a=0.9$        & $+28.2747$  & $+28.2743$  & $1\times10^{-5}$ \\
$S^{(1,1)}$, $a=0.9$        & $-92.1698$  & $-92.1720$  & $2\times10^{-5}$ \\
\end{tabular}
\end{ruledtabular}
\end{table}

All results quoted in the main text employ equilibration. The equilibration is
a purely numerical remedy; it does not extend the domain of validity of the
perturbative expansion itself. At $a=0.99$ the equilibrated solution remains
well behaved; at $50\times50$ the residuals of the collocated and
$r\chi$-equations are $10^{-9}$ and $10^{-8}$, and $\Omega_H^{(1,1)}$,
$\kappa^{(1,1)}$, and $A^{(1,1)}$ agree with their $30\times30$ values to one
part in $10^{4}$. Yet the mixed metric functions reach
$\lvert H_i^{(1,1)}\rvert \approx 3\times10^{4}$, two orders of magnitude
larger than at $a=0.9$. The expansion~\eqref{eq:gexp} is then no longer
controlled for any but the smallest couplings, and we exclude that point from
our quoted results (section~\ref{sec:extremal}). The Wald entropy integral is, in
addition, not convergent at this spin; its Chern--Simons term changes sign
between the two resolutions, so the first law cannot be tested there.

\section{The $r\chi$ Metric Component and Residual Gauge Freedom}
\label{app:gauge}

The metric ansatz~\eqref{eq:ansatz} employs four functions $H_i(r,\chi)$ and
takes the $r\chi$-block to be diagonal. This appendix justifies that choice
and explains the role of the $r\chi$-component of the field equations, which is
not imposed in our solution.

\subsection{Circularity}
\label{app:gauge-circ}

A stationary, axisymmetric spacetime with Killing vectors $\partial_t$ and
$\partial_\phi$ is called \emph{circular} (or orthogonally transitive) if the
two-dimensional planes orthogonal to both Killing vectors are tangent to a
family of two-surfaces~\cite{Papapetrou1966,Carter1969,Wald1984}. Coordinates
$r$ and $\chi$ can then be chosen on these surfaces such that the components
$g_{tr}$, $g_{t\chi}$, $g_{r\phi}$, and $g_{\chi\phi}$ vanish, and the metric
separates into a $t\phi$-block and a $r\chi$-block. Equivalently, the
metric is invariant under the simultaneous reflection
$(t,\phi)\to(-t,-\phi)$. In vacuum general relativity circularity follows from
the field equations~\cite{Papapetrou1966,Carter1969}. It is not automatic in
modified theories, but it has been proven for stationary, axisymmetric,
asymptotically flat solutions of a broad class of effective theories of gravity
whenever the solution is obtained perturbatively from one of general
relativity~\cite{Xie2021}, which is the situation considered here and in
Ref.~\cite{LamChungYunes2025}. Consistent with this, the $tr$-, $t\chi$-, $r\phi$-, and
$\chi\phi$-components of the mixed-order source $S^{(1,1)}_{\mu\nu}$ vanish
identically, so that the block structure of eq.~\eqref{eq:ansatz} is preserved
at mixed order.

\subsection{The $r\chi$-block and the $r\chi$-component}
\label{app:gauge-rchi}

A symmetric metric has three components in the $r\chi$-block: $g_{rr}$,
$g_{\chi\chi}$, and $g_{r\chi}$. The ansatz~\eqref{eq:ansatz} retains only the
first two, both parametrized by the single function $H_3$. This is not a
loss of generality. Any two-dimensional metric is locally conformally flat; the
two coordinate freedoms of the block can be used to remove the off-diagonal
component and to fix the ratio of the diagonal ones, leaving a single conformal
factor. At the linearized level this statement takes the familiar form of an
infinitesimal diffeomorphism; under
\begin{equation}
x^\mu \to x^\mu + \xi^\mu, \qquad
h_{\mu\nu} \to h_{\mu\nu} - \nabla_\mu\xi_\nu - \nabla_\nu\xi_\mu,
\label{eq:gaugetrans}
\end{equation}
a generator with components $\xi^r(r,\chi)$ and $\xi^\chi(r,\chi)$ can be chosen
to set $h_{r\chi} = 0$ and
$h_{rr}/g^{(0)}_{rr} = h_{\chi\chi}/g^{(0)}_{\chi\chi} \equiv H_3$ while
preserving the block structure. The metric component $h_{r\chi}$ therefore
carries no physical information; it is pure gauge.

The same is not true of the $r\chi$-component of the field equations. With
$h_{r\chi}=0$, the linearized Einstein tensor $\mathcal{E}[h]_{r\chi}$ built
from the four functions $H_i$ does not vanish, and neither does the source.
$S^{(1,1)}_{r\chi}$ is odd in $\chi$, as the $r\chi$-component of a tensor that
is symmetric under equatorial reflection must be, and its largest value is
$37\%$ of the largest component of $S^{(1,1)}_{\mu\nu}$ at $a=0.5$ and $10\%$
at $a=0.9$.
Equation~\eqref{eq:mixedEQ} therefore has six nontrivial components for four
unknown functions. They are not independent: the $r$- and $\chi$- components of
the contracted Bianchi identity relate them, leaving four independent equations
when the source is conserved. Reference~\cite{LamChungYunes2025} solves four
components and verifies that all six are satisfied. We collocate the five
$tt$-, $t\phi$-, $rr$-, $\chi\chi$-, and $\phi\phi$-components of
eq.~\eqref{eq:mixedEQ}, solve the resulting overdetermined system in the
least-squares sense, and do not impose the $r\chi$-component---so that we can use it as
an independent check (appendix~\ref{app:exp-audit}).

\subsection{Consequences for the observables}
\label{app:gauge-obs}

The gauge choice $h_{r\chi}=0$ does not restrict the observables reported in
this work.
\begin{itemize}
\item The horizon angular velocity $\Omega_H = -g_{t\phi}/g_{\phi\phi}$ and the
horizon area $A = \oint\sqrt{g_{\chi\chi}g_{\phi\phi}}$ involve only
$g_{\chi\chi}$ and components in the $t\phi$-block, making them invariant.
\item The surface gravity is built from the Killing vector
$\xi = \partial_t + \Omega_H\partial_\phi$ and its antisymmetrized derivative,
and the Wald entropy~\eqref{eq:wald} is an integral of scalars over the horizon
cross-section; both are invariant.
\item The orbital frequency and Lyapunov exponent of the equatorial circular
null orbit are invariant, and are determined by $g_{tt}$, $g_{t\phi}$,
$g_{\phi\phi}$, $g_{rr}$ and their radial derivatives at $\chi=0$. The ISCO and
photon-ring radii are the coordinate radii of invariantly defined orbits in the
gauge of eq.~\eqref{eq:ansatz}, as in Ref.~\cite{LamChungYunes2025}.
\end{itemize}

Since the $r\chi$-component of the field equations is not imposed, its
residual measures the accuracy of the solution and the consistency of the
source independently of the least-squares fit. It is quoted together with the
residual of the five collocated components in section~\ref{sec:validation}, and is
of comparable size; relative to the largest component of the source it is below
$3\times10^{-8}$ for $a\le0.76$ and $5\times10^{-6}$ at $a=0.9$ at
$30\times30$ resolution.

\section{Expansion of the Horizon Angular Velocity}
\label{app:omega}

This appendix derives eq.~\eqref{eq:omega11}, the mixed-order correction to the
horizon angular velocity, and explains why it contains cross terms built from
the pure-sector metric functions in addition to the linear response to
$h^{(1,1)}$.

\subsection{Multiplicative structure of the ansatz}
\label{app:omega-struct}

The horizon angular velocity of a stationary, axisymmetric black hole is the
angular velocity of the locally non-rotating observers at the horizon,
\begin{equation}
\Omega_H = -\,\frac{g_{t\phi}}{g_{\phi\phi}}\bigg|_{r=r_+}.
\label{eq:omegadef}
\end{equation}
For the Kerr metric this evaluates to $\Omega_H^{(0)} = a/(2Mr_+)$, which with
$M=1$ is $a/(2r_+)$.

The essential feature of the ansatz~\eqref{eq:ansatz} is that the deformations
of the $(t,\phi)$ sector are \emph{multiplicative}. The functions $H_2$ and
$H_4$ rescale the corresponding Kerr components rather than adding to them,
\begin{align}
g_{t\phi} &= g^{(0)}_{t\phi}\,\big(1 + H_2\big),
\label{eq:gtphi}\\
g_{\phi\phi} &= g^{(0)}_{\phi\phi}\,\big(1 + H_4\big),
\label{eq:gphiphi}
\end{align}
where $H_2$ and $H_4$ carry the full coupling dependence. From the expansion of
section~\ref{sec:mixed},
\begin{equation}
H_i = \zGB\,H_i^{(1,0)} + \zCS\,H_i^{(0,1)} + \zGB\zCS\,H_i^{(1,1)} + \dots
\label{eq:Hiexp}
\end{equation}
for each $i$. Substituting eqs.~(\ref{eq:gtphi}--\ref{eq:gphiphi}) into
eq.~\eqref{eq:omegadef}, the angular velocity becomes the Kerr value times a
\emph{ratio} of the two deformation factors,
\begin{equation}
\Omega_H = \Omega_H^{(0)}\,\frac{1 + H_2}{1 + H_4}\bigg|_{r_+}.
\label{eq:omegaratio}
\end{equation}
It is this ratio structure that generates the cross terms. Although the
underlying field equations are solved order by order---so that $h^{(1,1)}$ is a
genuinely linear response to the mixed source---the observable
$\Omega_H$ is a \emph{nonlinear} function of the metric, and its mixed-order
piece therefore receives contributions from products of lower-order quantities.

\subsection{Order-by-order expansion}
\label{app:omega-expand}

The functions $H_2$ and $H_4$ are themselves at least first order in the
couplings, so the ratio in eq.~\eqref{eq:omegaratio} is needed only through
second order in the $H_i$,
\begin{equation}
\frac{1+H_2}{1+H_4} = 1 + \big(H_2-H_4\big) - H_4\big(H_2-H_4\big) + \dots,
\label{eq:ratioexp}
\end{equation}
where the omitted terms are cubic in the $H_i$ and therefore do not contribute
at order $\zGB\zCS$. Inserting eq.~\eqref{eq:Hiexp}, the terms linear in a
single coupling come from $H_2-H_4$ alone and reproduce the pure-sector results,
\begin{equation}
\Omega_H^{(1,0)} = \frac{a}{2r_+}\big(H_2^{(1,0)}-H_4^{(1,0)}\big)\Big|_{r_+},
\label{eq:omega10app}
\end{equation}
and likewise for $\Omega_H^{(0,1)}$, in agreement with
Ref.~\cite{LamChungYunes2025}. The coefficient of $\zGB\zCS$ receives two
contributions. The term $H_2-H_4$ contributes $H_2^{(1,1)}-H_4^{(1,1)}$, the
linear response to the mixed metric correction. The product $-H_4(H_2-H_4)$
contributes the two ways of taking one first-order factor from each coupling,
$-H_4^{(1,0)}\big(H_2^{(0,1)}-H_4^{(0,1)}\big)
-H_4^{(0,1)}\big(H_2^{(1,0)}-H_4^{(1,0)}\big)$. Altogether,
\begin{align}
\Omega_H^{(1,1)} = \frac{a}{2r_+}\Big[
&\big(H_2^{(1,1)} - H_4^{(1,1)}\big) \nonumber\\
&- H_4^{(1,0)}\big(H_2^{(0,1)}-H_4^{(0,1)}\big) \nonumber\\
&- H_4^{(0,1)}\big(H_2^{(1,0)}-H_4^{(1,0)}\big)\Big]_{r_+},
\label{eq:omega11app}
\end{align}
which is eq.~\eqref{eq:omega11} with the products multiplied out. By
Eq.~\eqref{eq:omega10app}, the cross terms can equivalently be written as
$-H_4^{(1,0)}\,\Omega_H^{(0,1)} - H_4^{(0,1)}\,\Omega_H^{(1,0)}$ evaluated at the
horizon; each is the first-order angular-velocity correction from one coupling,
weighted by the $\phi\phi$-deformation from the other. At first order the ratio
contributes only the difference $H_2-H_4$; the cross terms are intrinsically a
mixed-order effect.

\subsection{Numerical evaluation and the zeroth-law check}
\label{app:omega-numerics}

Equation~\eqref{eq:omega11app} is evaluated directly from the spectral
coefficients. The horizon $z=+1$ is itself a collocation node of the
Chebyshev--Gauss--Lobatto grid, so the horizon values of the $H_i$ follow from
eq.~\eqref{eq:spectral} without extrapolation, and no derivatives of the $H_i$
are needed. This makes $\Omega_H^{(1,1)}$ the least numerically demanding of our
horizon observables.

The zeroth law of black hole thermodynamics requires $\Omega_H$ to be constant over
the horizon, so each coefficient in its expansion must be independent of
$\chi$. We use this as an internal check that needs no external benchmark. The
requirement is nontrivial, because the individual metric functions do vary over
the horizon; at $a=0.5$, for example, $H_2^{(1,0)}(r_+,\chi)$ varies by
$\approx0.27$, whereas the combination $H_2^{(1,0)}-H_4^{(1,0)}$ that enters
$\Omega_H^{(1,0)}$ is constant to $10^{-16}$, and the same holds for
$\Omega_H^{(0,1)}$. At mixed order the cancellation is between the two
contributions to eq.~\eqref{eq:omega11app}, neither of which is constant by
itself; at $a=0.5$ the linear response varies from $-0.18$ on the equator to
$-0.08$ toward the poles and the cross terms from $-0.16$ to $-0.26$, while
their sum is constant to $5\times10^{-9}$. We quote the value of
eq.~\eqref{eq:omega11app} at the equator and define the polar spread
$\Delta_\chi$ as the difference between its largest and smallest values over
the horizon nodes, omitting the pole and the two nodes adjacent to it at each
end. $\Delta_\chi$ is below $5\times10^{-8}$ for $a\le0.76$ and rises to
$10^{-5}$ at $a=0.9$ (Table~\ref{tab:omegadata}). Relative to
$|\Omega_H^{(1,1)}|$ it is below $10^{-6}$ at every spin, including the
neighborhood of the zero of $\Omega_H^{(1,1)}$.

Since the two contributions vary with $\chi$, the decomposition of
$\Omega_H^{(1,1)}$ into a linear response and cross terms is not itself
invariant; only their sum is. On the equator the cross terms are negative at
every spin, growing in magnitude from $-0.05$ at $a=0.1$ to $-1.6$ at $a=0.9$,
while the linear response changes sign at $a\approx0.57$. It is $-0.18$ at
$a=0.5$ and $+0.92$ at $a=0.7$. Away from the equator the cross terms are more
negative and the linear response reverses earlier, at $a\approx0.53$ near the
poles. This is the origin of the sign change discussed in
section~\ref{sec:signs}.

\subsection{Normalization and cross-validation}
\label{app:omega-xval}

Since the mixed horizon angular velocity is the one observable extracted from
the closed-form expression~\eqref{eq:omega11} rather than by finite differencing,
we state its normalization explicitly and validate it against a direct
extraction.

In the ansatz~\eqref{eq:ansatz} the deformation functions enter as
\begin{align}
g_{tt} &= g_{tt}^{(0)} + H_1, &
g_{t\phi} &= g_{t\phi}^{(0)}\,(1+H_2), \nonumber\\
g_{rr} &= g_{rr}^{(0)}\,(1+H_3), &
g_{\chi\chi} &= g_{\chi\chi}^{(0)}\,(1+H_3), \nonumber\\
g_{\phi\phi} &= g_{\phi\phi}^{(0)}\,(1+H_4),
\end{align}
with the Kerr background components (in units $M=1$)
$g_{t\phi}^{(0)} = -2ar(1-\chi^2)/\Sigma$ and
$g_{\phi\phi}^{(0)} = (1-\chi^2)[\,r^2+a^2+2a^2 r(1-\chi^2)/\Sigma\,]$. Thus
$H_2$, $H_3$, $H_4$ are \emph{fractional} (multiplicative) deformations of the
corresponding metric components and $H_1$ is the additive deformation of
$g_{tt}$; all four vanish on the Kerr background, $H_i^{(0)}=0$. In particular the
factor of two in $g_{t\phi}^{(0)}$ belongs to the Kerr metric itself; $H_2$
multiplies the full background $g_{t\phi}^{(0)}$, and carries no additional
numerical factor. This is the normalization according to which the cross-term coefficients
of eq.~\eqref{eq:omega11} are written.

To confirm that these coefficients are free of algebraic errors, we compare
eq.~\eqref{eq:omega11} against a direct second-mixed finite-difference
extraction of $\Omega_H = -g_{t\phi}/g_{\phi\phi}$ from the fully assembled
corrected metric, using the same stencil~\eqref{eq:kappaFD} employed for the
surface gravity in Appendix~\ref{app:kappa}, with step size $\epsilon=10^{-3}$.
Table~\ref{tab:omegaxval} shows that the two agree to five or six significant
figures across the spin range. The difference scales as $\epsilon^2$, as
expected for the truncation error of the finite difference, whereas a
misnormalized $t\phi$-term or an incorrect cross-term coefficient would produce
an $\mathcal{O}(1)$ discrepancy independent of $\epsilon$. This constitutes an
independent check of the closed-form expression and of its cross-term
structure.

\begin{table}[t]
\caption{Cross-validation of the closed-form mixed horizon angular velocity,
eq.~\eqref{eq:omega11}, against a direct second-mixed-difference extraction of
$\Omega_H$ from the corrected metric.}
\label{tab:omegaxval}
\begin{ruledtabular}
\begin{tabular}{cccc}
$a$ & eq.~\eqref{eq:omega11} & direct diff. & rel.\ diff. \\
\hline
0.1 & $-0.1475139$ & $-0.1475142$ & $1\times10^{-6}$ \\
0.3 & $-0.3773656$ & $-0.3773659$ & $9\times10^{-7}$ \\
0.5 & $-0.3367968$ & $-0.3367974$ & $2\times10^{-6}$ \\
0.7 & $+0.7959190$ & $+0.7959138$ & $6\times10^{-6}$ \\
0.9 & $+18.06736$  & $+18.06769$  & $2\times10^{-5}$ \\
\end{tabular}
\end{ruledtabular}
\end{table}

\section{Covariant Construction of the Surface Gravity}
\label{app:kappa}

This appendix describes our computation of the surface gravity and its
mixed-order correction. We reduce the covariant definition to an expression on
the horizon that is exact in the $H_i$, describe its numerical evaluation, and
document the validation against pure-sector benchmarks.

\subsection{Definition}
\label{app:kappa-def}

For a stationary black hole with a Killing horizon generated by
\begin{equation}
\xi = \partial_t + \Omega_H\,\partial_\phi,
\label{eq:killinggen}
\end{equation}
the surface gravity is defined through the norm of the gradient of $\xi$
evaluated on the horizon~\cite{Wald1984},
\begin{equation}
\kappa^2 = -\tfrac12\left(\nabla^\mu\xi^\nu\right)
\left(\nabla_\mu\xi_\nu\right)\bigg|_{r=r_+}.
\label{eq:kappadef}
\end{equation}
For the Kerr metric this yields the familiar result
$\kappa^{(0)} = (r_+-M)/(2Mr_+)$. Because $\xi$ is a Killing vector, its covariant
derivative is antisymmetric, and the Christoffel symbols drop out of the
antisymmetrized combination:
\begin{equation}
\nabla_\mu\xi_\nu = \partial_{[\mu}\xi_{\nu]}
\equiv K_{\mu\nu} = \tfrac12\left(\partial_\mu\xi_\nu
- \partial_\nu\xi_\mu\right).
\label{eq:Kdef}
\end{equation}
Equation~\eqref{eq:kappadef} then takes the form,
\begin{equation}
\kappa^2 = -\tfrac12\,g^{\mu\alpha}g^{\nu\beta}
K_{\mu\nu}K_{\alpha\beta},
\label{eq:kappa2}
\end{equation}
which requires only the metric, its inverse, and first derivatives of the
metric---no curvature tensors or other second derivatives.

\subsection{Rationale}
\label{app:kappa-why}

A closed-form expression for the first-order correction to $\kappa$, in terms of
the metric functions and their radial derivatives at the horizon, is given in
Ref.~\cite{LamChungYunes2025}. Such an expression is specific to first order. At
mixed order it would have to be supplemented by products of the pure-sector
functions, which arise from the nonlinearity of the definition in the same way
as the cross terms for $\Omega_H$ in appendix~\ref{app:omega}. We instead reduce
eq.~\eqref{eq:kappadef} to an expression on the horizon that is exact in the
$H_i$. It applies unchanged to the full corrected metric, so that the pure and
mixed corrections are obtained by the same procedure and validated by the same test;
and its first-order expansion reproduces the
pure-sector values of Ref.~\cite{LamChungYunes2025}
(apendix~\ref{app:kappa-valid}).

\subsection{Reduction to the horizon}
\label{app:kappa-numerics}

Equation~\eqref{eq:kappa2} cannot be evaluated directly at $r=r_+$, because the
inverse metric degenerates there in Boyer--Lindquist-type coordinates:
$g^{rr}\propto\Delta\to0$ while $g^{tt}$ diverges. We use instead the equivalent
form of the definition~\cite{Wald1984},
\begin{equation}
\kappa^2 = -\lim_{r\to r_+}
\frac{\nabla_\mu(\xi^2)\,\nabla^\mu(\xi^2)}{4\xi^{\nu}\xi_{\nu}},
\label{eq:kappalim}
\end{equation}
which follows from $\nabla_\mu(\xi^2) = -2\kappa\,\xi_\mu$ on the horizon. For
the metric~\eqref{eq:ansatz},
\begin{equation}
\xi^2 = -\frac{D}{g_{\phi\phi}} + g_{\phi\phi}\,(\Omega_H-\omega)^2,
\label{eq:xinorm}
\end{equation}
with $D \equiv g_{t\phi}^2 - g_{tt}\,g_{\phi\phi}$ and $\omega\equiv-g_{t\phi}/g_{\phi\phi}$. The horizon is the surface on
which $D$ vanishes, and it remains at $r=r_+$ for every $\chi$; in our solutions
the coefficient of $D(r_+,\chi)$ at each order is below $10^{-8}$ for
$a\le0.7$ and $5\times10^{-6}$ at $a=0.9$. Since $\omega-\Omega_H$ also
vanishes there, by the zeroth law, $\xi^2 = -(\partial_r D)\,(r-r_+)/g_{\phi\phi}$
to leading order in $r-r_+$. The radial component of the inverse metric is
$g^{rr} = \Delta/[(1+H_3)\Sigma]$, with $\Delta = \Delta'(r_+)\,(r-r_+)$ to the
same order, and the $\chi$ derivatives in eq.~\eqref{eq:kappalim} contribute
only at higher order. Equation~\eqref{eq:kappalim} therefore reduces to
eq.~\eqref{eq:kappahor} of the main text,
\begin{equation}
\kappa = \frac12\left[\frac{\Delta'(r_+)\,\partial_r D}
{(1+H_3)\,\Sigma\,g_{\phi\phi}}\right]^{1/2}_{r=r_+} .
\label{eq:kappahorapp}
\end{equation}
No expansion in the couplings has been made; the $H_i$ in
eq.~\eqref{eq:kappahorapp} are the exact correction functions.

The horizon $z=+1$ is a collocation node, so the values of the $H_i$ and of
their radial derivatives,
$\partial_r H_i = -[(1+z)^2/(2r_+)]\,\partial_z H_i$, are obtained there
directly from the spectral coefficients, and the Kerr functions and their
derivatives are known in closed form. Equation~\eqref{eq:kappahorapp} can thus be
evaluated on the horizon itself, separately at each $\chi$ node, with no
extrapolation. The alternative of evaluating eq.~\eqref{eq:kappa2} on the
interior nodes and extrapolating to $z=+1$ gives the same results at low spin
but loses accuracy at high spin: at $a=0.9$ it changes $\kappa^{(1,0)}$ by $4\%$
and $\kappa^{(1,1)}$ by $0.7\%$, and raises the polar spread of
$\kappa^{(1,1)}$ from $3\times10^{-5}$ to $0.1$.

\subsection{Extraction of the corrections}
\label{app:kappa-fd}

We extract the coupling corrections by finite differencing
eq.~\eqref{eq:kappahorapp}. Let $\kappa(\zGB,\zCS)$ denote the surface gravity
computed from the metric
\begin{equation}
\begin{aligned}
g_{\mu\nu}(\zGB,\zCS) ={}& g^{(0)}_{\mu\nu}
+ \zGB\,h^{(1,0)}_{\mu\nu} + \zCS\,h^{(0,1)}_{\mu\nu}\\
&+ \zGB\zCS\,h^{(1,1)}_{\mu\nu},
\end{aligned}
\label{eq:gofzeta}
\end{equation}
with $\zGB$ and $\zCS$ treated as two independent continuous parameters of
either sign. The pure and mixed corrections are then the derivatives of
$\kappa$ at $\zGB=\zCS=0$, which we evaluate by central differences with step
$\epsilon$:
\begin{align}
\kappa^{(1,0)} &= \frac{\kappa_{++} + \kappa_{+-}
- \kappa_{-+} - \kappa_{--}}{4\epsilon},\\
\kappa^{(0,1)} &= \frac{\kappa_{++} - \kappa_{+-}
+ \kappa_{-+} - \kappa_{--}}{4\epsilon},\\
\kappa^{(1,1)} &= \frac{\kappa_{++} - \kappa_{+-}
- \kappa_{-+} + \kappa_{--}}{4\epsilon^2},
\label{eq:kappaFD}
\end{align}
where $\kappa_{\pm\pm} \equiv \kappa(\zGB=\pm\epsilon,\,\zCS=\pm\epsilon)$, with
the first sign referring to $\zGB$. The mixed
coefficient~\eqref{eq:kappaFD} is the standard second mixed difference and
extracts the $\partial^2\kappa/\partial\zGB\,\partial\zCS$ term. We use
$\epsilon = 10^{-4}$. The truncation error is of relative order $\epsilon^2$
times the square of the pure-sector amplitudes; comparison with
$\epsilon=10^{-5}$ shows that it is $10^{-6}$ of $\kappa^{(1,1)}$ at $a=0.9$
and smaller at lower spins, whereas with $\epsilon=10^{-3}$ it would reach
$10^{-4}$. This approach requires no analytic
expansion of eq.~\eqref{eq:kappahorapp}, and automatically includes all cross
terms arising from the nonlinearity of the definition---the analogue, for
$\kappa$, of the ratio cross terms derived explicitly for $\Omega_H$ in
appendix~\ref{app:omega}.

\subsection{Validation}
\label{app:kappa-valid}

Three checks establish the reliability of this construction.
First, at zero coupling eq.~\eqref{eq:kappahorapp} reduces analytically to the
Kerr surface gravity. With $D^{(0)}=\Delta\,(1-\chi^2)$ and
$g^{(0)}_{\phi\phi}(r_+)=4r_+^2(1-\chi^2)/\Sigma$ it gives
$\kappa^{(0)}=\Delta'(r_+)/(4r_+)=(r_+-1)/(2r_+)$.

Second, the first differences reproduce the pure-sector values of
Ref.~\cite{LamChungYunes2025}. Table~\ref{tab:kappaval} lists a representative
comparison. The two agree to six significant digits for $a\le0.7$ and to five
at $a=0.9$, where the remaining difference of $5\times10^{-6}$ is the truncation
error of the step $\epsilon=10^{-4}$. The pure-sector coefficients are produced
by the same routine, in the same evaluation, as $\kappa^{(1,1)}$ at every spin,
so this comparison accompanies each mixed result.

Third, expanding eq.~\eqref{eq:kappahorapp} to first order in the $H_i$ gives,
for either pure sector and with $M=1$,
\begin{equation}
\begin{aligned}
\frac{\kappa^{(m,n)}}{\kappa^{(0)}} ={}& \Big\{H_2 - \tfrac12 H_3 - \tfrac12 H_4\\
&+ \frac{r_+^2}{(r_+-1)\,\Sigma^2}\Big[2(r_+-1)(H_4-2H_2)\\
&\quad+ \partial_r\big[a^2(1-\chi^2)(2H_2-H_4) - \Sigma\,H_1\big]\Big]\Big\}_{r_+} ,
\end{aligned}
\label{eq:kappalin}
\end{equation}
with $H_i\to H_i^{(m,n)}$. The corresponding expression in Ref.~\cite{LamChungYunes2025}
contains a transcription error, but the computed numerical values are correct and
agree with ours. Evaluated on the pure-sector functions,
eq.~\eqref{eq:kappalin} agrees with the first differences to the truncation
error of the latter, reproduces the values of Ref.~\cite{LamChungYunes2025} in
Table~\ref{tab:kappaval} to six significant digits, and is constant over the
horizon to $10^{-13}$.

\begin{table}[htbp]
\caption{Pure-sector surface gravity corrections computed from
eq.~\eqref{eq:kappahorapp} with $\epsilon=10^{-4}$, compared with the values of
Ref.~\cite{LamChungYunes2025}. The agreement validates the on-horizon
evaluation and the finite-difference extraction used in the mixed calculation.}
\label{tab:kappaval}
\begin{ruledtabular}
\begin{tabular}{ccccc}
$a$ & $\kappa^{(1,0)}$ (this work) & Ref.~\cite{LamChungYunes2025} &
$\kappa^{(0,1)}$ (this work) & Ref.~\cite{LamChungYunes2025} \\
\hline
0.1 & $0.150496$ & $0.150496$ & $0.0029843$ & $0.0029843$ \\
0.3 & $0.137925$ & $0.137925$ & $0.0281728$ & $0.0281728$ \\
0.5 & $0.114738$ & $0.114738$ & $0.0874487$ & $0.0874487$ \\
0.7 & $0.099370$ & $0.099370$ & $0.2158198$ & $0.2158200$ \\
0.9 & $0.385930$ & $0.385928$ & $0.6950605$ & $0.6950590$ \\
\end{tabular}
\end{ruledtabular}
\end{table}

Finally, as with $\Omega_H$, the zeroth law requires $\kappa$ to be constant over
the horizon, and we use the residual polar spread as an internal error estimate
(figure~\ref{fig:kappa}). For $\kappa^{(1,1)}$ it is below $2\times10^{-7}$ for
$a\le0.76$ and $3\times10^{-5}$ at $a=0.9$, two to three times that of
$\Omega_H^{(1,1)}$. The surface gravity is the more demanding of the two
because eq.~\eqref{eq:kappahorapp} involves radial \emph{derivatives} of the
metric functions at the horizon, whereas $\Omega_H$ is built from horizon
\emph{values} alone. This ordering is not
peculiar to the mixed sector; Ref.~\cite{LamChungYunes2025} observes the same
hierarchy for the pure sectors, where the $\chi$-dependence of the first-order
correction to $\kappa$ is likewise larger than that of the first-order
correction to $\Omega_H$.

\section{Circular and Non-Equatorial Orbits}
\label{app:orbits}

This appendix collects the geodesic machinery underlying the ISCO,
photon-ring, and non-equatorial photon-orbit observables discussed in
section~\ref{sec:orbital}, describes the numerical solution of the orbit conditions, and documents the validation against the exact Kerr results.

\subsection{Equatorial reduction}
\label{app:orbits-eq}

Both curvature invariants in the action~\eqref{eq:action} respect reflection
symmetry about the equatorial plane, so the corrected metric functions are even
in $\chi$ and the equatorial plane remains a totally geodesic surface; a
particle or photon released in the plane $\chi=0$ with no polar velocity remains
there. Equatorial orbits are therefore consistently described by the reduced
metric obtained by setting $\chi=0$ in eq.~\eqref{eq:ansatz}. With $M=1$ the
three relevant components are
\begin{align}
g_{tt} &= -\left(1 - \frac{2}{r}\right) + H_1,
\label{eq:eqgtt}\\
g_{t\phi} &= -\frac{2a}{r}\,\big(1 + H_2\big),
\label{eq:eqgtphi}\\
g_{\phi\phi} &= \left(r^2 + a^2 + \frac{2a^2}{r}\right)\big(1 + H_4\big),
\label{eq:eqgphiphi}
\end{align}
where each $H_i$ is understood to be evaluated at $\chi=0$ and to carry the full
coupling dependence of eq.~\eqref{eq:Hiexp}. Only these three components and
their radial derivatives are required; the $r\chi$-block enters no equatorial
circular-orbit condition (appendix~\ref{app:gauge}).

To evaluate eqs.~(\ref{eq:eqgtt}--\ref{eq:eqgphiphi}) we project the spectral
representation~\eqref{eq:spectral} onto the equator. Because $P_\ell(0)$ is known
in closed form, the double sum collapses to a single Chebyshev series in $z$,
\begin{equation}
H_i(z,\chi=0) = \sum_n \left[\sum_\ell c^{(i)}_{n\ell}P_\ell(0)\right]T_n(z)
\equiv \sum_n \tilde{c}^{(i)}_n\,T_n(z),
\end{equation}
so that $H_i$ and $dH_i/dr$ at any radius follow from a one-dimensional
Chebyshev evaluation, with $dH_i/dr = (dH_i/dz)(dz/dr)$ and
$dz/dr = -2r_+/r^2$ from eq.~\eqref{eq:zmap}. The orbits of interest lie outside
the horizon, $r_+ < r_{\rm ph} < r_{\rm ISCO}$, in the interior of the spectral
grid.

\subsection{Circular orbits}
\label{app:orbits-circ}

For a stationary, axisymmetric, circular spacetime the Killing vectors
$\partial_t$ and $\partial_\phi$ give two conserved quantities per geodesic, the
specific energy $E = -u_\mu(\partial_t)^\mu$ and angular momentum
$L = u_\mu(\partial_\phi)^\mu$. For a circular orbit at radius $r$ the angular
velocity $\Omega = \dot\phi/\dot t$ follows from the radial geodesic equation,
which for circular motion reduces to the algebraic condition
$g'_{tt} + 2\Omega g'_{t\phi} + \Omega^2 g'_{\phi\phi} = 0$,
with primes denoting $d/dr$. Solving this
quadratic gives eq.~\eqref{eq:orbfreq},
\begin{equation}
\Omega_\pm(r) = \frac{-\,g_{t\phi}' \pm
\sqrt{\left(g_{t\phi}'\right)^2 - g_{tt}'\,g_{\phi\phi}'}}
{g_{\phi\phi}'},
\label{eq:orbfreqapp}
\end{equation}
with the two roots corresponding to prograde ($+$)
and retrograde ($-$) motion. The two branches are distinguished only by the sign
of the square root. Because $g_{t\phi}'$ enters the $\pm\Omega_\pm$ with opposite
relative signs, a correction to the metric shifts the two senses of rotation
differently (section~\ref{sec:asymmetry}).

For a timelike circular orbit the normalization $u^\mu u_\mu = -1$ gives the
specific energy
\begin{equation}
E(r) = -\,\frac{g_{tt} + \Omega\,g_{t\phi}}
{\sqrt{-\left(g_{tt} + 2\Omega\,g_{t\phi} + \Omega^2 g_{\phi\phi}\right)}},
\label{eq:Ecirc}
\end{equation}
with $\Omega = \Omega_\pm(r)$ from eq.~\eqref{eq:orbfreqapp}. The quantity under
the square root is the negative of the norm of the orbital four-velocity direction and is
positive wherever circular timelike orbits exist; its vanishing marks the photon
orbit, as discussed below.

\subsection{The ISCO}
\label{app:orbits-isco}

The innermost stable circular orbit is the marginally stable circular orbit.
Equivalently---and this is the prescription we use for locating it---it is the radius at which the
energy of the circular-orbit sequence is extremal,
\begin{equation}
\frac{dE}{dr}\bigg|_{r=r_{\rm ISCO}} = 0.
\label{eq:iscocond}
\end{equation}
For $r > r_{\rm ISCO}$ the circular orbits
are stable and $E(r)$ increases outward, while at $r_{\rm ISCO}$ the sequence
turns over and stability is lost.

We evaluate $E(r)$ from eq.~\eqref{eq:Ecirc}, differentiate it numerically with a
centered difference of step size $\epsilon=10^{-5}$, scan $dE/dr$ on a fine radial grid
spanning from just outside the horizon to $r=12$, identify the innermost sign
change, and refine the root with Brent's method to a tolerance of $10^{-10}$.
Scanning for the innermost bracket, rather than seeding a local solver, ensures
that the same branch of the solution is tracked at all four coupling points
entering the finite-difference stencil---essential for the differences to be
meaningful.

\subsection{The photon ring}
\label{app:orbits-photon}

A circular null geodesic satisfies $u^\mu u_\mu = 0$, which for equatorial
circular motion reduces to the vanishing of the same combination that appears
under the square root in eq.~\eqref{eq:Ecirc}:
\begin{equation}
g_{tt} + 2\,\Omega_\pm\, g_{t\phi} + \Omega_\pm^2\, g_{\phi\phi} = 0,
\label{eq:photoncondapp}
\end{equation}
with $\Omega_\pm$ again from eq.~\eqref{eq:orbfreqapp}. This is the photon-sphere
condition quoted as eq.~\eqref{eq:photoncond}. It is solved by the same
scan-and-bracket procedure, over $r$ from just outside the horizon to $r=8$.

The contrast with the ISCO is worth emphasizing, as it explains the different
numerical behavior of the two observables. Equation~\eqref{eq:photoncondapp} is
an \emph{algebraic} condition on the metric components and their first
derivatives, whereas eq.~\eqref{eq:iscocond} requires an additional numerical derivative of
$E(r)$---effectively a second derivative of the metric. Each derivative amplifies
the spectral error, so the ISCO is intrinsically more sensitive to the accuracy
of $h^{(1,1)}$ than the photon ring. This is borne out numerically; between
$30\times30$ and $50\times50$ resolutions the mixed photon-ring shifts at $a=0.8$
and $a=0.9$ agree to $5\times10^{-7}$ or better, and the mixed ISCO shifts to
$2\times10^{-5}$. The ISCO is also the more sensitive of the two to the coupling
step (section.~\ref{app:orbits-fd}).

\subsection{Non-equatorial photon orbits}
\label{app:orbits-sph}
 
For the orbits of section~\ref{sec:sphph} we integrate Hamilton's equations for
$\mathcal{H}=\tfrac12 g^{\mu\nu}p_\mu p_\nu$ in the coordinates
$(t,r,\theta,\phi)$, with $p_t=-1$ and $p_\phi=b$ fixed. The metric functions
$H_i(r,\theta)$ are evaluated from their spectral coefficients on a $400\times401$
grid in $(r,\theta)$ and represented by quintic splines, which supply the first
derivatives entering the equations of motion; the integration uses an
eighth-order Dormand--Prince scheme with relative tolerance $10^{-12}$.
 
Photon orbits are unstable, so we do not integrate over many polar cycles.
Instead we start at the northern turning point, $\theta=\theta_{\rm t}$, with
$p_r=p_\theta=0$; the null condition then fixes $b$ as a function of the turning
radius $r_{\rm t}$, with the root chosen by the orbital sense. We integrate to the
first crossing of the equator and adjust $r_{\rm t}$ by Brent's method until $p_r$
vanishes there, to a tolerance of $10^{-14}$. By the two symmetries of
section~\ref{sec:sphph} this quarter orbit generates the full orbit; its time
reverse carries the photon to the southern turning point, and reflection through
the equator returns it to the north, with the azimuth advancing by four times the
quarter-orbit value per polar cycle. The illustrations in figure~\ref{fig:sphph}
are built in this way from a single integration, which avoids the exponential
growth of errors along an unstable orbit.
 
With the couplings set to zero the procedure recovers the Kerr spherical-orbit
radius $r_s(\theta_{\rm t})$~\cite{Teo2003} to $10^{-12}$, with $r_{\rm turn}=r_{\rm eq}$
to the same accuracy. The coefficients are extracted with the four-point stencil
of appendix~\ref{app:orbits-fd} at $\epsilon=10^{-3}$; repeating with
$\epsilon=2\times10^{-3}$ and Richardson extrapolating changes the mixed
coefficients by less than $2\times10^{-6}$. At the equatorial ends, extrapolating
from $\theta_{\rm t}=89^\circ$ and $89.5^\circ$ in $\cos^2\theta_{\rm t}$ gives
$r^{(1,1)}=-0.48441$ (prograde) and $-1.89685$ (retrograde), against $-0.4844$ and
$-1.8968$ from the photon-ring condition (Table~\ref{tab:photondata}), with the
same agreement for $r^{(0)}$, $r^{(1,0)}$, and $r^{(0,1)}$. This agreement also
confirms that the spline representation of the metric introduces no error at the
level of the quoted digits.

\subsection{Extraction of the corrections}
\label{app:orbits-fd}

As for the surface gravity, the coupling corrections are obtained by finite
differencing rather than analytic expansion. Writing $r_\bullet(\zGB,\zCS)$
for either root radius computed from the metric~\eqref{eq:gofzeta},
\begin{align}
r_\bullet^{(1,0)} &= \frac{r_{++} + r_{+-} - r_{-+} - r_{--}}{4\epsilon},\\
r_\bullet^{(0,1)} &= \frac{r_{++} - r_{+-} + r_{-+} - r_{--}}{4\epsilon},\\
r_\bullet^{(1,1)} &= \frac{r_{++} - r_{+-} - r_{-+} + r_{--}}{4\epsilon^2},
\end{align}
with $r_{\pm\pm} \equiv r_\bullet(\zGB=\pm\epsilon,\,\zCS=\pm\epsilon)$ and
$\epsilon = 10^{-3}$. The prograde and retrograde branches are differenced
independently. The step cannot be made much smaller, because the tolerance of
the root finding enters the second difference divided by $\epsilon^2$. The
truncation error of the mixed coefficient is of order $\epsilon^2$ and grows
with spin. At $a=0.5$ the prograde ISCO gives
$r_{\rm ISCO}^{(1,1)} = -4.2258$, $-4.2256$, and $-4.2249$ for
$\epsilon = 2\times10^{-3}$, $10^{-3}$, and $5\times10^{-4}$ respectively,
i.e.\ agreement to four significant digits. At $a=0.9$ the same steps give
$-3.152$, $-3.203$, and $-3.215$, and removing the leading truncation term by
Richardson extrapolation,
$[4\,r^{(1,1)}_\bullet(\epsilon) - r^{(1,1)}_\bullet(2\epsilon)]/3$, gives
$-3.219$. The value at $\epsilon=10^{-3}$ is thus $0.5\%$ too small in
magnitude, which is the largest truncation error among the orbital radii; for
the photon-ring shifts at $a=0.9$ it is below $0.1\%$ (section~\ref{sec:errors}).

\subsection{Validation}
\label{app:orbits-valid}

Setting $\zGB=\zCS=0$ reduces eqs.~(\ref{eq:eqgtt}--\ref{eq:eqgphiphi}) to
the equatorial Kerr metric, and the two root-finding procedures must then return
the exact Kerr radii. For the ISCO the analytic result is~\cite{Bardeen1972}
\begin{equation}
r_{\rm ISCO}^{\pm} = 3 + Z_2 \mp \sqrt{(3-Z_1)(3+Z_1+2Z_2)},
\end{equation}
with $Z_1 = 1 + (1-a^2)^{1/3}\left[(1+a)^{1/3}+(1-a)^{1/3}\right]$ and
$Z_2 = \sqrt{3a^2+Z_1^2}$, while the photon radii are
$r_{\rm ph}^{\pm} = 2\left\{1+\cos\left[\tfrac23\arccos(\mp a)\right]\right\}$.
Table~\ref{tab:orbitval} compares our zero-coupling output with these
expressions for two spins. At each of the $41$ spins of the sweep the two agree
to $4\times10^{-9}$ for the ISCO and $2\times10^{-11}$ for the photon ring, for
both orbital senses, which validates the equatorial projection, the root
finding, and the branch tracking---all of which are used unchanged in the
corrected case.

\begin{table}[htbp]
\caption{Zero-coupling validation of the equatorial orbit solver against the
exact Kerr results~\cite{Bardeen1972}. Radii are in units of $M$.}
\label{tab:orbitval}
\begin{ruledtabular}
\begin{tabular}{ccccc}
$a$ & orbit & sense & this work & exact \\
\hline
0.5 & ISCO   & prograde   & $4.2330$ & $4.2330$ \\
0.5 & ISCO   & retrograde & $7.5546$ & $7.5546$ \\
0.5 & photon & prograde   & $2.3473$ & $2.3473$ \\
0.5 & photon & retrograde & $3.5321$ & $3.5321$ \\
0.9 & ISCO   & prograde   & $2.3209$ & $2.3209$ \\
0.9 & ISCO   & retrograde & $8.7174$ & $8.7174$ \\
0.9 & photon & prograde   & $1.5579$ & $1.5579$ \\
0.9 & photon & retrograde & $3.9103$ & $3.9103$ \\
\end{tabular}
\end{ruledtabular}
\end{table}

\section{Eikonal Quasinormal Frequencies}
\label{app:qnm}

This appendix collects the geodesic construction underlying the eikonal
quasinormal-mode corrections of section~\ref{sec:qnm} and documents the validation
of the machinery against the exact Schwarzschild and Kerr results.

\subsection{The correspondence}
\label{app:qnm-corr}

In the eikonal (geometric-optics) limit, the quasinormal ringing of a black hole
admits an interpretation in terms of the unstable circular null geodesics of the
spacetime~\cite{Cardoso2009,Berti2009}. A wave packet of high angular momentum
$\ell$ is trapped near the photon sphere; its orbital frequency sets the real
part of the mode, while the rate at which the trapped packet leaks away---governed
by the instability timescale of the photon orbit---sets the imaginary part. For a
stationary, axisymmetric black hole the correspondence reads
\begin{equation}
\omega_{\ell n} \simeq \Omega_c\left(\ell+\tfrac12\right)
- i\left(n+\tfrac12\right)\lvert\lambda\rvert,
\label{eq:eikonalapp}
\end{equation}
where $\Omega_c$ is the orbital angular velocity of the circular null geodesic
and $\lambda$ its principal Lyapunov exponent, both evaluated at the photon
radius $r_{\rm ph}$. Equation~\eqref{eq:eikonalapp} is asymptotically exact as
$\ell\to\infty$ and provides a useful approximation for the low-$\ell$ modes of
astrophysical interest, with accuracy improving as $\ell$ grows.

Since both ingredients are properties of the equatorial photon orbit, they
follow directly from the machinery of appendix~\ref{app:orbits}. In particular
the mixed-order corrections $\Omega_c^{(1,1)}$ and $\lambda^{(1,1)}$ inherit the
conditioning of the photon-ring calculation, in which the photon radius is fixed
by the algebraic condition~\eqref{eq:photoncondapp}.
\subsection{Orbital frequency}
\label{app:qnm-freq}

The first ingredient requires no new machinery. The orbital frequency of the
photon orbit is simply eq.~\eqref{eq:orbfreqapp} evaluated at the photon radius,
\begin{equation}
\Omega_c = \Omega_\pm\!\left(r_{\rm ph}^{\pm}\right),
\label{eq:omegac}
\end{equation}
with $r_{\rm ph}^{\pm}$ the prograde and retrograde roots of
eq.~\eqref{eq:photoncondapp}. As with the orbital radii, the prograde and
retrograde branches are treated independently throughout.

\subsection{Lyapunov exponent}
\label{app:qnm-lyap}

The second ingredient measures the instability of the photon orbit against
radial perturbations. For a null geodesic in the equatorial plane, the conserved
energy $E$ and angular momentum $L$ define the impact parameter $b = L/E$, and
the radial motion is governed by
\begin{equation}
\dot r^2 = R(r)\equiv \frac{g_{\phi\phi} + 2b\,g_{t\phi} + b^2 g_{tt}}
{g_{rr}\left(g_{t\phi}^2 - g_{tt}\,g_{\phi\phi}\right)},
\label{eq:radialpot}
\end{equation}
where the overdot denotes differentiation with respect to an affine parameter and
we have normalized to $E=1$. The circular null orbit sits at a maximum of the
effective radial potential $-R(r)$, $R(r_{\rm ph}) = R'(r_{\rm ph}) = 0$ with
$R''(r_{\rm ph}) > 0$, and is therefore unstable, meaning that a small radial displacement
will grow exponentially.

The critical impact parameter of the photon orbit follows from the orbital
frequency,
\begin{equation}
b_c = -\,\frac{g_{t\phi} + \Omega_c\, g_{\phi\phi}}
{g_{tt} + \Omega_c\, g_{t\phi}}\bigg|_{r_{\rm ph}},
\label{eq:bcrit}
\end{equation}
and the growth rate of the radial deviation, measured in coordinate time, is the
principal Lyapunov exponent
\begin{equation}
\lambda = \sqrt{\frac{R''(r_{\rm ph})}{2\,\dot t^{\,2}}},
\qquad
\dot t = \frac{g_{\phi\phi} + b_c\,g_{t\phi}}
{g_{t\phi}^2 - g_{tt}\,g_{\phi\phi}}\bigg|_{r_{\rm ph}}.
\label{eq:lyapapp}
\end{equation}
The factor $\dot t^{\,2}$ converts the affine-parameter growth rate into one
measured by the asymptotic observer, which is the quantity appearing in
eq.~\eqref{eq:eikonalapp}. The second derivative $R''(r_{\rm ph})$ is evaluated
from eq.~\eqref{eq:radialpot} with the five-point centered stencil
\begin{equation}
R'' \approx \frac{-R_{+2} + 16R_{+1} - 30R_{0} + 16R_{-1} - R_{-2}}{12h^2},
\label{eq:fivepoint}
\end{equation}
where $R_{k} \equiv R(r_{\rm ph}+kh)$ and $h=10^{-2}$, for which both the
truncation and the round-off errors are
below $10^{-10}$. A wide stencil is needed because the error of $R''$ is divided
by $4\epsilon^2$ in the mixed coupling difference. With a three-point difference
of step size $10^{-4}$ the round-off error of $R''$, $\sim\!10^{-8}$, leaves noise of
several $10^{-3}$ in $\lambda^{(1,1)}$, comparable to the retrograde values
themselves. Since $R(r)$ requires $g_{rr}$---and hence the metric function
$H_3$---this is the one equatorial observable in our set that draws on the
$r\chi$-block, though still only through a diagonal component.

The corrections $\Omega_c^{(1,1)}$ and $\lambda^{(1,1)}$ are extracted by the
same second mixed finite difference used for the orbit radii
(appendix~\ref{app:orbits-fd}), with $\epsilon=10^{-3}$, treating the prograde
and retrograde branches separately. Comparison with the Richardson-extrapolated
values shows that the truncation error of this step is below $10^{-3}$ for
$a\le0.8$. At $a=0.9$ it is $2\times10^{-4}$ and $8\times10^{-4}$ for the
prograde $\Omega_c^{(1,1)}$ and $\lambda^{(1,1)}$, and $4\times10^{-3}$ and
$6\times10^{-3}$ for the retrograde ones, which are small in magnitude
(section~\ref{sec:errors}).

\subsection{Validation}
\label{app:qnm-valid}

At zero coupling the construction must reproduce the known Kerr eikonal
frequencies. Two independent checks are available.
The first is the static limit. For the Schwarzschild spacetime the photon sphere sits at
$r_{\rm ph}=3M$ and the eikonal quantities take the exact values
\begin{equation}
\Omega_c = \lambda = \frac{1}{3\sqrt3\,M} \approx 0.19245,
\end{equation}
the coincidence of the two reflecting the special structure of the
Schwarzschild potential. Our construction reproduces this to four digits: at
$a=0.05$ we obtain $\lambda = 0.19241$ (prograde) and $0.19242$ (retrograde),
converging to the exact value as $a\to0$. This is a nontrivial test of
eqs.~(\ref{eq:radialpot}--\ref{eq:lyapapp}), since $\Omega_c$ and $\lambda$ are
computed by entirely different routes---one from the orbital frequency, the other
from the curvature of the radial potential---yet must agree in this limit.

\begin{table}[htbp]
\caption{Zero-coupling eikonal quantities from the construction of
eqs.~(\ref{eq:omegac}--\ref{eq:lyapapp}). The Schwarzschild limit
$\Omega_c=\lambda=1/(3\sqrt3)\approx0.19245$ is recovered as $a\to0$, and the
Kerr spin dependence reproduces the known behavior, in particular the decrease
of the prograde damping rate toward extremality.}
\label{tab:qnmval}
\begin{ruledtabular}
\begin{tabular}{ccccc}
$a$ & sense & $r_{\rm ph}$ & $\Omega_c$ & $\lambda$ \\
\hline
0.05 & prograde   & $2.9417$ & $+0.19625$ & $0.19241$ \\
0.05 & retrograde & $3.0572$ & $-0.18884$ & $0.19242$ \\
0.5  & prograde   & $2.3473$ & $+0.24412$ & $0.18592$ \\
0.5  & retrograde & $3.5321$ & $-0.16292$ & $0.19009$ \\
0.9  & prograde   & $1.5579$ & $+0.35157$ & $0.13608$ \\
0.9  & retrograde & $3.9103$ & $-0.14636$ & $0.18655$ \\
\end{tabular}
\end{ruledtabular}
\end{table}

The second is the Kerr spin dependence. Table~\ref{tab:qnmval} lists the
zero-coupling values across the spin range. Both trends are the established
ones; the prograde frequency $\Omega_c$ increases with spin as frame dragging
speeds the co-rotating photon, while the prograde Lyapunov exponent
\emph{decreases}, reproducing the well-known lengthening of the prograde
ringdown as extremality is approached ($\lambda\to0$ in the extremal limit). The
retrograde quantities vary far more weakly, as expected, since the retrograde
photon orbit remains far from the horizon at all spins. At each of the $41$
spins of the sweep the zero-coupling values agree with those obtained from the
exact Kerr metric with analytic derivatives to $4\times10^{-12}$ for $\Omega_c$
and $5\times10^{-9}$ for $\lambda$.

The mixed corrections converge under grid refinement. At $a=0.9$---the
most demanding case---the $30\times30$ and $50\times50$ solutions give prograde
$\Omega_c^{(1,1)}$ and $\lambda^{(1,1)}$ that agree to $2\times10^{-7}$. The
prograde eikonal coefficients thus converge better than the prograde ISCO shift
at the same spin ($4\times10^{-6}$), despite the photon orbit sitting
closer to the horizon ($r_{\rm ph}=1.56$ versus $r_{\rm ISCO}=2.32$). The reason
is the one identified in appendix~\ref{app:orbits-photon}: that the photon radius is
fixed algebraically rather than by a derivative condition. The retrograde
coefficients are small in magnitude, $|\Omega_c^{(1,1)}|,|\lambda^{(1,1)}|<0.06$;
the retrograde $\Omega_c^{(1,1)}$ agrees to $2\times10^{-7}$ and the retrograde
$\lambda^{(1,1)}$ to $5\times10^{-5}$, the largest fractional difference between the two
resolutions among all the mixed observables.

\subsection{Scope}
\label{app:qnm-scope}

However, two limitations of eq.~\eqref{eq:eikonalapp} should be kept in mind when
interpreting our results. First, the correspondence is asymptotic in $\ell$; for
the low multipoles accessible to current ringdown observations the eikonal
formula captures the leading behavior but not the exact frequencies, and our
$\Omega_c^{(1,1)}$, $\lambda^{(1,1)}$ should correspondingly be read as leading
estimates of the mixed-order ringdown shift rather than exact mode corrections.

Second, and more fundamentally, the eikonal construction is blind to the mode
structure. In general relativity the axial and polar gravitational perturbations
of a Kerr black hole are isospectral, and the eikonal limit returns a single
frequency for both. For a non-rotating black hole the sGB coupling ties the scalar perturbation to
the polar sector and the dCS coupling ties the pseudoscalar perturbation to the
axial sector, and isospectrality is broken already in each pure theory. The
combined action is invariant under parity when $\vartheta$ transforms as a
pseudoscalar, and the corrected background shares that symmetry, so the two
parity sectors remain decoupled from one-another at mixed order. What is new
is that, for a rotating black hole, each sector couples the gravitational
perturbations to both scalar fields at once, and the frequency shift at
$\Oone$ receives contributions from second-order perturbation theory in the
two couplings in addition to those of the mixed background. Capturing these
requires solving the coupled perturbation equations on the corrected
background, which is beyond the scope of the present work.
\\

\section{Tabulated Results}
\label{app:data}

This appendix collects the numerical values underlying
figures~\ref{fig:omega}--\ref{fig:qnm}, at intervals of $0.1$ in spin. All
quantities are given in units $M=1$, at fixed mass and angular momentum,
and are computed at $30\times30$ spectral resolution with the equilibrated
Stage-II operator of appendix~\ref{app:conditioning}. The surface gravity and
the first-law entropy use the coupling step $\epsilon=10^{-4}$; the orbital and
eikonal coefficients use $\epsilon=10^{-3}$, whose truncation error is
quantified in appendices~\ref{app:orbits-fd} and~\ref{app:qnm-lyap}.
Pure-sector values are
listed alongside the mixed corrections, both because they are of independent
interest and because they constitute the running benchmark against
Ref.~\cite{LamChungYunes2025} described in section~\ref{sec:validation}.

\onecolumngrid
\begin{table*}[htbp]
\caption{Horizon angular velocity. The mixed correction $\Omega_H^{(1,1)}$ is
quoted at the equator, with $\Delta_\chi$ the polar spread over the horizon,
which vanishes by the zeroth law and serves as an internal error estimate
(appendix~\ref{app:omega}). The mixed correction changes sign at
$a\approx0.608$. The pure-sector values reproduce those of
Ref.~\cite{LamChungYunes2025}.}
\label{tab:omegadata}
\begin{ruledtabular}
\begin{tabular}{ccccc}
$a$ & $\Omega_H^{(1,1)}$ & $\Delta_\chi$ &
$\Omega_H^{(1,0)}$ & $\Omega_H^{(0,1)}$ \\
\hline
0.1 & $-0.1475$ & $2.1\times10^{-10}$ & $+2.614\times10^{-2}$ & $-3.968\times10^{-2}$ \\
0.2 & $-0.2796$ & $1.1\times10^{-10}$ & $+5.161\times10^{-2}$ & $-8.005\times10^{-2}$ \\
0.3 & $-0.3774$ & $3.5\times10^{-9}$ & $+7.551\times10^{-2}$ & $-1.220\times10^{-1}$ \\
0.4 & $-0.4130$ & $1.2\times10^{-9}$ & $+9.640\times10^{-2}$ & $-1.666\times10^{-1}$ \\
0.5 & $-0.3368$ & $4.6\times10^{-9}$ & $+1.116\times10^{-1}$ & $-2.162\times10^{-1}$ \\
0.6 & $-0.03852$ & $2.3\times10^{-8}$ & $+1.154\times10^{-1}$ & $-2.747\times10^{-1}$ \\
0.7 & $+0.7959$ & $3.3\times10^{-8}$ & $+9.406\times10^{-2}$ & $-3.522\times10^{-1}$ \\
0.8 & $+3.451$ & $6.8\times10^{-8}$ & $+4.395\times10^{-3}$ & $-4.788\times10^{-1}$ \\
0.9 & $+18.07$ & $9.5\times10^{-6}$ & $-3.665\times10^{-1}$ & $-8.035\times10^{-1}$ \\
\end{tabular}
\end{ruledtabular}
\end{table*}


\begin{table*}[htbp]
\caption{Surface gravity, in the same format as Table~\ref{tab:omegadata},
evaluated on the horizon (appendix~\ref{app:kappa}). The mixed correction
changes sign at $a\approx0.710$. The polar spread $\Delta_\chi$ is two to three
times that of $\Omega_H^{(1,1)}$.}
\label{tab:kappadata}
\begin{ruledtabular}
\begin{tabular}{ccccc}
$a$ & $\kappa^{(1,1)}$ & $\Delta_\chi$ &
$\kappa^{(1,0)}$ & $\kappa^{(0,1)}$ \\
\hline
0.1 & $+0.01925$ & $2.8\times10^{-9}$ & $0.15050$ & $2.984\times10^{-3}$ \\
0.2 & $+0.07570$ & $4.2\times10^{-9}$ & $0.14575$ & $1.215\times10^{-2}$ \\
0.3 & $+0.1646$ & $1.1\times10^{-8}$ & $0.13793$ & $2.817\times10^{-2}$ \\
0.4 & $+0.2744$ & $1.1\times10^{-8}$ & $0.12730$ & $5.243\times10^{-2}$ \\
0.5 & $+0.3768$ & $9.0\times10^{-9}$ & $0.11474$ & $8.745\times10^{-2}$ \\
0.6 & $+0.3945$ & $4.5\times10^{-8}$ & $0.10276$ & $1.382\times10^{-1}$ \\
0.7 & $+0.07380$ & $2.5\times10^{-8}$ & $0.09937$ & $2.158\times10^{-1}$ \\
0.8 & $-1.729$ & $4.0\times10^{-7}$ & $0.13484$ & $3.520\times10^{-1}$ \\
0.9 & $-14.77$ & $2.8\times10^{-5}$ & $0.38593$ & $6.951\times10^{-1}$ \\
\end{tabular}
\end{ruledtabular}
\end{table*}

\begin{table*}[htbp]
\caption{Horizon area. The uncorrected area $A^{(0)}$ agrees with the Kerr value
$8\pi r_+$ to $\sim\!10^{-14}$ at all spins. The mixed correction changes sign
at $a\approx0.888$.}
\label{tab:areadata}
\begin{ruledtabular}
\begin{tabular}{ccccc}
$a$ & $A^{(0)}$ & $A^{(1,0)}$ & $A^{(0,1)}$ & $A^{(1,1)}$ \\
\hline
0.1 & $50.13950$ & $-61.540$ & $-0.25783$ & $-3.0709$ \\
0.2 & $49.75770$ & $-61.431$ & $-1.0457$ & $-12.312$ \\
0.3 & $49.10785$ & $-61.232$ & $-2.4091$ & $-27.778$ \\
0.4 & $48.16728$ & $-60.908$ & $-4.4310$ & $-49.412$ \\
0.5 & $46.89833$ & $-60.384$ & $-7.2452$ & $-76.652$ \\
0.6 & $45.23893$ & $-59.486$ & $-11.055$ & $-107.20$ \\
0.7 & $43.08111$ & $-57.745$ & $-16.142$ & $-132.83$ \\
0.8 & $40.21239$ & $-53.548$ & $-22.704$ & $-122.83$ \\
0.9 & $36.08785$ & $-37.939$ & $-28.422$ & $+28.275$ \\
\end{tabular}
\end{ruledtabular}
\end{table*}

\begin{table*}[htbp]
\caption{Entropy. $S^{(0)}=A^{(0)}/4$ is the Kerr value, and $S^{(1,0)}$,
$S^{(0,1)}$, and $S^{(1,1)}$ are the Wald entropy coefficients of
section~\ref{sec:entropy}. The last column, $S^{(1,1)}_{\rm FL}$, is the mixed coefficient implied by
$\kappa$ and $\Omega_H$ through the first law, eq.~\eqref{eq:Sfirstlaw}. The
mixed coefficient changes sign at $a\approx0.762$.}
\label{tab:entropydata}
\begin{ruledtabular}
\begin{tabular}{cccccc}
$a$ & $S^{(0)}$ & $S^{(1,0)}$ & $S^{(0,1)}$ & $S^{(1,1)}$ & $S^{(1,1)}_{\rm FL}$ \\
\hline
0.1 & $12.5349$ & $+7.6307$ & $+0.05004$ & $+0.15815$ & $+0.15815$ \\
0.2 & $12.4394$ & $+7.5890$ & $+0.2041$ & $+0.63104$ & $+0.63104$ \\
0.3 & $12.2770$ & $+7.5198$ & $+0.4752$ & $+1.4097$ & $+1.4097$ \\
0.4 & $12.0418$ & $+7.4244$ & $+0.8888$ & $+2.4595$ & $+2.4594$ \\
0.5 & $11.7246$ & $+7.3081$ & $+1.492$ & $+3.6623$ & $+3.6622$ \\
0.6 & $11.3097$ & $+7.1883$ & $+2.373$ & $+4.6185$ & $+4.6184$ \\
0.7 & $10.7703$ & $+7.1238$ & $+3.723$ & $+3.8215$ & $+3.8215$ \\
0.8 & $10.0531$ & $+7.3474$ & $+6.038$ & $-6.1085$ & $-6.1085$ \\
0.9 & $9.0220$ & $+9.2848$ & $+11.38$ & $-92.170$ & $-92.169$ \\
\end{tabular}
\end{ruledtabular}
\end{table*}

\begin{table*}[htbp]
\caption{ISCO radii and their corrections, for prograde and retrograde orbits.
The uncorrected radii $r^{(0)}$ reproduce the Bardeen values~\cite{Bardeen1972}
to the digits shown. The mixed coefficients carry root-finding noise of about
$4\times10^{-4}$ (prograde) and up to $5\times10^{-3}$ (retrograde), so their
last digit is uncertain; at $a=0.9$ the prograde value has in addition a
truncation error of $0.5\%$ (appendix~\ref{app:orbits-fd}).}
\label{tab:iscodata}
\begin{ruledtabular}
\begin{tabular}{ccccccccc}
& \multicolumn{4}{c}{prograde} & \multicolumn{4}{c}{retrograde} \\
\cline{2-5}\cline{6-9}
$a$ & $r^{(0)}$ & $r^{(1,0)}$ & $r^{(0,1)}$ & $r^{(1,1)}$
    & $r^{(0)}$ & $r^{(1,0)}$ & $r^{(0,1)}$ & $r^{(1,1)}$ \\
\hline
0.1 & $5.66930$ & $+0.4554$ & $+0.01732$ & $-0.065$ & $6.32289$ & $+1.1545$ & $-0.10318$ & $-0.341$ \\
0.2 & $5.32944$ & $+0.2212$ & $-0.04155$ & $-0.543$ & $6.63904$ & $+1.6505$ & $-0.30813$ & $-1.050$ \\
0.3 & $4.97862$ & $+0.0484$ & $-0.17109$ & $-1.425$ & $6.94927$ & $+2.2714$ & $-0.64111$ & $-2.034$ \\
0.4 & $4.61434$ & $-0.0718$ & $-0.36842$ & $-2.677$ & $7.25427$ & $+3.0420$ & $-1.14727$ & $-3.069$ \\
0.5 & $4.23300$ & $-0.1453$ & $-0.63178$ & $-4.226$ & $7.55458$ & $+3.9849$ & $-1.91055$ & $-3.598$ \\
0.6 & $3.82907$ & $-0.1738$ & $-0.95682$ & $-5.871$ & $7.85069$ & $+5.0938$ & $-3.10754$ & $-1.996$ \\
0.7 & $3.39313$ & $-0.1461$ & $-1.32423$ & $-7.032$ & $8.14297$ & $+6.2060$ & $-5.18617$ & $+7.830$ \\
0.8 & $2.90664$ & $-0.0023$ & $-1.64610$ & $-5.931$ & $8.43176$ & $+6.2511$ & $-9.68242$ & $+60.81$ \\
0.9 & $2.32088$ & $+0.5832$ & $-1.42978$ & $-3.203$ & $8.71735$ & $-4.4829$ & $-26.1167$ & $+621.3$ \\
\end{tabular}
\end{ruledtabular}
\end{table*}

\begin{table*}[htbp]
\caption{Photon-ring radii and their corrections. The uncorrected radii
reproduce the exact Kerr photon-sphere values to the digits shown.}
\label{tab:photondata}
\begin{ruledtabular}
\begin{tabular}{ccccccccc}
& \multicolumn{4}{c}{prograde} & \multicolumn{4}{c}{retrograde} \\
\cline{2-5}\cline{6-9}
$a$ & $r^{(0)}$ & $r^{(1,0)}$ & $r^{(0,1)}$ & $r^{(1,1)}$
    & $r^{(0)}$ & $r^{(1,0)}$ & $r^{(0,1)}$ & $r^{(1,1)}$ \\
\hline
0.1 & $2.88219$ & $-0.1535$ & $+0.07946$ & $+0.1671$ & $3.11335$ & $+0.1078$ & $-0.09958$ & $-0.2895$ \\
0.2 & $2.75919$ & $-0.2294$ & $+0.14506$ & $+0.2084$ & $3.22281$ & $+0.3015$ & $-0.22710$ & $-0.6885$ \\
0.3 & $2.63003$ & $-0.2715$ & $+0.20260$ & $+0.1229$ & $3.32885$ & $+0.5465$ & $-0.39366$ & $-1.1637$ \\
0.4 & $2.49336$ & $-0.2787$ & $+0.25844$ & $-0.0967$ & $3.43184$ & $+0.8515$ & $-0.61692$ & $-1.6343$ \\
0.5 & $2.34730$ & $-0.2472$ & $+0.32072$ & $-0.4844$ & $3.53209$ & $+1.2254$ & $-0.92852$ & $-1.8968$ \\
0.6 & $2.18891$ & $-0.1682$ & $+0.40159$ & $-1.1655$ & $3.62985$ & $+1.6668$ & $-1.39419$ & $-1.3566$ \\
0.7 & $2.01333$ & $-0.0241$ & $+0.52171$ & $-2.5988$ & $3.72535$ & $+2.1168$ & $-2.18021$ & $+2.2316$ \\
0.8 & $1.81109$ & $+0.2228$ & $+0.72088$ & $-6.7354$ & $3.81876$ & $+2.1795$ & $-3.85834$ & $+21.822$ \\
0.9 & $1.55785$ & $+0.6622$ & $+1.08456$ & $-25.560$ & $3.91027$ & $-1.7698$ & $-9.98603$ & $+230.86$ \\
\end{tabular}
\end{ruledtabular}
\end{table*}

\begin{table*}[htbp]
\caption{Eikonal quasinormal-mode data. The columns $\Omega_c$ and $\lambda$ are
the uncorrected Kerr values of the photon-orbit frequency and Lyapunov exponent;
$\Omega_c^{(1,1)}$ and $\lambda^{(1,1)}$ are the mixed-order corrections. The
Schwarzschild limit $\Omega_c=\lambda=1/(3\sqrt3)\approx0.19245$ is recovered as
$a\to0$ (appendix~\ref{app:qnm}).}
\label{tab:qnmdata}
\begin{ruledtabular}
\begin{tabular}{ccccccccc}
& \multicolumn{4}{c}{prograde} & \multicolumn{4}{c}{retrograde} \\
\cline{2-5}\cline{6-9}
$a$ & $\Omega_c$ & $\Omega_c^{(1,1)}$ & $\lambda$ & $\lambda^{(1,1)}$
    & $\Omega_c$ & $\Omega_c^{(1,1)}$ & $\lambda$ & $\lambda^{(1,1)}$ \\
\hline
0.1 & $+0.2003$ & $-0.0043$ & $0.1923$ & $+0.0329$ & $-0.1854$ & $-0.0072$ & $0.1923$ & $-0.0229$ \\
0.2 & $+0.2091$ & $-0.0025$ & $0.1917$ & $+0.0761$ & $-0.1790$ & $-0.0153$ & $0.1920$ & $-0.0370$ \\
0.3 & $+0.2190$ & $+0.0105$ & $0.1907$ & $+0.1283$ & $-0.1732$ & $-0.0227$ & $0.1915$ & $-0.0431$ \\
0.4 & $+0.2306$ & $+0.0439$ & $0.1888$ & $+0.1828$ & $-0.1679$ & $-0.0286$ & $0.1908$ & $-0.0425$ \\
0.5 & $+0.2441$ & $+0.1151$ & $0.1859$ & $+0.2161$ & $-0.1629$ & $-0.0325$ & $0.1901$ & $-0.0363$ \\
0.6 & $+0.2605$ & $+0.2650$ & $0.1813$ & $+0.1489$ & $-0.1583$ & $-0.0341$ & $0.1893$ & $-0.0260$ \\
0.7 & $+0.2812$ & $+0.6207$ & $0.1739$ & $-0.3133$ & $-0.1541$ & $-0.0344$ & $0.1884$ & $-0.0146$ \\
0.8 & $+0.3089$ & $+1.7866$ & $0.1612$ & $-2.5465$ & $-0.1501$ & $-0.0372$ & $0.1875$ & $-0.0098$ \\
0.9 & $+0.3516$ & $+9.8431$ & $0.1361$ & $-17.336$ & $-0.1464$ & $-0.0596$ & $0.1865$ & $-0.0438$ \\
\end{tabular}
\end{ruledtabular}
\end{table*}

\end{document}